\documentclass[prd,amsmath,%showkeys,
twocolumn,floatfix,amssymb, preprintnumbers, nofootinbib, superscriptaddress]{revtex4}

\usepackage{amssymb, amsmath, amsfonts, bm}
\usepackage{euscript}
\usepackage{float}

\usepackage[colorlinks=true, pdfstartview=FitV, linkcolor=blue, citecolor=blue, urlcolor=blue]{hyperref}

\usepackage{pstricks,pst-node,pst-text,pst-3d}
\usepackage{graphicx}
\usepackage{color}
\usepackage[normalem]{ulem}
\usepackage{relsize} 
\usepackage{dsfont}
\usepackage[version=4]{mhchem}
\usepackage[makeroom]{cancel}
\usepackage{mathtools}
\usepackage{mathrsfs}
\usepackage{booktabs,tabulary}
\usepackage{mciteplus}
\usepackage{siunitx}

\DeclareMathOperator{\im}{Im}
\DeclareMathOperator{\re}{Re}

\DeclareMathOperator{\Tr}{Tr}
\newcommand*\diff{\mathop{}\!\mathrm{d}}

\newcommand{\be}{\begin{equation}}
\newcommand{\ee}{\end{equation}}
\newcommand{\bml}{\begin{multline}}
\newcommand{\eml}{\end{multline}}
\newcommand{\BM}{\begin{pmatrix}}
\newcommand{\EM}{\end{pmatrix}}

\begin{document}

\title{Momentum-space second-order pion-nucleus potential including medium effects in the $\Delta(1232)$ region}

\author{V. Tsaran}
\email{vitsaran@uni-mainz.de}
\author{M. Vanderhaeghen}
\address{Institut f\"ur Kernphysik \& PRISMA$^+$  Cluster of Excellence, Johannes Gutenberg Universit\"at,  D-55099 Mainz, Germany}

\date{\today }

\begin{abstract}

In this work, we develop an updated model for pion-nucleus scattering in the framework of the distorted wave impulse approximation in momentum space.  
We construct the second-order pion-nucleus potential, which involves analysis of pion-nucleus elastic scattering as a solution of the Lippmann-Schwinger equation. 
The potential is based on the individual pion-nucleon scattering amplitudes extracted from SAID, and its second-order correction is presented in detail. We estimate optimal energy-independent parameters of the potential by a multienergy fit of the pion-${}^{12}$C total, reaction, and differential elastic cross sections.
We show the predictive power by applying it to pion elastic scattering on ${}^{16}$O, ${}^{28}$Si, and ${}^{40}$Ca.
\end{abstract}

\maketitle

\section{Introduction}
\label{sec:intro}

The study of the pion-nucleus interaction has a long history filled with various theoretical approaches~\cite{Ericson:1988gk} and has seen a renewed interest in very recent years~\cite{DUET:2016yrf,PinzonGuerra:2018rju,LArIAT:2021yix}.
While the earlier works were concentrated on the pion-nucleus scattering and pionic atoms, modern experiments open new perspectives and challenges in applying the pion-nucleus reactions knowledge base.
The pion production experiments in photon (electron)- and neutrino-nucleus scattering serve as examples of utmost importance. 
They are related to the extraction of neutron skin and neutrino oscillation measurements, respectively.
The final-state interaction between the outgoing pion and nucleus in these two processes is non-negligible at the energies considered, and it is particularly significant in the $\Delta(1232)$ resonance region~\cite{Drechsel:1999vh, Nakamura:2009iq}.
Moreover, the $\Delta(1232)$ excitation is the dominant mechanism of single-pion production, implying the significance of studying modifications of the resonance in the nuclear medium.
For the neutrino experiments, a good understanding of the pion final-state interaction is paramount to interpret the measurements to the level of precision required \cite{Balantekin:2022jrq, Ankowski:2022thw}.

After the initial studies on pion-nucleus elastic scattering and energy levels of pionic atoms using the simple first-order potential, it became evident that higher-order effects are required for a consistent description of experimental data~\cite{Ericson:1966fm}. 
There are essentially two types of existing theoretical models. 
The first is based on multiple-scattering theory and provides terms beyond first order to the pion-nucleus optical potential, treating the pion-nucleon amplitudes phenomenologically.
The second approach is the isobar-doorway model, which considers the $\Delta$ resonance as an elementary particle modified by various medium corrections. 
Our work is inspired by both of these approaches.

The optical potential formalism effectively describes the many-body pion-nucleus scattering process by a one-particle equation for the pion interacting with a complex phenomenological potential. 
The Kisslinger optical potential~\cite{Kisslinger:1955zz}, built on Watson's theoretical basis~\cite{Watson:1953zz}, was introduced more than half a century ago and has been continuously improved over the years by including various corrections~\cite{Ericson:1966fm,Landau:1977th,Stricker:1979de,Carr:1982zz,Gmitro:1985kj,Gmitro:1987un}.
The Kerman-McManus-Thaler formulation of the multiple scattering theory~\cite{Kerman:1959fr}, treatment of the Fermi motion, and relativistic kinematics have been taken into account. 
The addition of the phenomenological term proportional to the squared nuclear density, which covers beyond-first-order effects and real pion absorption, has resulted in a much-improved agreement between theory and pion-nucleus scattering data for a large set of nuclei.

However, the properties of the $\Delta(1232)$ isobar in the nuclear medium are essential in understanding pion-nucleus interaction and have been the subject of numerous investigations, especially in the framework of the $\Delta$-hole model~\cite{Hirata:1977hg,Hirata:1978wp,Horikawa:1980cv,Oset:1979bi,Oset:1987re,Carrasco:1989vq}. 
This resonance is particularly important for pion-nucleus interaction because its excitation drives the dominant $p$-wave spin-isospin-$\frac32$ ($P_{33}$) channel in the elementary pion-nucleon scattering. 
However, strong scalar and vector fields affect the $\Delta$-isobar propagating through the nuclear many-body system. 
The many-body medium effects are incorporated in the complex effective $\Delta$ self-energy $\Sigma_\Delta$, which shifts the $\Delta$ mass and width. 
The treatment of pion-nuclear reactions within the framework of the $\Delta$-hole model is done by means of a phenomenological spreading potential, the parameters of which are fitted to the data.

The aim of the present work is to develop the second-order pion-nuclear potential in momentum space.
In addition to the first-order part of the potential, which has a standard form~\cite{Gmitro:1985kj}, our second-order part involves more realistic two-body correlation functions than have been used in earlier works.
In addition, we account for nuclear medium effects, which affect the resonant $P_{33}$ pion-nucleon scattering amplitude.
The pion-bound nucleon amplitude in our approach relies on the relativistic $\Delta$-isobar model~\cite{Oset:1981ih} with modified $\Delta$-propagator.
The effective $\Delta$ self-energy is considered as a parameter in our model, which is fixed by a multienergy fit to $\pi^\pm$-${}^{12}$C scattering data in the energy range 80–\SI{180}{MeV} laboratory kinetic energy. 
In addition to describing pion-nucleus scattering, our work aims to develop a model that can be applied directly to the processes of pion photoproduction and neutrino-induced pion production on spin-zero nuclei.

The paper is organized as follows:
In Sec.~\ref{sec:multiple-scattering}, we present the main aspects of the multiple scattering formalism.
Then, in Sec.~\ref{sec:pion-nucl-ampl}, we consider the pion-nucleon elementary amplitudes and the dominant $P_{33}$ channel.
In Sec.~\ref{sec:Uopt}, we derive the second-order pion-nucleus potential and introduce in-medium modifications to the scattering amplitudes.
Next, in Sec.~\ref{sec:fits-results}, we fit the obtained potential to the data on pion-${}^{12}$C scattering and apply it to the ${}^{16}$O, ${}^{28}$Si, and ${}^{40}$Ca data.
Finally, in Sec.~\ref{sec:conclusion}, we provide our conclusions.

\section{Multiple scattering formalism}
\label{sec:multiple-scattering}
In multiple scattering theory, the overall pion-nuclear transition amplitude $\hat T$ is a symmetric sum of amplitudes over all $A$ individual nucleons
\begin{multline}
\hat T(E) = 
\sum_{i=1}^A \hat \tau_i(E) +  \sum_{i=1}^A \sum_{j \ne i}^A \hat \tau_i(E) \hat G(E) \hat \tau_j(E) \\ +\sum_{i=1}^A \sum_{j \ne i}^A \sum_{k \ne j}^A \hat \tau_i(E) \hat G(E) \hat \tau_j(E) \hat G(E)  \hat \tau_k(E) + \cdots,
\label{T-series}
\end{multline}
where $E$ is the reaction energy and $\hat G(E)$ is the Green's function of the noninteracting pion-nuclear system. 
The pion-nucleon transition amplitude describing scattering to all orders on a single nucleon bound inside the nucleus is
\be
\hat \tau_i(E) = \hat v_i +  \hat v_i \hat G(E)  \hat \tau_i(E),
\label{tau_i-def}
\ee
where $\hat v_i$ denotes the pion-single nucleon potential.
Subsequently, we are going to replace the potential $\hat v_i$ with the corresponding free-space pion-nucleon amplitude $\hat t_i$, which may be more easily parametrized from the experiment (see Sec.~\ref{sec:pion-nucl-ampl}):
\be
\hat t_i(W) = \hat v_i +  \hat v_i \hat g(W)  \hat t_i(W).
\label{t_i-def}
\ee
The scattering series for $\hat t_i$ with the pion-nucleon reaction energy $W$ differs from Eq.~(\ref{tau_i-def}) by the Green's function of the pion-free nucleon system $\hat g(W)$.

A determination of the transition amplitude $\hat T$ from  Eq.~(\ref{T-series}) is difficult due to the presence of all possible intermediate nuclear excited states in the series.
Moreover, $\hat T$, $\hat G$ and $\hat \tau_i$ are $(A+1)$-particle operators, so nucleon degrees of freedom must be integrated out.
Further simplification of the problem is possible by separating the equation involving only the ground state matrix elements from the one containing excited states.
For this purpose, we introduce projection operators, which distinguish the ground state from the excited states of the target nucleus:
\be 
\hat P_0 = |\Psi_0 \rangle \langle \Psi_0|
\qquad\text{and}\qquad
\hat P_\emptyset = \sum_{\alpha^* \ne 0} |\Psi_{\alpha^*} \rangle\langle \Psi_{\alpha^*}|,
\ee
where $|\Psi_0 \rangle$ and $|\Psi_{\alpha^*} \rangle$ correspond to the nuclear ground state and all possible excited states, respectively. Also we assume $\hat P_\emptyset =  \hat{ \mathds{1} } - \hat P_0$.
Following the Kerman-McManus-Thaler formulation of the multiple scattering theory, Eqs.~(\ref{T-series})–(\ref{t_i-def}) are equivalent to the system of integral equations \cite{Kerman:1959fr}:\\
\begin{subequations}
\begin{align}
&\hat T(E) = \hat U(E) + \frac{A-1}A  \hat U(E) \hat G(E) \hat P_0 \hat T(E),
\label{T-series-KMT}\\
&\hat U(E) = A \, \hat \tau(E) + (A-1) \hat \tau(E) \hat G(E) \hat P_\emptyset \hat U(E),
\label{U-series}\\
&\hat \tau(E) = \hat t(W) + \hat t(W) \left[ \hat G(E) -  \hat g(W)\right]\hat \tau(E).
\label{tau(t)KMT}%
\end{align}
\end{subequations}
Here and further, we drop the index of $\hat t_i$ when there is no need to distinguish nucleons.
The above scattering equation on $\hat T(E)$ ($\hat U(E)$) resembles the  Lippmann-Schwinger equation, with the additional factor $(A-1)/A$ and projector $\hat P_0$ ($\hat P_\emptyset$), which forbids intermediate nuclear excited (ground) states, respectively. 
The factor $(A-1)/A$ prevents double counting of pion rescattering on the same nucleon since all possible rescatterings on a single nucleon are already included in the pion-nucleon amplitude $\hat \tau$.

The many-body process of pion-nucleus elastic scattering is completely determined by the nuclear ground state expectation value of $\langle \Psi_0 | \hat T | \Psi_0 \rangle$, defined by the scattering equation
\begin{multline}
\langle \Psi_0 | \hat T(E) | \Psi_0 \rangle = \langle \Psi_0 | \hat U(E) | \Psi_0 \rangle \\
+ \frac{A-1}A  \langle \Psi_0 | \hat U(E) | \Psi_0 \rangle \hat G_0(E) \langle \Psi_0 | \hat T(E) | \Psi_0 \rangle,
\label{T-series-KMT-ground}
\end{multline}
where we have used the property of $\hat G_0(E)$: $\langle \Psi_0 | \hat G(E) | \Psi_\alpha \rangle = \hat G_0(E) \delta_{0\alpha}$.
Note, Eq.~(\ref{T-series-KMT-ground}) contains only the terms diagonal in the nuclear ground state. 
As a result, this equation is not necessarily rapidly convergent. 
However, it can be solved numerically if the effective potential  $\langle \Psi_0 | \hat U | \Psi_0 \rangle$ is known.
As follows from Eq.~(\ref{U-series}), the scattering equation for the potential $\langle \Psi_0 | \hat U | \Psi_0 \rangle$ contains two nondiagonal matrix elements in the second term and is expected to converge rapidly.
This is a consequence of the fact that all influence of the excited states is contained in $\hat U$.
A detailed consideration of the effective potential is presented in Sec.~\ref{sec:Uopt}.

It is convenient to consider the pion-nucleus scattering in the center-of-mass (c.m.) frame of the pion-nucleus system. 
The reaction energy is then defined as $E = E(k_0) = \omega(k_0) + E_A(k_0)$, where $\omega(k_0)$ and $E_A(k_0)$ are the energies of the pion and nucleus defined relativistically, and $k_0$ is the on-shell momentum.
As was discussed above, Eq.~(\ref{T-series-KMT}) contains only the diagonal in the nuclear ground-state relativistic propagator of the pion-nucleus system $\hat G_0(E) = \langle \Psi_0 | \hat G(E)| \Psi_0 \rangle$. 
In pion momentum space, it becomes 
\be
\langle \pi(\bm k') | \hat G_0(E) | \pi(\bm k) \rangle = (2\pi)^3 \delta(\bm k' - \bm k) G_0(k) 
\ee
where $k = |\bm k|$ and
\be
G_0(k) = \frac1{E - \omega(k) - E_A(k) + i \, \varepsilon},
\label{G0(E)}
\ee
We can write Eq.~(\ref{G0(E)}) in the pseudononrelativistic form

\be
G_0(k) = \frac{2 \mathscr{M}(k)}{k_0^2 - k^2 + i \, \varepsilon},
\ee
with an off-shell analog of the relativistic reduced mass,
\begin{multline}
\mathscr{M}(k) \equiv \\ \frac{[E + \omega(k) + E_A(k)][\omega(k_0)E_A(k_0) + \omega(k)E_A(k)]}{2 \left(E^2 + (\omega(k) + E_A(k))^2\right)}.
\end{multline}
Taking into account the equality $\mathscr{M}(k_0) = \omega(k_0) E_A(k_0) / (\omega(k_0) + E_A(k_0))$, we introduce the elastic scattering amplitude in the momentum space, defined as
\begin{multline}
F(\bm k', \bm k) = - \frac{\sqrt{\mathscr{M}(k') \mathscr{M}(k)}}{2\pi} \\
\times \langle \pi(\bm k'), \Psi_0 | \hat T(E) | \pi(\bm k), \Psi_0 \rangle,
\end{multline}
where $\bm k$ and $\bm k'$ are the pion c.m. momenta in the initial and final states, respectively.
Then, in accordance with Eq.~(\ref{T-series-KMT}), the elastic scattering amplitude is calculated by solving the integral equation 
\begin{multline}
F(\bm k^\prime, \bm k) = V(\bm k^\prime, \bm k) \\
- \frac{A - 1}A  \int \frac{\diff \bm k^{\prime\prime}}{2\pi^2} \frac{V(\bm k', \bm k'') F(\bm k'', \bm k)}{k_0^2 - {k''}^2 + i \, \varepsilon},
\label{LSh-pseudo-classical}
\end{multline}
where the momentum space potential of the pion-nuclear interaction is defined as:
\be
V(\bm k', \bm k) = - \frac{\sqrt{\mathscr{M}(k') \mathscr{M}(k)}}{2\pi} U(\bm k', \bm k),
\label{V-nucl-def}
\ee
with $U(\bm k', \bm k) =  \langle \pi(\bm k'), \Psi_0 | \hat U (E) | \pi(\bm k), \Psi_0 \rangle$.

Note that the formulas given above and developed in the following sections are derived for the case of nuclear interaction.
The modification that is needed for the inclusion of the Coulomb interaction is discussed in Appendix~\ref{sec:Coulomb}.

%%%%%%%%%%%%%%%%%%%%%%%%%%%%%%%%%%%%%%%%%%%%%%%%%%%%%%%%%%%
\section{Pion-nucleon elementary scattering amplitude}
\label{sec:pion-nucl-ampl}
%%%%%%%%%%%%%%%%%%%%%%%%%%%%%%%%%%%%%%%%%%%%%%%%%%%%%%%%%%%
The potential $\hat U$ relies on the knowledge of the pion-nucleon scattering amplitude.
To describe the scattering on a single bound nucleon, we assume that the contribution from the second term of Eq.~(\ref{tau(t)KMT}) can be neglected. 
In this way, we impose  $\hat \tau(E) \approx \hat t(W)$, which is known as the \textit{impulse approximation}.
However, the c.m. energy of the pion-nucleon subsystem $W$  is a dynamical variable~\cite{Faddeev:1960su, Hirata:1978wp}.
An optimal approach for choosing $W$ would be minimizing the second term of Eq.~(\ref{tau(t)KMT}), describing binding correction to $\hat \tau$.
There are several prescriptions with various motivations for choosing the optimal value for $W$~\cite{Gmitro:1985kj, Landau:1977th, mach1983galileo}.
We will follow the arguments of Gurvitz~\cite{Gurvitz:1986zza} and set 
\be
W(\bm k, \bm p) = \sqrt{\left(\omega(k)+E_N(p) \right)^2 - (\bm k + \bm p)^2},
\label{W-def}
\ee
where $\bm k$ and $\bm p$ are the pion and target nucleon momenta in the pion-nucleus c.m. frame, and $\omega(k = |\bm k|)$ and $E_N(p = |\bm p|)$ are the corresponding relativistic energies.
The choice of the effective value of $\bm p$ will be discussed in Sec.~\ref{sec:U-1st}. 
We note that the freedom in choosing $W$ can be absorbed in the model parameters when studying the medium effects (see Sec.~\ref{subsec-bound-nucl}).

While we require the pion-nucleon transition amplitude in the pion-nucleus c.m. frame, it is more convenient to consider the pion-single-nucleon interaction in the pion-nucleon c.m. system.
All quantities denoted by the subscript "2cm" refer to the pion-nucleon frame in order to distinguish both systems.
The pion momenta in both reference frames are related by the Lorentz transformation
\begin{align}
\begin{split}
&\bm k_\text{2cm}(\bm k, \bm p) = \bm k + \alpha \, (\bm k + \bm p),  \\
&\alpha = \frac1{W(\bm k, \bm p)}\left(\frac{(\bm k + \bm p) \cdot \bm k}{W(\bm k, \bm p) + \omega(k) + E_N(p)} - \omega(k) \right),
\end{split}
\label{Lorentz-transform}%
\end{align}
and an analogous relation for $\bm k_\text{2cm}'$.
Also, we assume that the transformation in Eq.~(\ref{Lorentz-transform}) is justified for virtual particles, which is the approach of relativistic potential
theory \cite{Ernst:1980my, Heller:1976wd, Bakamjian:1953kh}. 

The free pion-nucleon scattering matrix in the pion-nucleus and pion-nucleon c.m. frames is then related through
\begin{multline}
\langle \pi(\bm k^\prime), N(\bm p^\prime) | \hat t | \pi(\bm k), N(\bm p) \rangle \\ = 
(2\pi)^3
\delta(\bm k^\prime + \bm p^\prime - \bm k - \bm p) 
 \gamma \, t_\text{2cm}(\bm k'_\text{2cm}, \bm k_\text{2cm}),
 \label{<t>-t()}
\end{multline}
with the usual M{\o}ller phase-space factor~\cite{moller1945general}
\be
\gamma = \sqrt{\frac{\omega(\bm k_\text{2cm}) \omega(\bm k_\text{2cm}^\prime)}{\omega(\bm k) \omega(\bm k^\prime)}\frac{E_N(\bm k_\text{2cm})E_N(\bm k_\text{2cm}^\prime)}{E_N(\bm p) E_N(\bm p^\prime)}}
\label{gamma-moller}
\ee
due to the noncovariant normalization convention used to calculate $t_\text{2cm}$.
In Eq.~(\ref{<t>-t()}) and further, we imply that the transition amplitude is calculated at the pion-nucleon reaction energy calculated according to Eq.~(\ref{W-def}) for the on-shell process.
The notation $t(\bm k', \bm k)$ indicates that the momentum-conserving $\delta$ function was explicitly separated.

The pion-nucleon on-shell $T$ matrix is related to the elastic scattering amplitude $f$ as
\be
t_\text{2cm}(\bm k'_{0, \text{2cm}}, \bm k_{0, \text{2cm}}) = - \frac{4\pi}{2\bar \omega} f(\bm k'_{0, \text{2cm}}, \bm k_{0, \text{2cm}}),
\label{t(f)-def}
\ee 
where $\bar \omega = \omega(k_{0, \text{2cm}}) E_N(k_{0, \text{2cm}}) / W$ is the pion-nucleon relativistic reduced mass, $W = \omega(k_{0, \text{2cm}}) + E_N(k_{0, \text{2cm}})$, and $|\bm k'_{0, \text{2cm}}| = |\bm k_{0, \text{2cm}}| = k_{0, \text{2cm}}$.
We consider further in this section only the most relevant properties of the scattering amplitude for the $\pi(\bm k_\text{2cm}) + N(- \bm k_\text{2cm}) \longrightarrow \pi(\bm k'_\text{2cm}) + N(- \bm k'_\text{2cm})$ process and refer to Ref.~\cite{Ericson:1988gk} for a more detailed review.

Assuming the isospin conservation, we can explicitly represent the spin-isospin structure of the amplitude as
\be
\hat t = \hat t^{(0)} + \hat t^{(1)} \ \hat{\bm t} \cdot \hat{\pmb \tau} + 
\left(\hat t^{(2)} +
\hat t^{(3)} \ \hat{\bm t} \cdot \hat{\pmb \tau} \right)\hat{\pmb \sigma} \cdot \bm n,
\label{t-isospin-struct}
\ee
where $\hat{\bm t}$ and $\hat{\pmb \tau}$ are the pion and nucleon isospin operators, $\hat{\pmb \sigma}$ is the nucleon Pauli spin operator, and $\bm n = \bm k_\text{2cm} \times \bm k_\text{2cm}^\prime/|\bm k_\text{2cm} \times \bm k_\text{2cm}^\prime|$ is the normal to the scattering plane.
The same notation also holds for $t_\text{2cm}(\bm k'_\text{2cm}, \bm k_\text{2cm})$ and $f(\bm k'_\text{2cm}, \bm k_\text{2cm})$.

The $P_{33}$ partial wave is the only resonant one at low and intermediate energies, peaking at about the pion laboratory kinetic energy $T_\text{lab} \approx \SI{190}{MeV}$.
Correspondingly, within the energy range under our consideration, $T_\text{lab} \lesssim \SI{300}{MeV}$, only the $s$- and $p$-wave contributions are dominant. 
As a result, the pion-nucleon scattering amplitude can be written as
\begin{multline}
f(\bm k'_{0, \text{2cm}}, \bm k_{0, \text{2cm}}) \approx b_0 + b_1 \, \hat{\bm t} \cdot \hat{\pmb \tau}  \\
+ ( c_0 + c_1 \ \hat{\bm t} \cdot \hat{\pmb \tau}) \, \bm k'_{0, \text{2cm}} \cdot \bm k_{0, \text{2cm}} \\
+ i ( s_0 + s_1 \ \hat{\bm t} \cdot \hat{\pmb \tau}) \, \hat{\pmb \sigma} \cdot [\bm k_{0, \text{2cm}}' \times \bm k_{0, \text{2cm}}],
\label{f-low-energy}
\end{multline}
where $b_{0,1}$, $c_{0,1}$, and $s_{0,1}$ are energy-dependent complex $s$- and $p$-wave coefficients.

The multipole expansion allows us to express the parameters $b_{0,1}$, $c_{0,1}$, and $s_{0,1}$ through the partial wave amplitudes $f^l_{2T \, 2J}$ as
\begin{subequations}
\begin{align}
b_0 &= \frac13 \left[ f^0_{1 \, 1} + 2 f^0_{3 \,1} \right],  \label{f-sp-coefs-b0} \\
b_1 &= \frac13 \left[ f^0_{3 \, 1} - f^0_{1 \, 1} \right], \label{f-sp-coefs-b1}%
\\
c_0 &= \frac1{3k_{0, \text{2cm}}^2} \left[ f^1_{1 \, 1} + 2 f^1_{3 \, 1} + 2f^1_{1 \, 3} + 4 f^1_{3 \, 3} \right],\\
c_1 &= \frac1{3k_{0, \text{2cm}}^2} \left[ f^1_{3 \, 1} - f^1_{1 \, 1} + 2 f^1_{3 \, 3} - 2f^1_{1 \, 3} \right],\\
s_0 &= \frac1{3k_{0, \text{2cm}}^2} \left[ f^1_{1 \, 1} + 2 f^1_{3 \, 1} - f^1_{1 \, 3} - 2 f^1_{3 \, 3} \right] , \\
s_1 &= \frac1{3k_{0, \text{2cm}}^2} \left[ f^1_{3 \, 1} - f^1_{1 \, 1} - f^1_{3 \, 3} + f^1_{1 \, 3} \right] .
\end{align}
\label{f-sp-coefs}%
\end{subequations}
Here $l$, $T$, and $J$ are, respectively, the orbital angular momentum, isospin, and total angular momentum of the pion-nucleon system.
The partial-wave amplitudes are related to the measured pion-nucleon phase shifts as
\be
f^l_{2T \, 2J} = \frac1{2ik_{0, \text{2cm}}} \left(e^{2i \delta^l_{2T \, 2J}} - 1 \right).
\label{f(delta)-def}
\ee
In this work, we take the complex scattering phase shifts $\delta^l_{2T \, 2J}$ as extracted from the state-of-the-art phase shift analysis (WI08) by the SAID Collaboration~\cite{Workman:2012hx}.

As can be seen from Eq.~(\ref{LSh-pseudo-classical}), explicit knowledge of the off-energy-shell behavior of the potential $V$ is required to solve the scattering equation.
Whereas the on-shell behavior is directly defined by the partial wave amplitudes $f^l_{2T \, 2J}$, Eq.~(\ref{f(delta)-def}), the off-shell extrapolation needs a model specification.
We assume that for the on-shell momentum $k_{0, \text{2cm}}$ the dependence of the amplitude $f^l_{2T \, 2J}$ on the off-shell momenta $\bm k_\text{2cm}$ and $\bm k'_\text{2cm}$ is defined by the separable form
\begin{multline}
f^l_{2T \, 2J}(k'_\text{2cm}, k_\text{2cm}) = f^l_{2T \, 2J}(k_{0, \text{2cm}}, k_{0, \text{2cm}})  \\
\times 
\left( \frac{k'_\text{2cm} k_\text{2cm}}{k_{0, \text{2cm}}^2} \right)^l
\frac{v(k'_\text{2cm})v(k_\text{2cm})}{v^2(k_{0, \text{2cm}})}, 
\label{f-off-shell}
\end{multline}
with the off-shell vertex factor for $s$ and $p$ waves
\be
v(k) = \frac{1}{\Lambda^2 - (\omega^2(k_{0, \text{2cm}}) - k^2)},
\label{off-vertex-fact}
\ee
where $\Lambda = 1.25 \text{ GeV}$ is taken.
Note that including the second-order part of the potential $\hat U$ (see Sec.~\ref{sec:Uopt}) reduces the model sensitivity to the off-shell behavior of the pion-nucleon amplitude. 
The alteration of the off-shell parameter $\Lambda$ does not significantly impact the model's predictions after fitting. However, the variation of $\Lambda$ does result in changes to the free parameters, which will be described in the following sections.

An important feature of pion-nucleon scattering is the relative weakness of the $s$-wave interaction. It makes the $p$-wave part of the amplitude not only dominant at intermediate energies but also significant at low energies, even close to the threshold. 
As a result, an accurate description of the $p$-wave interaction is essential for the pion scattering on both free and bound nucleons.
The starting point should be a model which effectively describes the basic dynamical features of the free pion-nucleon process.
In our work, we adopt the \textit{relativistic $\Delta$-isobar model} by Oset, Toki and Weise \cite{Oset:1981ih}, which successfully reproduces the $p$-wave pion-nucleon phase shifts at low and intermediate energies, especially the resonant $P_{33}$ channel.
The model is based on the $K$-matrix formalism in which we express the elastic-scattering partial amplitudes as
\be
f^l_{2T \, 2J} = \frac{K^l_{2T \, 2J}}{1- i  k_{0, \text{2cm}} K^l_{2T \, 2J}}.
\ee
When the $K$ matrix is real, the unitarity is automatically incorporated.
In general, the phase shifts and, correspondingly, the $K$ matrix remains real only below the pion production threshold ($\pi N \rightarrow \pi\pi N$), which is approximately at \SI{170}{MeV} pion laboratory kinetic energy.
However, even when the inelastic channel is open, the inelasticity parameters for the $p$ wave remain close to 1 with high accuracy.
As a result, the $p$-wave pion-nucleon interaction can be described by the real crossing symmetric $K$ matrix. 
According to the relativistic $\Delta$-isobar model, the pion-nucleon $K$ matrix is based entirely on the pion-baryon effective Lagrangian and contains direct and crossed contributions from nucleon $N$, $\Delta(1232)$-isobar, and  Roper resonance $N^*(1440)$. 
The resulting $K$-matrix in the dominant $P_{33}$ channel is given by
\begin{multline}
K^1_{33} = \frac13 \frac{k_{0,\text{2cm}}^2}{4\pi m_{\pi}^2} \frac{m_N}{\sqrt{s}} 
\left[
4 f_N^2 \frac{2 m_N}{m_N^2 - \bar u} +  4 f_{N^*}^2 \frac{2 m_{N^*}}{m_{N^*}^2 - \bar u} 
\right. \\
\left.
+ f_\Delta^2 \left( \frac{2 m_\Delta}{m_\Delta^2 - s} + \frac19 \frac{2 m_\Delta}{m_\Delta^2 - \bar u} \right)
\right],
\label{K33-RDM}
\end{multline}
where
$s = W^2$, $m_\pi$ is the pion mass and the approximate $u$-channel Mandelstam variable is $\bar u = u + 2 \bm k'_\text{2cm} \cdot \bm k_\text{2cm} = m_N^2 + m_\pi^2 - 2 \omega(k_{0,\text{2cm}}) E_N(k_{0,\text{2cm}})$.
The masses and coupling constants used are~\cite{Oset:1981ih}:
\begin{align*}
&m_N = \SI{939}{MeV}, &  &f_N^2/4\pi = 0.079,\\
&m_\Delta = \SI{1232}{MeV}, &  &f_\Delta^2/4\pi = 0.37, \\
&m_{N^*} = \SI{1450}{MeV}, &  &f_{N^*}^2/4\pi = 0.015.
\end{align*}
The primary role of the Roper resonance $N^*(1440)$ in this model is providing the correct behavior in the $P_{11}$ channel.
The contributions of the $u$-channel $N^*(1440)$ and $\Delta$ to the $P_{33}$ channel are on the order of a few percent.
In contrast, the nucleon $u$-channel term is not negligible and becomes particularly significant at lower energies, e.g., making an approximately 50\% contribution at the pion laboratory kinetic energy of $\SI{50}{MeV}$.

\begin{figure}[!htb]
\center
\begin{minipage}[h]{0.99\linewidth}
\center{\includegraphics[width=1\linewidth]{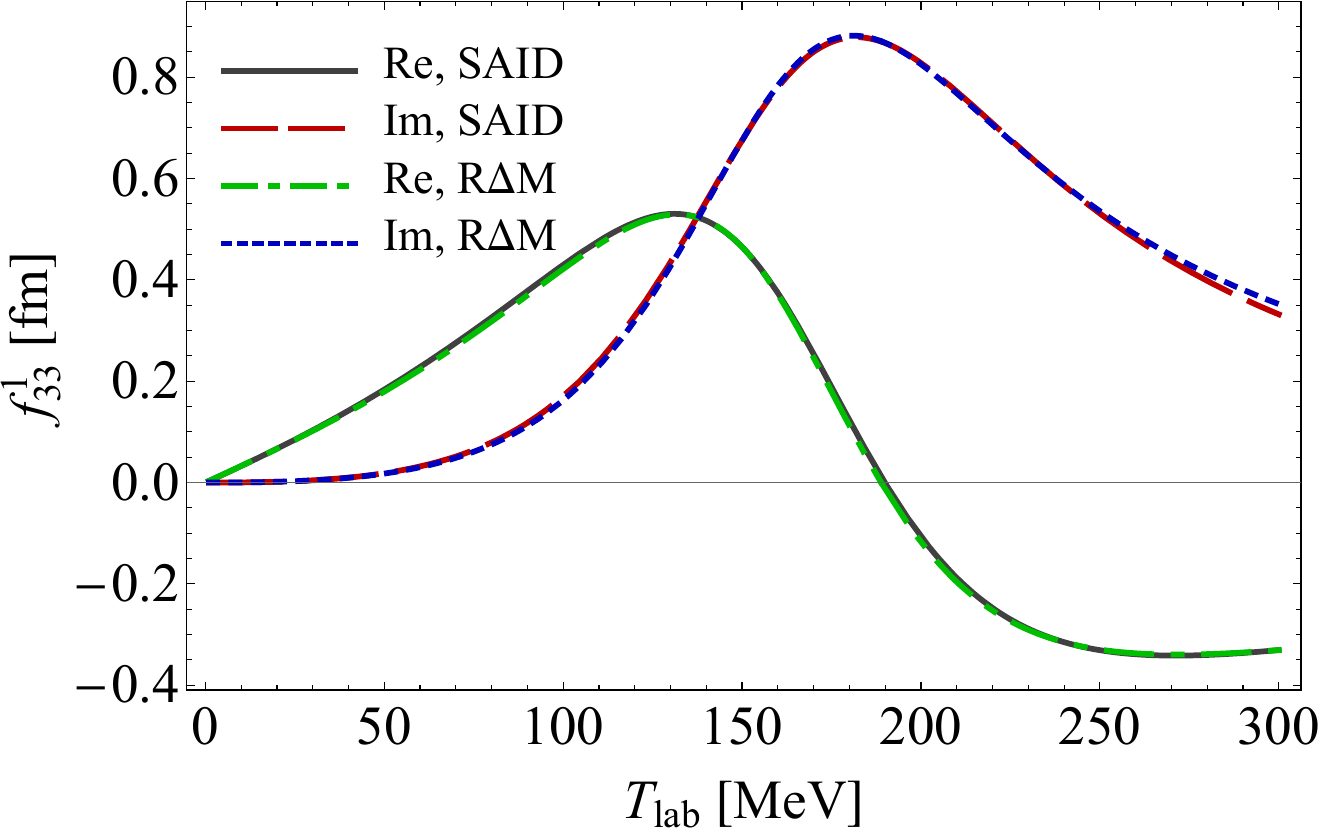}} % {\center c)} \\
\end{minipage}
\caption{The theoretical $f^1_{33}$ amplitude obtained with the relativistic $\Delta$-isobar model (R$\Delta$M) as a function of pion laboratory kinetic energy compared with SAID phase shift analysis~\cite{Workman:2012hx}.
The solid black (long-dashed red) curve represents the real (imaginary) part of the amplitude taken from SAID, while the dot-dashed green (short-dashed blue) curve corresponds to the R$\Delta$M calculation.
}
\label{isobar-SAID-comp}
\end{figure}

In Fig.~\ref{isobar-SAID-comp}, we compare the $P_{33}$ partial amplitude taken from the SAID phase shift analysis with the relativistic $\Delta$-isobar model results.
The corresponding curves in the plot are almost indistinguishable, showing excellent agreement between the theoretical model and experiment.

The dominant term in Eq.~(\ref{K33-RDM}) comes from the direct ($s$-channel) $\Delta$-pole contribution.
This resonant part of the $K^1_{33}$ can be written as
\be
K^{1 (\Delta)}_{33} = \frac1{k_{0,\text{2cm}}} \frac{m_\Delta \Gamma_\Delta}{m_\Delta^2 - W^2}, 
\label{K33-Delta}
\ee
where we have introduced the $\Delta$ decay width
\be
\Gamma_\Delta = \frac23 \frac{f_\Delta^2}{4\pi} \frac{k_{0, \text{2cm}}^3}{m_\pi^2} \frac{m_N}{W}.
\label{Gamma_Delta}
\ee
The width at resonance ($W = m_\Delta$) is $\Gamma_\Delta \approx \SI{115}{MeV}$.
This separation of the $s$-channel $\Delta$ term in the form of Eq.~(\ref{K33-Delta}) will be useful in the following introducing the medium modifications.

%%%%%%%%%%%%%%%%%%%%%%%%%%%%%%%%%%%%%%%%%%%%%%%%%%%%%%%%%%%
\section{Derivation of the pion-nucleus potential}
\label{sec:Uopt}
%%%%%%%%%%%%%%%%%%%%%%%%%%%%%%%%%%%%%%%%%%%%%%%%%%%%%%%%%%%
We are now in the position to construct the effective pion-nucleus potential used in scattering equation~(\ref{LSh-pseudo-classical}).
We assume the potential $\hat U(E)$ is approximated by the first two terms of the iterative series for Eq.~(\ref{U-series}):
\be
\hat U(E) \approx \hat U^{(1)} + \hat U^{(2)},
\ee
where within the impulse approximation, the first-order part has the simple form
\be
\hat U^{(1)} = A \, \hat t
\label{U1st-def}
\ee
and the second-order part is given by
\be
\hat U^{(2)} = A(A-1) \hat t \hat G(E) \hat P_\emptyset \hat t.
\label{U2nd-def}
\ee

In the following, we will express Eqs.~(\ref{U1st-def}) and~(\ref{U2nd-def}) for the effective potential into more practical forms.

%##################################################################################
\subsection{The first-order potential}
\label{sec:U-1st}
%##################################################################################
The first-order potential in momentum space can be written as:
\begin{multline}
U^{(1)}(\bm k', \bm k) = \int \frac{\diff \bm p'}{(2\pi)^3} \frac{\diff \bm p}{(2\pi)^3} \\
\times
\Tr \left[
 \langle \pi(\bm k'), N(\bm p') | \, \hat t \, | \pi(\bm k), N(\bm p) \rangle \, 
\rho(\bm p'; \bm p)
\right],
\label{U1st_mom_statr} 
\end{multline}
where $\bm p$ and $\bm p'$ are the initial and final momenta of the target nucleon under consideration, $\Tr$ represents summation over all nucleon spin and isospin projections as:
\begin{multline}
\Tr \left[ \hat t \, \rho(\bm p^\prime; \bm p) \right]   \equiv
\sum_{\sigma, \sigma'} \sum_{\tau, \tau'}  \langle \sigma_{1z}', \tau_{1z}' | \, \hat t \, | \sigma_{1z}, \tau_{1z} \rangle \\
\times
\rho(\bm p', \sigma', \tau'; \bm p, \sigma, \tau),
\label{trace-explicit}
\end{multline}
and the one-body density matrix for the target nucleus is
\begin{multline}
\rho(\bm p', \sigma', \tau'; \bm p, \sigma, \tau) = A  \mathlarger{ \int} \left( \prod_{i= 2}^A \diff x_i  \right) \diff \bm r_1 \diff \bm r_1^\prime \\
\times e^{i ( \bm p' \cdot \bm r_1^\prime - \bm p \cdot \bm r_1)}
\Psi_0^*(x_1^\prime, x_2, \ldots, x_A)  \Psi_0(x_1, \ldots, x_A).
\end{multline}
Here, the notation $x_i = \{\bm r_i, \sigma_i, \tau_i \}$ covers nucleon spin and isospin, and $\int \diff x_i (\cdots) = \sum_{\sigma_i} \sum_{\tau_i} \int \diff \bm r_i (\cdots) $. 
The spin and isospin variables are suppressed in what follows.

As a result, the first-order potential in the impulse approximation including the recoil of the struck nucleon is given by the Fermi motion integral:
\begin{multline}
U^{(1)}(\bm k', \bm k) =
\mathlarger{ \int} \frac{\diff \bm p}{(2\pi)^3} \gamma
\Tr \left[ \rho(\bm p - \bm q /2; \bm p + \bm q /2) \right. \\
\times \left.
 t_\text{2cm}(\bm k_\text{2cm}^\prime(\bm k^\prime, \bm p - \bm q /2), \bm k_\text{2cm}(\bm k, \bm p + \bm q /2) \right],
\label{U1st-Fermi-int}
\end{multline}
where $\bm q = \bm k' - \bm k$ and $\gamma$ is given by Eq.~(\ref{gamma-moller}).
The integration over $\bm p$ in Eq.~(\ref{U1st-Fermi-int}) requires nondiagonal elements of the one-body density matrix which are model dependent. 
Moreover, the proper treatment of the Fermi averaging should also take into account the binding effects. 
To simplify the problem, one can treat the nucleon Fermi motion approximately by evaluating the pion-nucleon amplitude at the effective initial and final nucleon momenta
\be
\bm p_\text{eff} =  \frac{\bm q}{2} - \frac{\bm k^\prime + \bm k}{2A}, \quad\text{and}\quad
\bm p_\text{eff}^\prime = - \frac{\bm q}{2} - \frac{\bm k^\prime + \bm k}{2A},
\ee 
respectively. 
This result was obtained for elastic nucleon-deuteron scattering (for $A = 2$) in Ref.~\cite{kowalski1963elastic}.
The terms proportional to $A^{-1}$ arrive due to the correct treatment of the target recoil.
In this so-called \textit{optimized factorization approximation} we arrive at
% \be
\begin{multline}
U^{(1)}(\bm k', \bm k)  \\ = \gamma
 \Tr \left[ \rho(\bm q)  
 t_\text{2cm}(\bm k_\text{2cm}^\prime(\bm k^\prime, \bm p_\text{eff}'), \bm k_\text{2cm}(\bm k, \bm p_\text{eff}) 
\right],
\label{t-rho-potential}
\end{multline}
% \ee
with the nuclear form factor
\be
\rho(\bm q) = A \mathlarger{ \int} \left( \prod_{i=1}^{A-1} \diff \pmb \xi_i \right) e^{i \frac{A-1}A \bm q \cdot \pmb \xi_{A-1}} |\Psi_0(\pmb \xi_1, \ldots, \pmb \xi_{A-1})|^2 
\label{form-factor-Jacobi}
\ee
normalized to $\rho(0) = A$.
Galilean-invariant Jacobi coordinates, $\pmb \xi_i$, were introduced in order to eliminate the motion of the nucleus as a whole, as the form factor characterizes the internal structure of the nucleus (see Ref.~\cite{Mach:1973du} for details).
The factorization approximation is justified by the compensation between the binding potential of the nucleon and the Fermi motion kinetic energy~\cite{Gurvitz:1986zza}.

Finally, the pion-nuclear potential, Eq.~(\ref{V-nucl-def}), is expressed through the pion-nucleon scattering amplitude as: 
\be
V^{(1)}(\bm k', \bm k) = \mathscr{W}(\bm k', \bm k)
 \Tr \left[ \rho(\bm q)  
f\left(\bm k_\text{2cm}', \bm k_\text{2cm}\right) 
\right],
\ee
with the phase space factor
\be
\mathscr{W}(\bm k', \bm k) =  \sqrt{ \frac{\mathscr{M}(k') \mathscr{M}(k)}{\mu(\bm k', \bm p_\text{eff}') \mu(\bm k, \bm p_\text{eff})}},
\ee
where $\mu(\bm k, \bm p) = {\omega(k)E_N(p)}/{W(\bm k, \bm p)}$ and we imply $\bm k_\text{2cm} = \bm k_\text{2cm}(\bm k, \bm p_\text{eff}) $.

For spin- and isospin-zero nuclei, only the spin- and isospin-independent part of the scattering amplitude, Eq.~(\ref{f-low-energy}), contributes to the first-order potential:  
\be
V^{(1)}(\bm k', \bm k) = \tilde{\mathscr{W}}(\bm k', \bm k)
\left[ 
b_0 + c_0 \, \bm k_\text{2cm}' \cdot \bm k_\text{2cm}
\right] \rho(\bm q),
\label{U1st-fin}
\ee
where $\tilde{\mathscr{W}}(\bm k', \bm k) = \mathscr{W}(\bm k', \bm k) v(k'_\text{2cm})v(k_\text{2cm}) / v^2(k_{0, \text{2cm}})$.
Note that the scattering parameters $b_0$ and $c_0$ are derived at the pion-nucleon c.m. energy given by Eq.~(\ref{W-def}) for on-shell momenta and thus depend on the scattering angle.
We extract the nuclear form factor  $\rho(\bm q)$ from the corresponding nuclear charge form factor determined through elastic electron scattering (see Appendix~\ref{sec:FF-and-correlation} for details).

Note, in our calculation, besides the most important $s$- and $p$-wave terms, we also include the $d$-wave contribution in the same manner.

%##################################################################################
\subsection{The second-order correction}
\label{sec:U-2nd}
\begin{figure}[!t]
\center{\includegraphics[width=0.8\linewidth]{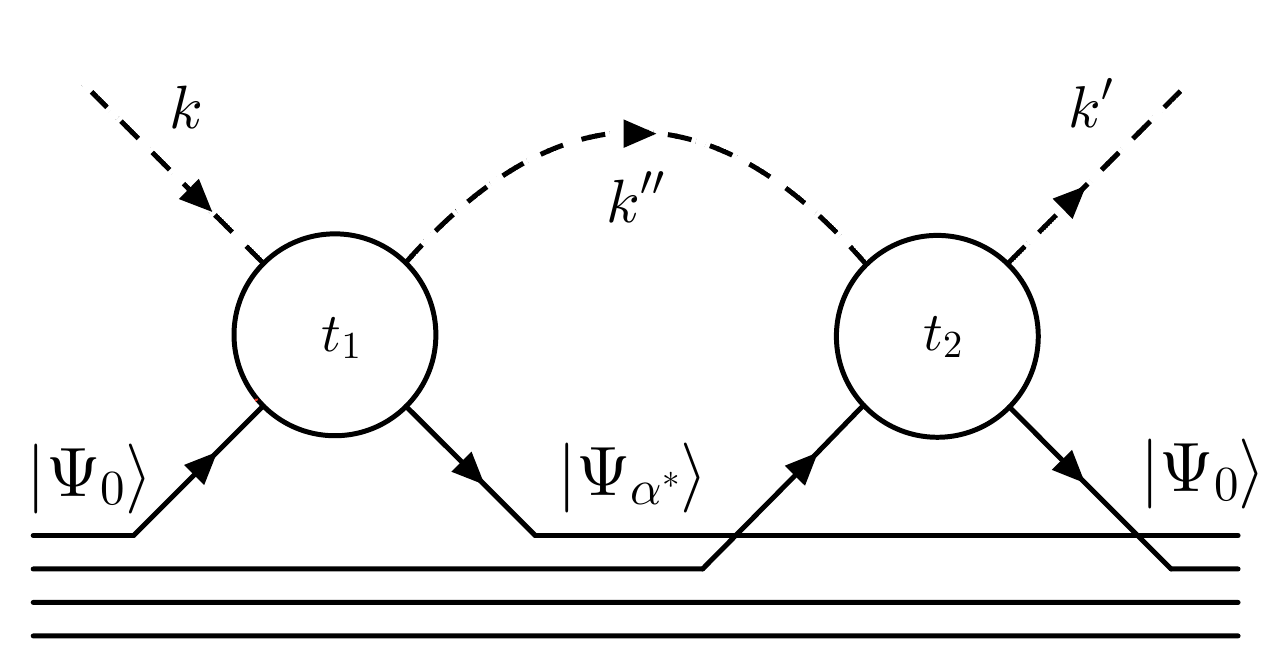}}
\caption{Diagram representation of the second-order part of the pion-nuclear potential.
}
\label{U-second-ord-plt}
\end{figure}

%##################################################################################
The second-order part of the potential, Eq.~(\ref{U2nd-def}), describes scattering to all orders from one nucleon, after which the nucleus makes a transition into an excited state followed by propagation and then scattering to all orders on a second nucleon, summed over all nucleons (see Fig.~\ref{U-second-ord-plt}). 
We use the subscripts "1" and "2" in this section to distinguish the initial and final nucleons involved in the second-order scattering process.

In calculating the second-order correction for the kinetic energies larger than around $\SI{30}{MeV}$ considered in this work, we neglect the nuclear excitation energies in comparison with energies of the pion-nucleus system intermediate states. 
In this way, the excited system propagator is approximated by the ground state one, $\hat G_{\alpha^*} \approx \hat G_0$.
Correspondingly, the second-order part of the pion-nucleus becomes
\be
\langle \Psi_0 | \hat U^{(2)}  | \Psi_0 \rangle = A(A - 1) \langle \Psi_0 | \hat t_2 \hat G_0 \hat P_\emptyset \hat t_1  | \Psi_0 \rangle.
\ee
Substituting the projection operator explicitly, we arrive at
\begin{multline}
\langle \Psi_0 | \hat U^{(2)} | \Psi_0 \rangle = 
 A (A - 1) \left[\langle \Psi_0 | \hat t_2 \hat G_0 \hat t_1 | \Psi_0 \rangle
 \right.\\\left.
 - \langle \Psi_0 | \hat t_2 | \Psi_0 \rangle \hat G_0 \langle \Psi_0 | \hat t_1 | \Psi_0 \rangle \right].
\label{second-ord-term-general}
\end{multline}
According to Eq.~(\ref{t-rho-potential}) for the first-order potential, the second term of Eq.~(\ref{second-ord-term-general}) in momentum space becomes:
\begin{multline}
\langle \pi(\bm k'), \Psi_0 |\hat t_2 | \Psi_0 \rangle \hat G_0 \langle \Psi_0 |\hat t_1 | \pi(\bm k), \Psi_0 \rangle  \\
= \frac1{A^2}  \int \frac{\diff\bm k^{\prime\prime}}{(2\pi)^3} 
\Tr\left[ t_{2}(\bm k^{\prime} ,\bm k^{\prime\prime})
\rho(\bm k^\prime - \bm k^{\prime\prime}) \right]
G_0(\bm k^{\prime\prime}) \\
\times
\Tr\left[ t_{1}(\bm k^{\prime\prime} ,\bm k) 
\rho( \bm k^{\prime\prime} - \bm k) \right],
\label{U2-1st-part}
\end{multline}
Similarly, the first term in Eq.~(\ref{second-ord-term-general}) acquires the form
\begin{multline}
% \be
\langle \pi(\bm k') \Psi_0 |\hat t_2 \hat G_0 \hat t_1| \pi(\bm k) \Psi_0 \rangle  = 
\frac1{ A(A-1)}\int \frac{\diff\bm k^{\prime\prime}}{(2\pi)^3} G_0(\bm k^{\prime\prime}) 
\\
\times \Tr\left[
t_2(\bm k^{\prime} ,\bm k^{\prime\prime})  
 t_1(\bm k^{\prime\prime} ,\bm k) 
\rho_2(\bm k^\prime - \bm k^{\prime\prime},  \bm k^{\prime\prime} - \bm k)
\right],
\label{U2-2nd-part}
% \ee
\end{multline}
where $\rho_2(\bm q_1,  \bm q_2)$ is the Fourier transform,
\be
\rho_2(\bm q_1, \bm q_2) = 
\int \diff \bm r_1 \diff \bm r_2 \, e^{- i (\bm q_1 \cdot \bm r_1 + \bm q_2 \cdot \bm r_2)} \rho_2(\bm r_1, \bm r_2),
\label{2b-density-mom-sp}
\ee
of the two-body density function
\begin{multline}
\rho_2(\bm r_1, \bm r_2) =
A(A-1) \mathlarger{ \int} \left( \prod_{i=3}^A \diff \bm r_i \right) 
\\
\times \Psi_0^\dag(\bm r_1, \ldots, \bm r_A) 
\Psi_0(\bm r_1, \ldots, \bm r_A).
\label{two-body-density-matrix-def}
\end{multline}
In Eqs.~(\ref{U2-1st-part}) and~(\ref{U2-2nd-part}), we imply the same convention as in Eq.~(\ref{trace-explicit}), omitting spin and isospin variables.
The nuclear two-body density $\rho_2(x_1, x_2)$ characterizes the probability of finding one nucleon with $\sigma_1$ and $\tau_1$ at $\bm r_1$ and another nucleon with $\sigma_2$ and $\tau_2$ at $\bm r_2$, while all the other nucleons have arbitrary positions, spins, and isospins.
We imply that $\rho_2$ is normalized to $A(A-1)$.

\begin{figure}[!htb]
\center
\begin{minipage}{0.99\linewidth}
\center{\includegraphics[width=1\linewidth]{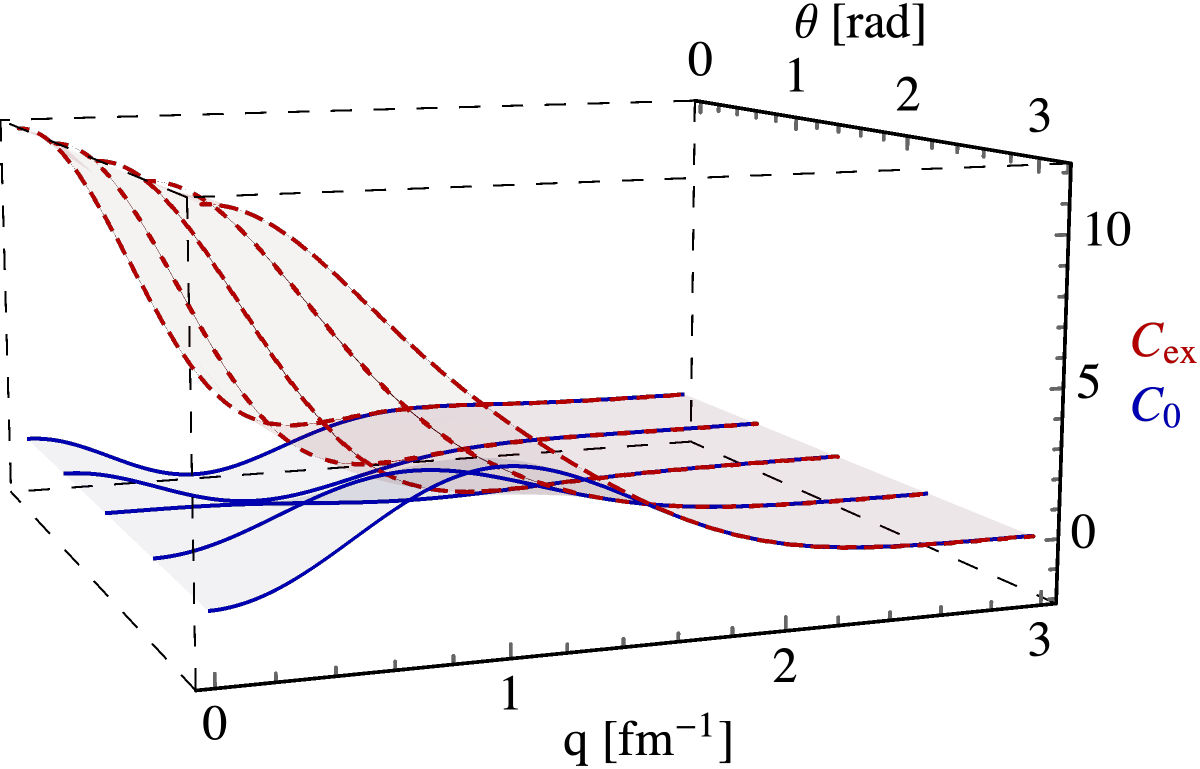}}
\end{minipage}
\caption{The correlation functions $C_\text{ex}$ and $C_0$ for ${}^{12}$C in momentum space given by the harmonic oscillator shell model. $C_\text{ex}(\bm q_1, \bm q_2)$ and $C_0(\bm q_1, \bm q_2)$ are plotted for $|\bm q_1| = |\bm q_2| = q$ as functions of $q$ and the relative angle $\theta$ between $\bm q_1$ and $\bm q_1$.
The red dashed curves correspond to $C_\text{ex}$, while the blue solid curves correspond to $C_0$.
}
\label{C-D-plot}
\end{figure}

As can be seen from Eqs.~(\ref{second-ord-term-general})–(\ref{U2-2nd-part}), the second-order part of the optical potential depends directly on the nucleon-nucleon correlations within the nucleus.
We introduce two  correlation functions:
\begin{subequations}
\begin{align}
&C_\text{ex}(\bm r_1, \bm r_2) = \rho(\bm r_1) \rho(\bm r_2) - \rho_2(\bm r_1, \bm r_2), 
\label{rho_ex(r)-def}\\
&C_0(\bm r_1, \bm r_2) = 
C_\text{ex}(\bm r_1, \bm r_2) - \frac1A \rho(\bm r_1) \rho(\bm r_2),
\label{C(r)-def}
\end{align}
\label{rho_ex-and-C(r)}%
\end{subequations}
which were considered in Ref.~\cite{Jackson:1970zz}.\footnote{Note that different normalizations are used in this work compared to Ref.~\cite{Jackson:1970zz}.}
The function $C_\text{ex}(\bm r_1, \bm r_2)$ can be referred to as the "exchange correlation function" because, as demonstrated below, it accounts for the spin and isospin exchange contributions to pion-nucleon scattering.
It is expressed as the exchange sum in terms of individual nucleon wave functions, Eq.~(\ref{rho_ex-shell}).
Both correlation functions are employed in our calculations, as we do not neglect terms of order $A^{-1}$ appearing in the calculation.
Both correlation functions in momentum space are then obtained by the Fourier transform:
\be
C_{\text{ex},0}(\bm q_1, \bm q_2) = \int \diff \bm r_1 \diff \bm r_2 \, e^{-i (\bm q_1 \cdot \bm r_1 + \bm q_2 \cdot \bm r_2)} C_{\text{ex},0}(\bm r_1, \bm r_2).
\label{D-and-C(q)}
\ee

\begin{figure}[htb]
\center{\includegraphics[width=0.8\linewidth]{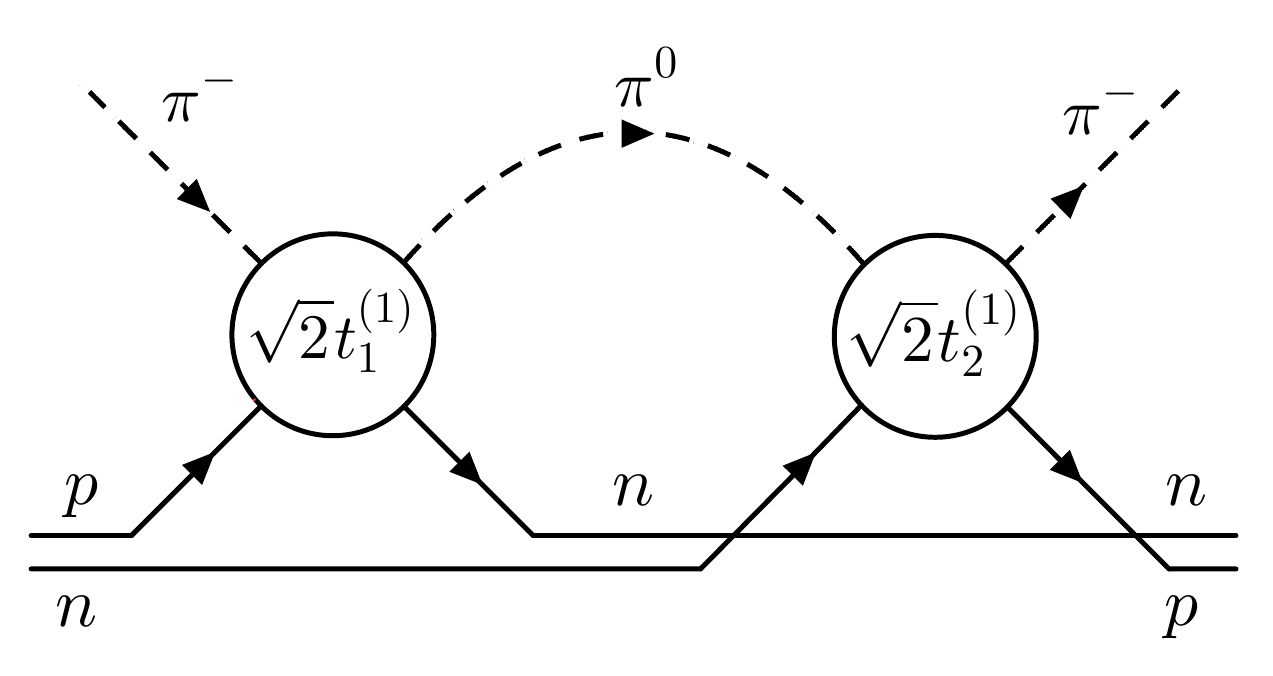}}
\caption{Diagrammatic representation of the second-order isospin exchange for negative pion scattering.
}
\label{U-second-ord-exchange-plt}
\end{figure}

None of the two-body correlation functions is directly measurable, and a model is required to calculate the second-order correction.
A common approach is using the Fermi gas approximation to evaluate the second-order part of the potential~\cite{Ericson:1966fm, Krell:1969xn,Stricker-thesis}. In our calculation, we employ  the more realistic harmonic oscillator nuclear shell model (see Appendix~\ref{sec:FF-and-correlation}).
The explicit forms of $C_\text{ex}(\bm q_1, \bm q_2)$ and $C_0(\bm q_1, \bm q_2)$, summed over spin and isospin, are given by Eqs.~(\ref{C-D-12C})–(\ref{C-D-40Ca}).
The correlation functions for ${}^{12}$C in momentum space in the case of $|\bm q_1| = |\bm q_2| = q$ are shown in Fig.~\ref{C-D-plot}.
While the difference between $C_\text{ex}$ and $C_0$ is less noticeable in coordinate space, Fig.~\ref{C-D-plot} demonstrates their different behavior in the case of small momenta transfer, which is especially important for the pion-nucleus scattering process.

Even for a nucleus with zero spin and isospin, the trace operator in Eq.~(\ref{U2-2nd-part}) yields a nontrivial result containing spin- and isospin-dependent parts of the scattering amplitude, Eq.~(\ref{t-isospin-struct}).
A direct calculation for spin-isospin-zero nuclei yields the following spin and isospin sums entering Eq.~(\ref{U2-2nd-part}):

\begin{widetext}

\begin{multline}
\sum\limits_{s,s',\tau,\tau' = -1/2}^{1/2} 
\chi_1^\dag(s) \chi_2^\dag(s') \eta_1^\dag(\tau) \eta_2^\dag(\tau^\prime) 
\left[ 
 \hat t^{(0)}_2 + \hat t^{(1)}_2 \ \hat{\bm t} \cdot \hat{\pmb \tau}_2 + 
\left(\hat t^{(2)}_2 +
\hat t^{(3)}_2 \ \hat{\bm t} \cdot \hat{\pmb \tau}_2 \right)\hat{\pmb \sigma}_2 \cdot \bm n_2
\right] \\
\times
\left[ 
 \hat t^{(0)}_1 + \hat t^{(1)}_1 \ \hat{\bm t} \cdot \hat{\pmb \tau}_1 + 
\left(\hat t^{(2)}_1 +
\hat t^{(3)}_1 \ \hat{\bm t} \cdot \hat{\pmb \tau}_1 \right)\hat{\pmb \sigma_1} \cdot \bm n_1
\right]
\eta_1(\tau^\prime) \eta_2(\tau) \chi_1(s') \chi_2(s)  \\
 = 4 \left[ \hat t_2^{(0)} \hat t_1^{(0)} + 2 \, \hat t_2^{(1)} \hat t_1^{(1)} + 
\left( \hat t_2^{(2)} \hat t_1^{(2)} + 2 \, \hat t_2^{(3)} \hat t_1^{(3)} \right)  \bm n_1 \cdot \bm n_2
\right],
\label{U2-isospin-struct}
\end{multline}
where $\chi(s)$ ($\eta(\tau)$) is the nucleon spinor (isospinor). 

The first term on the right-hand side of Eq.~(\ref{U2-isospin-struct}) consists of the spin-isospin averaged part $\hat t^{(0)}$ of the scattering amplitudes. 
The remaining terms involve the spin- and isospin-dependent parts and describe intermediate spin and isospin exchange. 
In Fig.~\ref{U-second-ord-exchange-plt}, we show a diagrammatic representation of the isospin exchange for negative pion scattering.
In the following, we include the global factor 4, which arises due to spin-isospin summation, in the correlation functions.

Finally, combining the above results, we express the second-order part of the potential for spin- and isospin-zero nuclei in terms of the correlation functions:
\begin{multline}
U^{(2)}(\bm k^{\prime} ,\bm k) 
= - \int \frac{\diff \bm k^{\prime\prime}}{(2\pi)^3} G_0(\bm k^{\prime\prime}) 
\left[ 
t^{(0)}(\bm k^{\prime} ,\bm k^{\prime\prime}) 
t^{(0)}(\bm k^{\prime\prime} ,\bm k) 
C_0(\bm k^\prime - \bm k^{\prime\prime},  \bm k^{\prime\prime} - \bm k)
\right.\\
\left.
+
\left(
2 t^{(1)}(\bm k^{\prime} ,\bm k^{\prime\prime}) 
t^{(1)}(\bm k^{\prime\prime} ,\bm k) 
+
\left(
t^{(2)}(\bm k^{\prime} ,\bm k^{\prime\prime}) 
t^{(2)}(\bm k^{\prime\prime} ,\bm k) 
+
2 t^{(3)}(\bm k^{\prime} ,\bm k^{\prime\prime}) 
t^{(3)}(\bm k^{\prime\prime} ,\bm k) 
\right)
\,\bm n_1 \cdot \bm n_2
\right)
C_\text{ex}(\bm k^\prime - \bm k^{\prime\prime},  \bm k^{\prime\prime} - \bm k)
\right]
\label{U2nd-corr-functions}
\end{multline}
or equivalently 
\begin{multline}
V^{(2)}(\bm k' ,\bm k) 
= \int \frac{\diff \bm k''}{2\pi^2}   \frac{\mathscr{W}(\bm k', \bm k'') \mathscr{W}(\bm k'', \bm k)  }{k_0^2 - {k''}^2+ i\varepsilon} \left[ 
f^{(0)}(\bm k^{\prime} ,\bm k^{\prime\prime}) 
f^{(0)}(\bm k^{\prime\prime} ,\bm k) 
C_0(\bm k^\prime - \bm k^{\prime\prime},  \bm k^{\prime\prime} - \bm k)
\right.\\
\left.
+
\left(
2 f^{(1)}(\bm k^{\prime} ,\bm k^{\prime\prime}) 
f^{(1)}(\bm k^{\prime\prime} ,\bm k) 
+
\left(
f^{(2)}(\bm k^{\prime} ,\bm k^{\prime\prime}) 
f^{(2)}(\bm k^{\prime\prime} ,\bm k) 
+
2 f^{(3)}(\bm k^{\prime} ,\bm k^{\prime\prime}) 
f^{(3)}(\bm k^{\prime\prime} ,\bm k) 
\right)
\,\bm n_1 \cdot \bm n_2
\right)
C_\text{ex}(\bm k^\prime - \bm k^{\prime\prime},  \bm k^{\prime\prime} - \bm k)
\right].
\label{V2-final}
\end{multline}
\end{widetext}
The first term in Eq.~(\ref{V2-final}) describing spin-isospin averaged individual nucleon scattering on two nucleons is similar to Eq.~(6.5) of Foldy and Walecka~\cite{Foldy:1969rk}.
The term proportional to $\hat f^{(2)}_1 \hat f^{(2)}_2$ ($\hat f^{(1)}_1 \hat f^{(1)}_2$) corresponds to spin (isospin) exchange between the intermediate pion and two nucleons, keeping the scattered nucleus in the ground state (see Fig.~\ref{U-second-ord-exchange-plt} for an example).
Similarly, the term $\hat f^{(3)}_1 \hat f^{(3)}_2$ describes the simultaneous exchange of both spin and isospin.

At the initial step of our calculation, the Pauli principle was included in the pion-nucleus potential through the antisymmetric nature of the nucleon wave functions. 
However, for the first-order potential,~Eq.~(\ref{t-rho-potential}),
this property was lost after the integration over nucleon momenta within the factorization approximation~\cite{goldberger1948interaction, Landau:1974vk}.
The obtained structure of the second-order correction, Eq.~(\ref{V2-final}), explicitly involves two types of the two-nucleon correlation function and arises primarily from the Pauli principle.
This can, e.g., be understood by considering the process with zero momentum transfer to each of the nucleons involved in the second-order scattering. 
As shown in Fig.~\ref{C-D-plot}, for this situation, $C_0(0, 0) = 0$ and $C_\text{ex}(0, 0) = A$.
As a result, the first term in Eq.~(\ref{V2-final}), describing the process which does not change the nucleon quantum numbers, makes zero contribution to the integral.
In contrast, the second term is proportional to the nonzero $C_\text{ex}$ correlation function. 
It corresponds to the situation when, after the pion scattering on the first nucleon, this nucleon changes its spin and/or isospin, acquiring quantum numbers already occupied by another nucleon.
In this way, the second-order part of the potential, Eq.~(\ref{V2-final}), introduces the Pauli corrections to the model.

\begin{figure}[!htb]
\begin{minipage}[h]{0.99\linewidth}
\center{\includegraphics[width=1\linewidth]{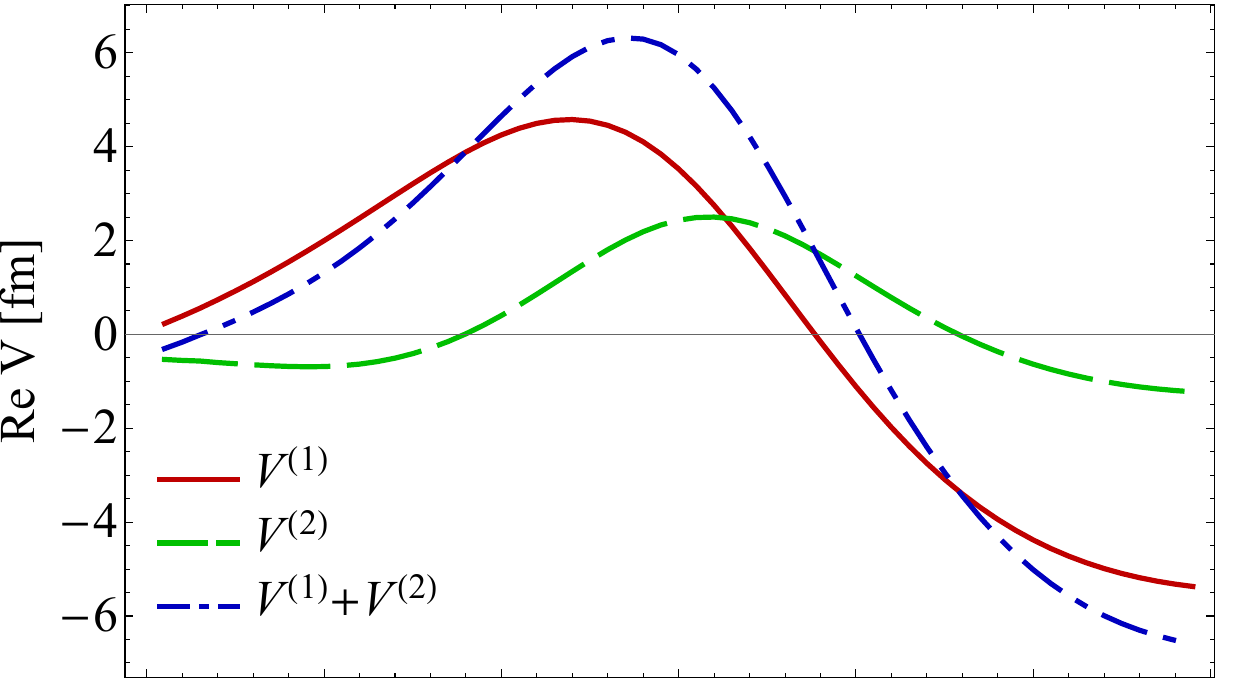}} \\
\end{minipage}
\vfill
\begin{minipage}[h]{0.999\linewidth}
\center{\includegraphics[width=1\linewidth]{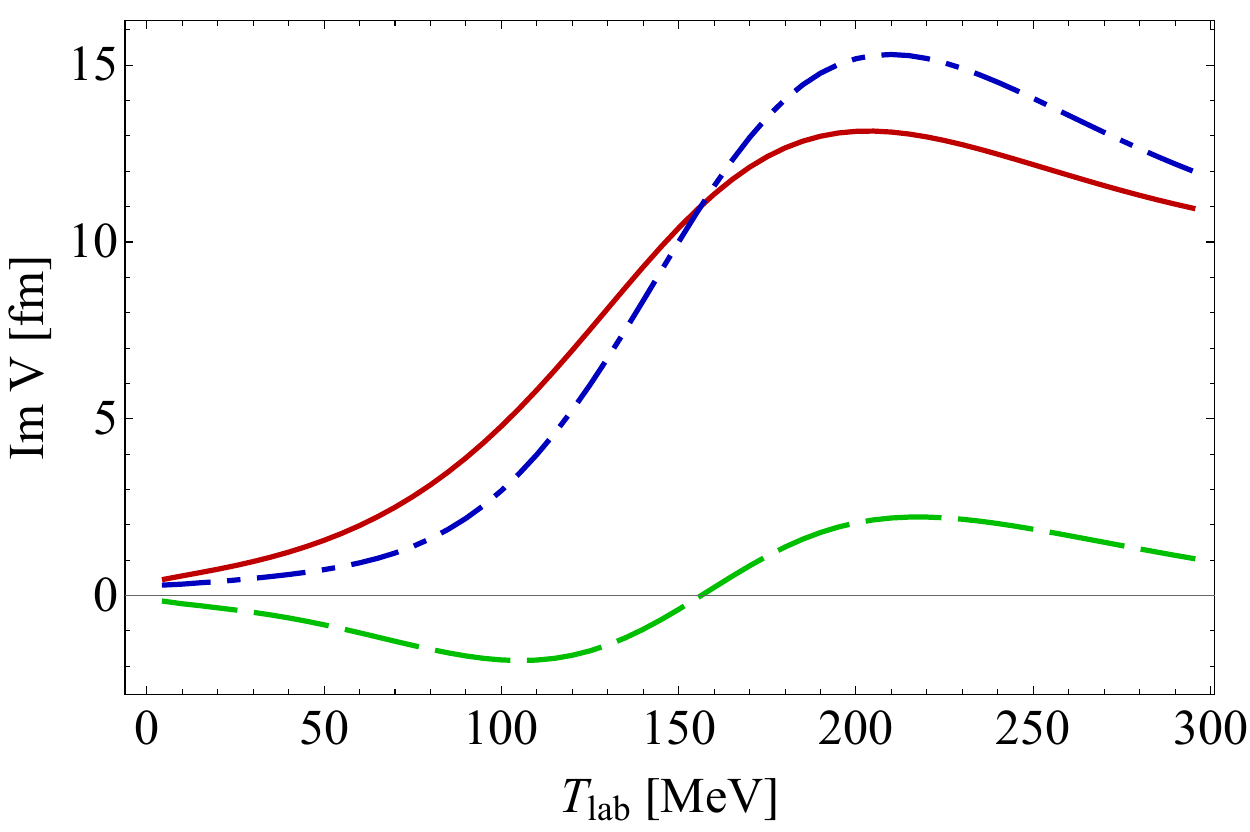}} \\
\end{minipage}
\caption{The on-shell forward pion-nucleus potential for ${}^{12}$C as a function of pion laboratory kinetic energy for parameters given by fit~1 in Table~\ref{tabl:fit-Sigma}. 
The upper and lower panels are for real and imaginary parts, respectively.  
The solid red curves represent the first-order part, $V^{(1)}(\bm k_0, \bm k_0)$ given by Eq.~(\ref{U1st-fin}), with the on-shell momentum $\bm k_0$ corresponding to $T_\text{lab}$.
The dashed green curves correspond to the second-order part, $V^{(2)}(\bm k_0, \bm k_0)$ given by Eq.~(\ref{V2-final}), and the dash-dotted blue are the sums of these two contributions.}
\label{fig-U1st-2nd-comp}
\end{figure}

Figure~\ref{fig-U1st-2nd-comp} demonstrates the first- and second-order parts of the pion-nucleus potential for on-shell forward scattering on ${}^{12}$C. 
As Pauli blocking limits the phase space available to the struck nucleon, the second-order correction to the potential leads to a reduction of the imaginary part of the potential. 
Around $T_\text{lab} = 160$~MeV, the struck nucleon on-shell momentum becomes close to the Fermi momentum, $p_F \approx \SI{1.36}{fm^{-1}}$, and the imaginary part of Eq.~(\ref{V2-final}) changes sign.

 In Appendix~\ref{sec:2nd-explicit}, we further discuss the second-order correction, Eq.~(\ref{V2-final}).

%%%%%%%%%%%%%%%%%%%%%%%%%%%%%%%%%%%%%%%%%%%%%%%%%%%%%%%%%%%%%%%%%%%%%%%%%%%%%%%%%%%%%%%%%%%%%%%%%%%%%
\subsection{Medium modifications}
\label{subsec-bound-nucl}
%%%%%%%%%%%%%%%%%%%%%%%%%%%%%%%%%%%%%%%%%%%%%%%%%%%%%%%%%%%

An essential part of the pion-nucleus total cross section for all energies up to \SI{300}{MeV} comes from pion absorption~\cite{Ashery:1981tq}.
In the nuclear medium, the pion can be absorbed by one or more nucleons, which indicates that intermediate states without a pion should also contribute to the pion-nucleus effective potential. 
This mechanism is usually referred to as "true absorption" to distinguish it from the flux loss due to scattering through many open inelastic channels.
However, even zero-energy pion absorption on a single nucleon results in a momentum  $\sqrt{2m_N m_\pi} \approx \SI{2.6}{fm^{-1}}$ to be carried off by the nucleon.
This value is very large for a nucleon within a nucleus, which means the single-nucleon absorption is significantly suppressed~\cite{Bassalleck:1978iz}.
As a result, the true absorption originates from many-body mechanisms.

Early models of pion absorption hypothesized dominance of two-nucleon pion absorption~\cite{Ashery:1986nt}, where the pion is scattered on one nucleon and then absorbed by another.
Following this assumption, the pioneering work of Ref.~\cite{Ericson:1966fm} introduced additional phenomenological terms proportional to the square of the nuclear density in the pion-nucleus potential to allow for true absorption. 
However, it was shown both experimentally~\cite{Weyer:1990ye} and theoretically~\cite{Oset:1986yi} that the absorption process is more complicated and the three-nucleon mechanism yields a significant fraction of the total absorption cross section in the resonance region and above. 

As a result of the above, the pion-nucleon interaction is significantly modified in the presence of surrounding nucleons.
In general, this means that the medium-modified scattering coefficients $b_{0,1}$, $c_{0,1}$, and $s_{0,1}$ are not only functions of the reaction energy but also acquire a dependence on nuclear density $\rho(r)$. 
Even if the exact form of this dependence were known, its inclusion in the momentum space approach would not be trivial.
To solve this difficulty, we need to use the fact that the pion interacts mainly with a limited part of the nucleus due to strong absorption, which results in the existence of an effective nuclear density. 
Reference~\cite{Seki:1983sh} studied the correlations between the $\rho(r)$ and $\rho^2(r)$ terms of the pion-nucleus optical potential from the threshold to $T_\text{lab} = \SI{50}{MeV}$.
It was proven that an effective density $\rho_e$ could be defined such that the substitution $\rho^2(r) \longleftrightarrow \rho_e \rho(r)$ would result in approximately the same binding energies and scattering amplitudes for various nuclei in the range from ${}^{12}$C to ${}^{208}$Pb.

Even though we only fit in the range 80–\SI{180}{MeV} pion kinetic energy, we still wish to check our model predictions at lower energies.
For this reason, the in-medium influence on the $s$-wave scattering should be considered.
Moreover, due to $s$-$p$-wave interference in the second-order part of the pion-nucleus potential constructed in Sec.~\ref{sec:U-2nd}, both isoscalar $b_0$ and isovector $b_1$ parts of the $s$-wave pion-nucleon scattering amplitude are substantial even at high energies (see Appendix~\ref{sec:2nd-explicit} for details).
In the following, we subsequently describe modifications of both $s$- and $p$-wave pion-nucleon scattering.
The primary effect of the Pauli exclusion principle, which reduces the phase space accessible to the struck nucleon, is incorporated by explicitly calculating the second-order correction to the pion-nuclear potential as described in Sec.~\ref{sec:U-2nd}.

%%%%%%%%%%%%%%%%%%%%%%%%%%%%%%%%%%%%%%%%%%%%%%%%%%%%%%%%%%%
% \subsubsection{$P_{33}$ modification}
% \label{subsec-P33-mod}
\subsubsection{\texorpdfstring{$P_{33}$}{P} modification}
\label{subsec-P33-mod}
%%%%%%%%%%%%%%%%%%%%%%%%%%%%%%%%%%%%%%%%%%%%%%%%%%%%%%%%%%%
In our approach, we assume that for the $p$-wave interaction, only the resonant $P_{33}$ channel is changed in the nuclear medium, keeping all other small partial-wave amplitudes at their free values taken from SAID.

The interaction of the $\Delta$ isobar with the surrounding nucleons significantly modifies the $f^1_{33}$ partial amplitude.
A comparably long lifetime of $\Delta$ on the nuclear scale and its mean free path within a nucleus of around \SI{1}{fm} suggest that the $\Delta$ is a nuclear quasiparticle that may still be treated effectively as a separate baryonic species without considering the intrinsic quark dynamics. 
The open inelastic channels involving many-body interactions, e.g., the two-body absorption ($\pi NN \rightarrow \Delta N \rightarrow NN$) and three-body absorption~\cite{Oset:1986yi}, considerably affect the $\Delta$-resonance decay width inside nuclear matter.
As a result, we consider the in-medium interactions effectively by a renormalization of the intermediate $\Delta$ propagator by the complex self-energy $\Sigma_\Delta$ function:
\be
K^{1 (\Delta)}_{33} (\Sigma_\Delta) =
\frac1{k_{0, \text{2cm}}} \frac{m_\Delta}{m_\Delta + W} \frac{ \Gamma_\Delta}{m_\Delta + \Sigma_\Delta - W}.
\label{K-in-medium}
\ee
In this approach, the dressed resonance leads to a complex $K^1_{33}$ matrix element in which the effective many-body $p$-wave absorption is automatically included in the model.

The $\Delta$ self-energy $\Sigma_\Delta$ in a finite nucleus is, in general, nonlocal~\cite{Oset:1979bi}.
However, we are looking for a simple phenomenological parametrization of $\Sigma_\Delta$, which would still provide a reasonable description of the data.
Since the real part of $\Sigma_\Delta$ has a weak energy dependence~\cite{GarciaRecio:1989xa}, it is often approximated to be constant.
In contrast, the imaginary part of $\Sigma_\Delta$ is regularly considered as a function of the pion energy~\cite{Nieves:1991ye}.
However, we have found that including the second-order part of the pion-nucleus potential, Eq.~(\ref{V2-final}), allows us to neglect the energy dependence of $\im\Sigma_\Delta$.
As a result, we treat $\re \Sigma_\Delta$ and $\im \Sigma_\Delta$ as two energy-independent $p$-wave model parameters determined by fitting the experimental data for pion-carbon scattering in Sec.~\ref{sec:fit}.  

The pion absorption process by a nucleus, unlike scattering, can occur even at pion energies below its mass.
While the $\Delta$ width, Eq.~(\ref{Gamma_Delta}), starts at $\omega = m_\pi$, the imaginary part of the $\Delta$ self-energy inside nuclear matter starts at $\omega = 0$~\cite{HjorthJensen:1993rm}. 
As a result, we expect the constant $\im \Sigma_\Delta$ assumption to be applicable not only in the $\Delta$-resonance region but also at low energies.

%%%%%%%%%%%%%%%%%%%%%%%%%%%%%%%%%%%%%%%%%%%%%%%%%%%%%%%%%%%
\subsubsection{Isoscalar \texorpdfstring{$s$}{s}-wave modification}
\label{subsec-b0}
%%%%%%%%%%%%%%%%%%%%%%%%%%%%%%%%%%%%%%%%%%%%%%%%%%%%%%%%%%%

The $s$- and $p$-wave true absorption within the optical potential formalism~\cite{Ericson:1966fm} is typically characterized by two complex parameters denoted as $B_0$ and  $C_0$, respectively. 
It is assumed to be based on a two-nucleon mechanism.
As was pointed out above, we effectively take into account various inelastic in-medium $p$-wave channels by introducing the $\Delta$ self-energy. 
Thereby, we expect $\Sigma_\Delta$ to incorporate absorption corrections associated with $C_0$. 
Furthermore, we limit our consideration of analyses based on the optical model of Ref.~\cite{Ericson:1966fm} to only the $s$-wave part of the potential:
\be
U^{(s)}(r) \propto b_0 \rho(r) + B_0 \rho^2(r),
\label{U-absorption}
\ee
where phase space factors were omitted for simplicity, and the first term here corresponds to the Fourier-transformed first term in Eq.~(\ref{U1st-fin}).
Due to the correlation between $b_0$ and $B_0$ pointed out in Ref.~\cite{Seki:1983sh}, the two terms can be lumped together, resulting in an effective modification of the isoscalar parameter $b_0$:
\be
U^{(s)}(r) \propto (b_0 + \Delta b_0) \rho(r),
\label{U-absorption-lin}
\ee
where $\Delta b_0 = B_0 \rho_e$.

In our model, we assume the following in-medium modification of the isoscalar scattering parameter: 
\be
b_0^\text{bound}( T_\text{lab}) = b_0^\text{free}( T_\text{lab}) + \Delta b_0( T_\text{lab}),
\label{b0-bound}
\ee
where $b_0^\text{free}(T_\text{lab})$ is given by Eq.~(\ref{f-sp-coefs-b0}), and the complex parameter $\Delta b_0$ effectively takes into account not only true absorption but also all possible in-medium modifications.

Comparing Eqs.~(\ref{U-absorption-lin}) and~(\ref{b0-bound}), we see that pionic atom analyses with the $s$-wave part of the potential given by Eq.~(\ref{U-absorption}) can provide us with information about the threshold value of $\Delta b_0$ (see more detailed discussion in Appendix~\ref{sec:2nd-explicit}).
Using the value $B_0 = 0.189 \text{ fm}^4$ from Ref.~\cite{piAF:2022gvw}, we arrive at the following result for the imaginary part of the in-medium isoscalar correction: 
\be
\im \Delta b_0(0) = \frac{1 + m_\pi/2m_N}{1 + m_\pi/m_N} \rho_e  \im B_0(0) = \SI{0.017}{fm},
\label{b_0mod(0)}
\ee
where we restore the phase space factor and use the $s$-wave effective density $\rho_e = 0.6 \rho_0 \approx \SI{0.1}{fm^{-3}}$ deduced from the overlapping of pion and nucleus densities for pionic atoms~\cite{Kienle:2004hq}.

The resulting imaginary part of $\Delta b_0$ is assumed to be
\be
\im\Delta b_0( T_\text{lab}) = \im\Delta b_0(0) + \alpha_{b_0} k_{0,\text{2cm}}(T_\text{lab}),
\label{b_0mod(T)}
\ee
where $\alpha_{b_0}$ is the effective $s$-wave isoscalar slope parameter, determined by the fitting procedure, and $k_{0,\text{2cm}}(T_\text{lab})$ is the on-shell pion-nucleon c.m. momentum corresponding to $T_\text{lab}$.

Performing fitting with various parametrizations of the real part of $\Delta b_0$, we conclude that while $\im \Delta b_0$ is an important parameter of our model, the resulting $\re \Delta b_0$ is always close to zero and can be neglected. 
For this reason, we assume $\Delta b_0$ is purely imaginary, given by Eqs.~(\ref{b_0mod(T)}) and~(\ref{b_0mod(0)}).

%%%%%%%%%%%%%%%%%%%%%%%%%%%%%%%%%%%%%%%%%%%%%%%%%%%%%%%%%%%
\subsubsection{Isovector \texorpdfstring{$s$}{s}-wave modification}
\label{subsec-b1}
%%%%%%%%%%%%%%%%%%%%%%%%%%%%%%%%%%%%%%%%%%%%%%%%%%%%%%%%%%%

Our approach includes the in-medium modification of the $s$-wave amplitude $b_1$, as it was successfully applied for the $s$-wave pionic atom~\cite{Friedman:2002sys,Suzuki:2002ae} and low-energy pion-nucleus~\cite{Friedman:2002sys,Friedman:2004jh} potentials.

To lowest order in the chiral expansion, the parameter $b_1$ for the scattering of a pion on a free nucleon in the threshold region is given by the Tomozawa-Weinberg expression~\cite{Weinberg:1966kf}:
\be
b_1^\text{TW} = - \frac1{8\pi f_\pi^2} \frac{m_\pi m_N}{m_\pi + m_N} \approx \SI{-0.11}{fm},
\ee
where $f_\pi = $ \SI{92.2}{MeV} is the free-space pion decay constant~\cite{Holstein:1990ua,*Descotes-Genon:2005wrq}.
The value for $b_1$ obtained in this way is very close to the empirical one not only at low energies but also in the resonance region.
According to the suggestion by Weise~\cite{Weise:2001sg, Weise:2000xp}, the medium dependence of the pion decay constant $f_\pi$, which is related to the quark condensate, is in the simplest approximation given by a linear function of the nuclear density
\be
{f_\pi^*}^2(\rho) = f_\pi^2 - \frac{\sigma}{m_\pi^2} \rho,
\ee
where $\sigma$ is the pion-nucleon sigma term~\cite{Gasser:1990ce}.
As a result, the in-medium threshold parameter $b_1$ is obtained as
\be
b_1^\text{bound} = \frac{b_1^\text{free}}{1 - \sigma \rho / m_\pi^2 f_\pi^2}.
\ee
This simple model successfully described both pionic atoms~\cite{Friedman:2001tp, Friedman:2002ix, Kienle:2004hq} and low-energy pion-nucleus scattering~\cite{Friedman:2005pt}.

For energies above the threshold, $b_1$ is not constant but a slowly varying function of energy, which is, however, still close to its threshold value even in the $\Delta$-resonance region.
In our analysis, we assume the following weak energy dependence of $b_1$:
\be
b_1^\text{bound}(T_\text{lab}) = b_1^\text{free}(T_\text{lab}) + \Delta b_1, 
\ee
where $b_1^\text{free}(T_\text{lab})$ is given by Eq.~(\ref{f-sp-coefs-b1}) and the energy-independent in-medium correction is taken from the pionic atom
\be
\Delta b_1 = b_1^\text{free}(0) \frac{\sigma \rho_e / m_\pi^2 f_\pi^2}{1 - \sigma \rho_e / m_\pi^2 f_\pi^2} = \SI{-0.044}{fm},
\ee
where following Ref.~\cite{Friedman:2019zhc} $\sigma = \SI{57}{MeV}$ is taken and $b_1^\text{free}(0) \approx \SI{-0.122}{fm}$~\cite{Baru:2010xn}. 
The resulting value of $b_1$ at the effective density $\rho_e$ is in quantitative agreement with microscopic~\cite{Doring:2007qi} and chiral calculations~\cite{Meissner:2001gz}, and the recent deeply bound pionic atoms analysis~\cite{piAF:2022gvw}. 
The effect of double scattering to higher order was shown to be a minor correction~\cite{Kaiser:2001bx}.

\section{Results and discussion}
\label{sec:fits-results}

In this section, we apply the model developed in Sec.~\ref{sec:Uopt} to fit $\pi^\pm$-${}^{12}$C scattering data.
As a result of the fit, we determine our model's three energy-independent real parameters:  the real and imaginary parts of the effective $\Delta$-resonance self-energy, $\re\Sigma_\Delta$ and $\im\Sigma_\Delta$ entering Eq.~(\ref{K-in-medium}), and the slope of the imaginary $s$-wave isoscalar amplitude, $\alpha_{b_0}$ in Eq.~(\ref{b_0mod(T)}).
Subsequently, the same fixed parameters are used to compare our predictions for the pion scattering on ${}^{16}$O, ${}^{28}$Si, and ${}^{40}$Ca with available experimental data.

\subsection{Observables}
\label{sec:observables}

The Coulomb interaction significantly influences the charged pion scattering process. The differential elastic cross section is given by
\be
\frac{d\sigma}{d \Omega}(\theta) = \left| F_{C,p}(\theta) + F_{NC}(\theta) \right|^2,
\label{diff-cross-section}
\ee
where we have separated  the
Coulomb distorted strong-interaction amplitude, $F_{NC}$, from the singular point-charge Coulomb amplitude
\be
F_{C,p}(\theta) = - \frac{\eta_c}{2 k_0 \sin^2(\theta/2)} \exp\{2i [\sigma_0 - \eta_c \log\sin(\theta/2)]\},
\label{point-Coul-amp}
\ee
with the Lorentz-invariant Sommerfeld parameter $\eta_c = \alpha Z Z_\pi \omega_\text{lab}/k_\text{lab}$,
where $Z$($Z_\pi$) is the nucleus (pion) charge.
The Coulomb phase shifts $\sigma_l$ are defined as
\be
e^{2i\sigma_l} = \frac{\Gamma(1 + l + i \eta_c)}{\Gamma(1 + l - i \eta_c)}, 
\ee
with Euler's gamma function $\Gamma$.

The Coulomb-nuclear interference term is split in partial waves as:
\be
F_{NC}(\theta) =  \sum_l (2l + 1) e^{2i \sigma_l} F_l \, P_l(\cos \theta), 
\label{F_NC-expansion}
\ee
where 
\be
F_l = \frac12 \int \diff\cos\theta \, F(\bm k', \bm k) P_l(\cos\theta)
\label{F-l-def}
\ee
and $\cos\theta = \bm k' \cdot \bm k / (k'k)$.

The full partial-wave amplitudes $F_l$ depend not purely on the hadronic interaction, Eq.~(\ref{LSh-pseudo-classical}), but also on the short-range part of the Coulomb potential due to the nuclear charge distribution and long-range Coulomb effects.   
To account for this nuclear-Coulomb interference, we apply the matching method of Vincent and Phatak~\cite{Vincent:1974zz} and an effective Coulomb modification of the reaction energy (see Appendix~\ref{sec:Coulomb} for details).

In addition to differential cross sections, experimental measurements also provide Coulomb-subtracted angle-integrated elastic and total cross sections.
The direct calculation provides the angle-integrated elastic cross section in the form
\be
\sigma^\text{El} = 4\pi  \sum_l (2l+1) \left| F_{l}  \right|^2.
\ee
Due to the optical theorem, the total cross section can be derived as
\be
\sigma^\text{Tot} = \frac{4\pi}{k_0}  \sum_l (2l+1) \im[F_{l}].
\label{sigma-Tot}
\ee
The pion-nuclear potential is a non-Hermitian operator giving rise to the reaction channel with the corresponding cross section, which can be calculated as
\be
\sigma^\text{R} = \sigma^\text{Tot} - \sigma^\text{El}.
\label{sigmaR-def}
\ee
The reaction cross section $\sigma_R$ includes quasielastic scattering, charge exchange, and true pion absorption.

The total cross section is significant for our analysis since it has a different sensitivity to the imaginary part of the potential as compared with the differential elastic cross section.

\subsection{Fit to \texorpdfstring{\ce{^{12}C}}{Lg} data}
\label{sec:fit}

\begin{table}[t]
\caption{Summary of the $\pi^\pm$-${}^{12}$C data}
% \begin{tabular*}{\textwidth}[t]{@{\extracolsep{\fill}}lclll@{}}
\begin{tabular}{clccc}
\hline\hline
Ref.   &    Facility         &    & $T_\text{lab}$ {[}MeV{]} & Observable  \\
\hline
\cite{Wright:1987qi}  &  LAMPF  & $\pi^-$ & $30$                  &  \\
\cite{Preedom:1981zz}  &  LAMPF   & $\pi^+$ & $30,50$                  & \\
\cite{Seth:1990rp}  &  LAMPF     & $\pi^\pm$ & $30, 50$   &  \\
\cite{Sobie:1984js}  &  TRIUMF     & $\pi^\pm$ & $50$    &  $d\sigma^\text{El}/d\Omega$  \\
\cite{Ritchie:1990vf}  &  LAMPF     & $\pi^\pm$ & $50$      &  \\
\cite{Blecher:1984ng}  &  LAMPF     & $\pi^\pm$ & $65,80$                  &  \\
\cite{Antonuk:1983gy}  &  SIN  & $\pi^\pm$ & $100$  &   \\
\cite{Binon:1970ye}  &  CERN     & $\pi^-$   & 120–280   &  \\
\hline
\cite{Binon:1970ye}  &  CERN     & $\pi^-$   & 90–280   &      $\sigma^\text{Tot}, \sigma^\text{Re}, \sigma^\text{El}$             \\
\cite{Clough:1974qt}  &  RAL      & $\pi^\pm$ & 89–855                 &  $\sigma^\text{Tot}$               \\
\cite{Ashery:1981tq}  &  SIN     & $\pi^\pm$ & 85–245                 &      $\sigma^\text{Tot}, \sigma^\text{Re}, \sigma^\text{El}$       \\
\cite{Saunders:1996ic}  &  TRIUMF   & $\pi^\pm$ & 42–65   &     $\sigma^\text{Tot}, \sigma^\text{Re}$           
\\
\cite{Meirav:1988pn}  &  TRIUMF   & $\pi^\pm$ & 50–80                      &      $\sigma^\text{Re}$   \\
\cite{Friedman:1991it}  &  TRIUMF  & $\pi^\pm$ & 20, 30                       &     $\sigma^\text{Re}$  \\
\hline\hline
\end{tabular}
\label{tabl:12C}
\end{table}

Various groups intensively studied pion scattering on carbon from the 1970s through the 1990s. 
Table~\ref{tabl:12C} summarizes the $\pi^\pm$-${}^{12}$C scattering data used in our analysis.  
The dataset includes measurements of the total, angle-integrated elastic, reaction, and differential elastic cross sections done at different facilities: Schweizerisches Institut fur Nuklearforschung (SIN), Canada's particle accelerator center (TRIUMF), Los Alamos Meson Physics Facility (LAMPF), Rutherford Appleton Laboratory (RAL), and the European Organization for Nuclear Research (CERN).

As our aim is the extraction of the effective $\Delta$-resonance self-energy, in the fitting procedure, we only use the data having strong sensitivity to the $\Delta$ properties. 
We choose to fit the data in the energy range of 80–\SI{180}{MeV} pion laboratory kinetic energy, corresponding with the region up to the $\Delta$-resonance excitation energy on a nucleon. 
Furthermore, our treatment of the Coulomb interaction (the Coulomb energy shift, Eq.~(\ref{nuclear-Coulomb-pot})) relies on the small momentum transfer approximation. 
Thereby, we limit the fitting of the differential cross section data to momentum transfers $q \le \SI{1.5}{fm^{-1}}$.
Since $\sigma^\text{Tot}$, $\sigma^\text{R}$, and $\sigma^\text{El}$ are related through Eq.~(\ref{sigmaR-def}), we include in the fit only $\sigma^\text{Tot}$ and $\sigma^\text{El}$ if all three observables are provided.

The best fit is found by minimizing the $\chi^2$ defined as
\begin{multline}
\chi^2 = \sum_i \sum_j^{n_i} \left[ \frac1{n_i} \left( \frac{ d\sigma_j^{\text{Data}_i} - N_i^{-1}  d\sigma_j}{\Delta d\sigma_j^{\text{Data}_i}} \right)^2 \right. \\
\left.
+ \left( \frac{N_i - 1}{\Delta N_i} \right)^2
\right]
+ \sum_i \sum_j^{n_i} \left( \frac{ \sigma_j^{\text{Data}_i} -  \sigma_j}{\Delta \sigma_j^{\text{Data}_i}} \right)^2,
\label{chi-tot}
\end{multline}
where the first (second) term represents a sum over differential (angle-integrated) cross section data sets and $n_i$ is  the number of data points in the dataset "$i$". 
Every differential cross section dataset $d\sigma^{\text{Data}_i}$ consists of correlated measurements taken at individual energies and is treated as a single uncorrelated point of the fit.
Since $\Delta d\sigma_j^\text{Data}$ contains only the sum of the statistical and the measured background errors, the normalization parameters $N_i$ are included to account for a fully correlated component between the data points of each differential cross-section data set (instrumental error).
The normalization parameters are allowed to vary, keeping the number of degrees of freedom (ndf) the same.

In our formalism, only three energy-independent fitting parameters are entering Eq.~(\ref{chi-tot}): the $\Delta$ self-energy parameters $\re\Sigma_\Delta$ and $\im \Sigma_\Delta$ in Eq.~(\ref{K-in-medium}), and the $s$-wave isoscalar slope parameter $\alpha_{b_0}$ in Eq.~(\ref{b_0mod(T)}).
We also tested the possibility of improving our fit by adding model parameters that modify the energy dependence of $\Sigma_\Delta$ and $b_0$.
We found that the resulting $\chi^2/\text{ndf}$ value can be improved only slightly in this way. 
However, the strong correlation between the parameters results in large uncertainties, making it impossible to determine the fitted parameters precisely. 
Moreover, introducing additional parameters does not improve our model predictions beyond the fitting range and for other nuclei.

As was mentioned in Sec.~\ref{sec:U-1st}, we also include the $d$-wave contribution to the first-order potential in addition to the traditional $s$- and $p$-wave terms.
This small component does not change the overall energy and momentum behavior of observables in a significant way. 
However, including the $d$-wave amplitude improves the resulting minimal $\chi^2$ of the fit by about 10\%. 
Note that the observables and fitting parameters are sensitive to the value of the effective bound nucleon mass, which in our calculation is taken as the average of the proton and neutron masses, $m_N = \SI{938.92}{MeV}$.

\begin{table}[h]
\caption{Potential parameters from fits to $\pi^\pm$-${}^{12}$C scattering data. For both fits $\text{ndf}=32$.}
\begin{tabular}{lcccccc}
% \begin{tabular*}{\textwidth}[t]{@{\extracolsep{\fill}}lcccccc@{}}
\hline\hline

fit  & $\re\Sigma_\Delta$ [MeV] & $\im\Sigma_\Delta$ [MeV] & $\alpha_{b_0}$ [$\SI{}{fm^2}$]  & $\chi^2$ & $\chi^2$/ndf  \\
\hline
1   & $12.9 \pm 1.3$ & $-33.2 \pm 0.8$ & $0.039 \pm 0.006$ & 53.4 & 1.67 \\
2   & $12.8 \pm 1.4$ & $-33.3 \pm 0.9$ & $0.040 \pm 0.006$ &  47.9 & 1.50  \\
\hline
\end{tabular}
% \end{tabular*}
% \caption{The self-energy parameters and the minimum $\chi^2$}
\label{tabl:fit-Sigma}
\end{table}

\begin{table}[h]
\caption{Fitted normalization parameters for fit 2. The first column indicates the pion energy, the second (fourth) column -- the experimental normalization uncertainties for $\pi^-$ ($\pi^+$) data, and the third (fifth) column -- the values of $\Delta N_i^\text{fit} = (N_i - 1)\, \%$ for $\pi^-$ ($\pi^+$) data obtained from the fit.}
\begin{tabular}{ccccc}
\hline\hline
 &  \multicolumn{2}{c}{$\pi^-$} & \multicolumn{2}{c}{$\pi^+$} \\
$T_\text{lab}$ [MeV] & $\Delta N_i \ [\%]$ & $\Delta N_i^\text{fit} \ [\%]$ & $\Delta N_i \ [\%]$ & $\Delta N_i^\text{fit} \ [\%]$  \\
\hline
80  & $\pm 5$ &  $0.5 \pm 4.8$ & $\pm 7$ &  $3.9 \pm 4.0$ \\
100 & $\pm 8$ &  $-5.6 \pm 4.5$ & $\pm 8$ &  $-1.8 \pm 4.5$ \\
120 & $\pm 5$ &  $1.1 \pm 3.4$  &   &  \\
150 & $\pm 4$ &  $0.1 \pm 2.6$  &   &  \\
180 & $\pm 3$ &  $-1.0 \pm 2.2$ &   &  \\
\hline
\end{tabular}
\label{tabl:fit-N}
\end{table}

\begin{table}[h]
\caption{The correlation matrix for fit 1.}
\begin{tabular}{c|ccc}
\hline\hline
 & $\re\Sigma_\Delta$ & $\im\Sigma_\Delta$ & $\alpha_{b_0}$ \\
\hline
$\re\Sigma_\Delta$   & $1$ & $0.53$ & $0.22$ \\
$\im\Sigma_\Delta$   & $0.53$ & $1$ & $-0.4$ \\
$\alpha_{b_0}$   & $0.22$ & $-0.4$ & $1$ \\
\hline
\end{tabular}
\label{tabl:fit-error}
\end{table}

\begin{figure*}[!ht]
\includegraphics[width=0.45\textwidth]{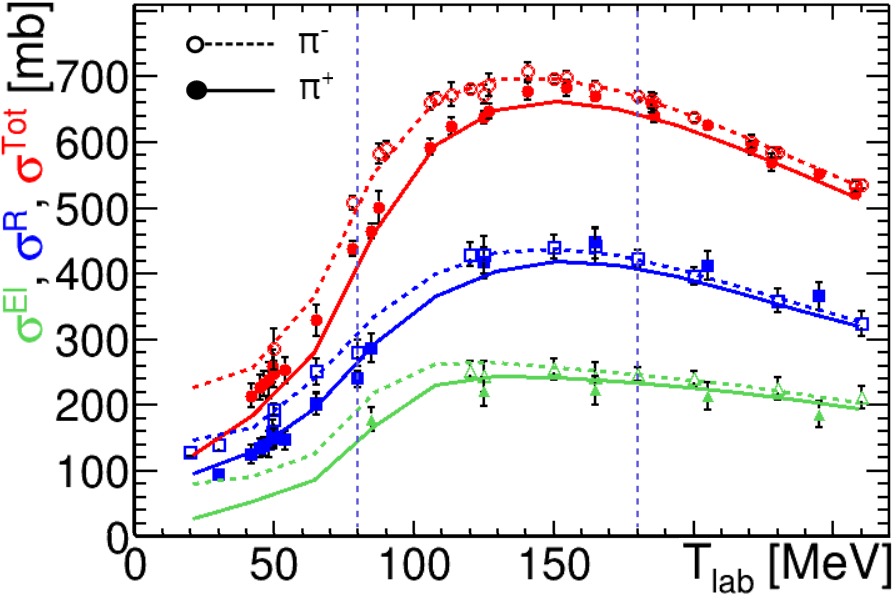} \includegraphics[width=0.45\textwidth]{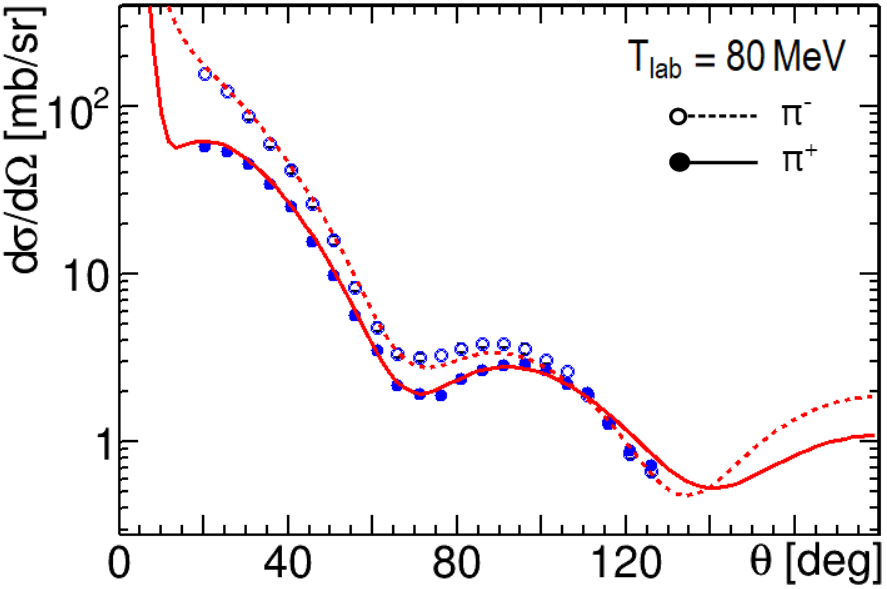}
\includegraphics[width=0.45\textwidth]{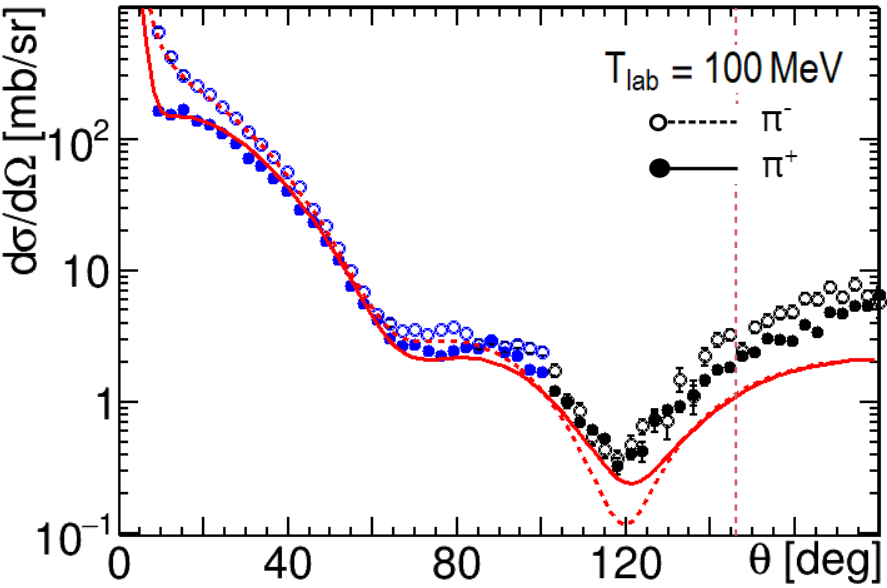}
\includegraphics[width=0.45\textwidth]{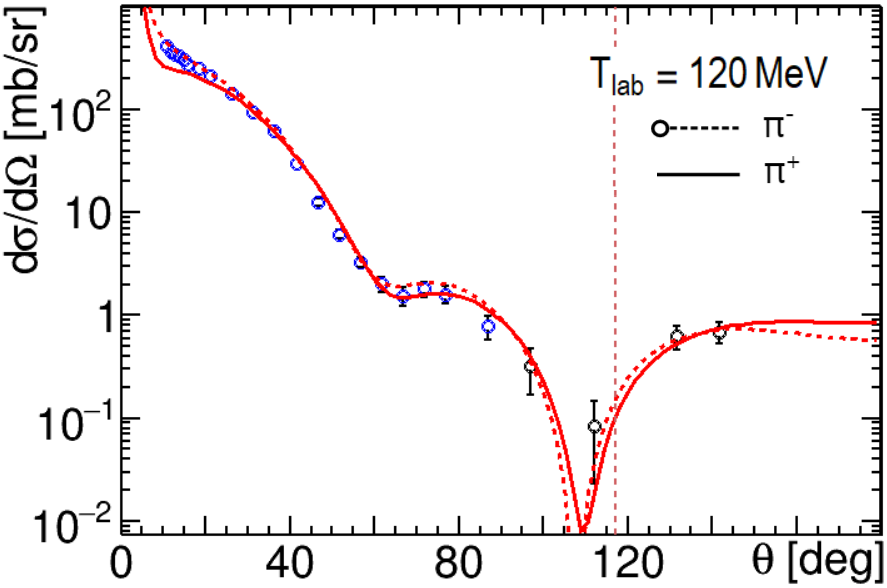}
\includegraphics[width=0.45\textwidth]{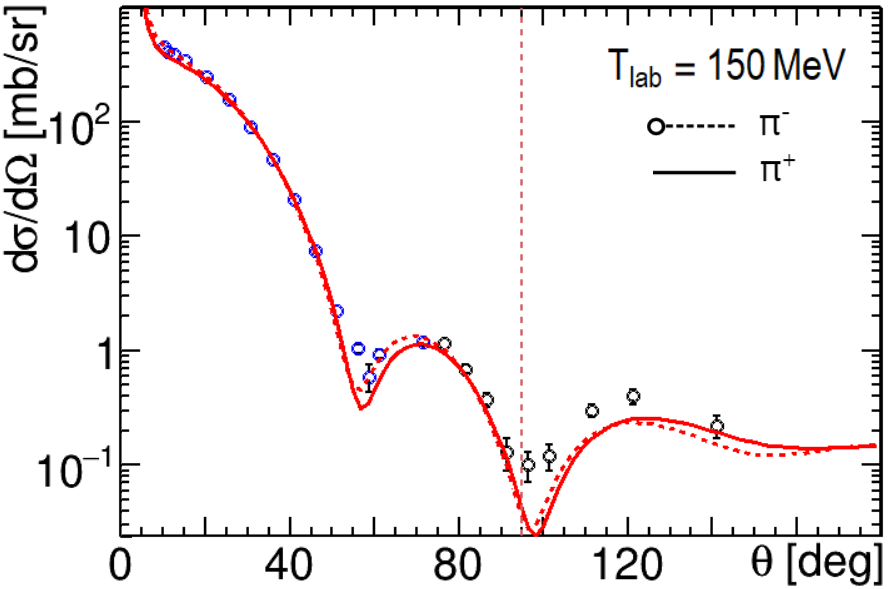}
\includegraphics[width=0.45\textwidth]{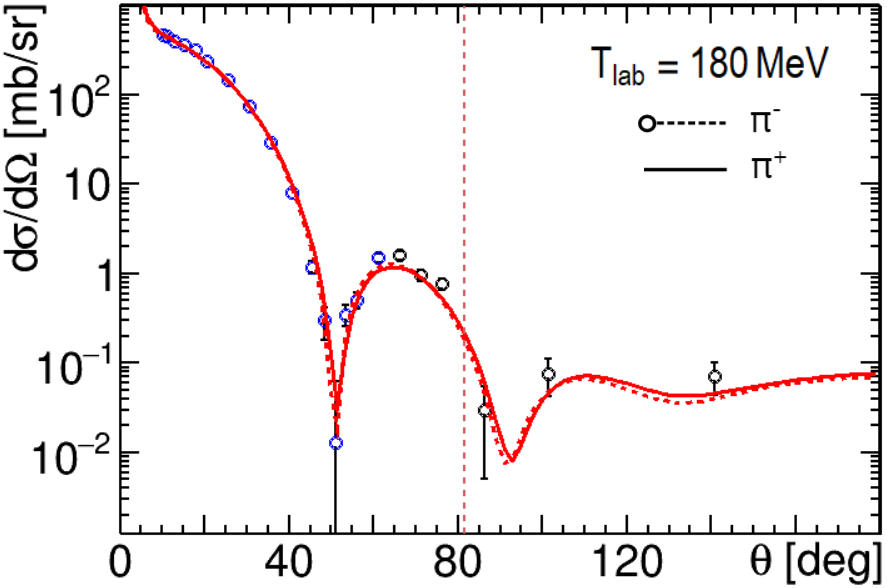}
\caption{
Fit to $\pi^\pm$-${}^{12}$C scattering data using the full second-order potential. 
The top left panel demonstrates the total (red curves and circles), integrated reaction (blue curves and squares), and elastic (green curves and triangles) cross sections. 
Solid curves and closed markers stand for $\pi^+$; dashed and open markers for $\pi^-$. 
The vertical dashed lines on the top left panel indicate the fitted energy range. Differential cross sections in the 80–\SI{180}{MeV} range as functions of the scattering angle in the c.m. frame are shown on other panels. 
The blue circles on the differential cross section plots correspond to the $q < \SI{1.5}{fm^{-1}}$ range, which was fitted; the black circles were not included in $\chi^2$. The dashed vertical lines on $d\sigma/d\Omega$ plots indicate the zero position of the form factor.
Table~\ref{tabl:12C} lists the experimental data presented in the plots.
}
 \label{12C-fit-pot}
\end{figure*}

Tables~\ref{tabl:fit-Sigma}–\ref{tabl:fit-error} summarize the fitting results.
Two fits were performed: fit~1 with fixed normalization parameters and fit~2 with $N_i$ also being fitted. 
The obtained model and normalization parameters are collected in Tables~\ref{tabl:fit-Sigma} and~\ref{tabl:fit-N}, respectively.  
The covariance matrix for fit~1 is given in Table~\ref{tabl:fit-error}.
As can be seen from Table~\ref{tabl:fit-Sigma}, letting $N_i$ free improves the resulting $\chi^2$ by about 10\%, keeping the fitted parameters almost unchanged.
The obtained normalization factors in Table~\ref{tabl:fit-N} are well within the provided experimental normalization uncertainties.
The consistency of the results strengthens the reliability of derived results and the robustness of the method.
In the following calculations, we will use the parameter set corresponding to fit~1.

\begin{figure*}[!ht]
\includegraphics[width=0.45\textwidth]{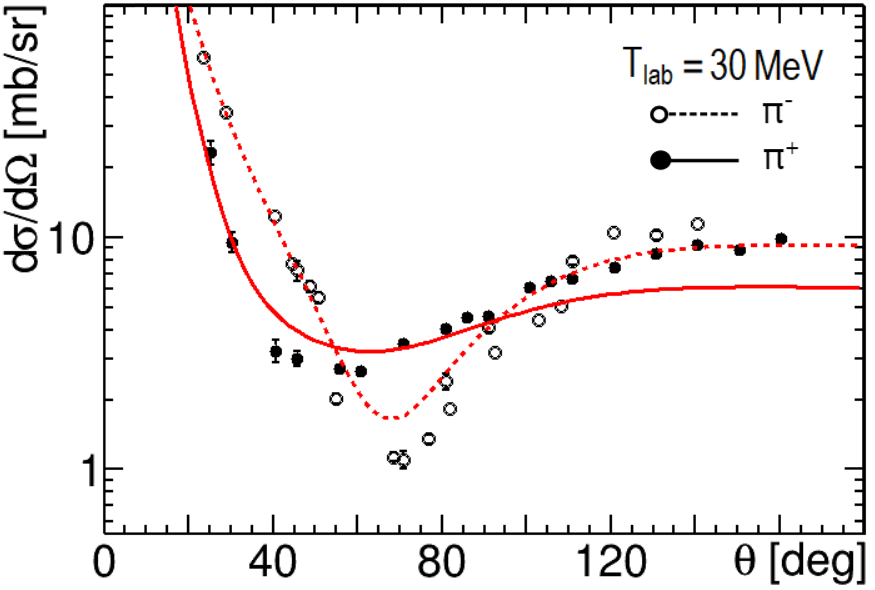}
\includegraphics[width=0.45\textwidth]{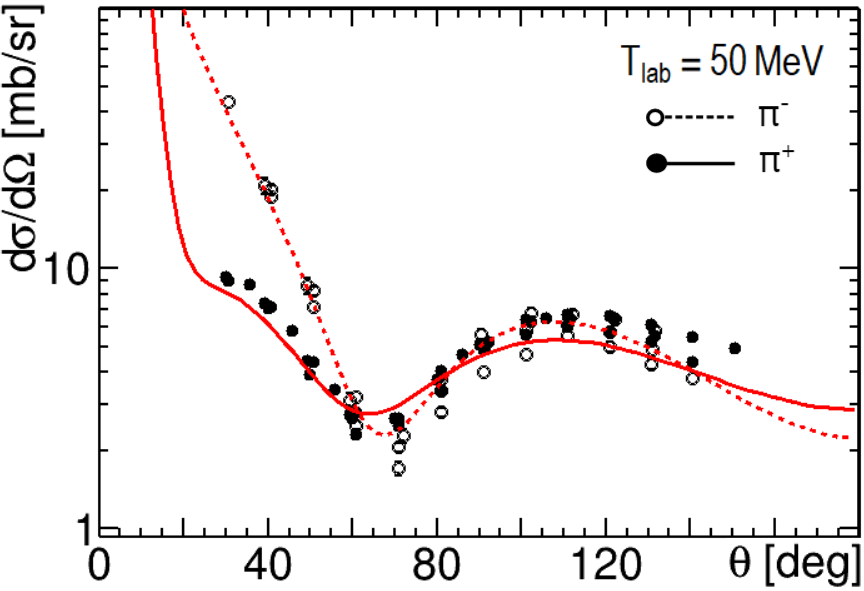}
\includegraphics[width=0.45\textwidth]{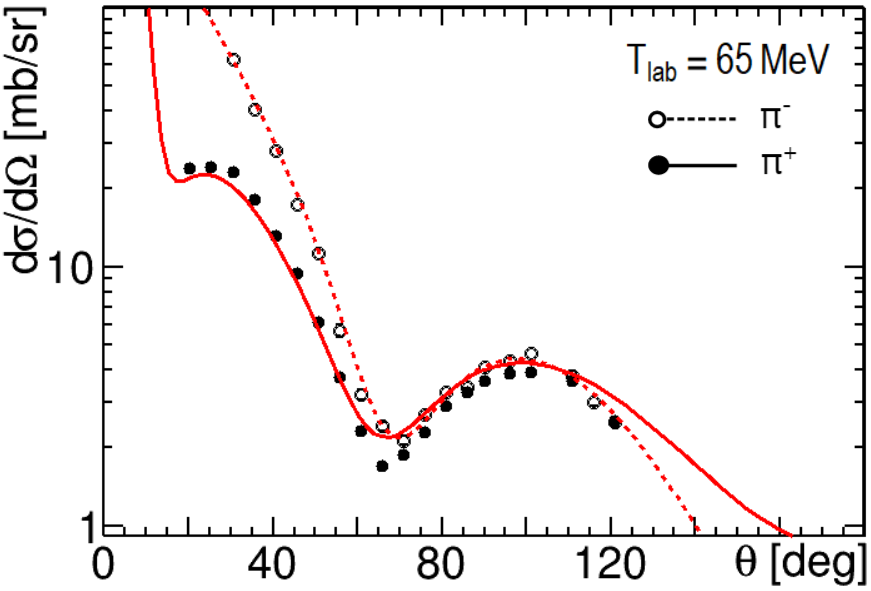}
\includegraphics[width=0.45\textwidth]{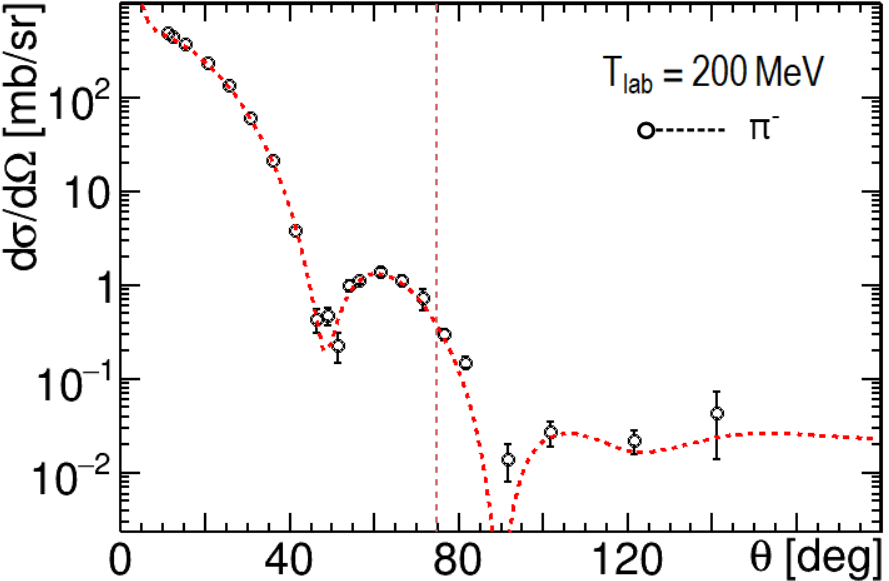}
\caption{
Comparison of the theoretical prediction based on fit 1 with the $\pi^\pm$-${}^{12}$C scattering data at kinetic energies outside the fitting range.
The meaning of the curves is the same as in Fig.~\ref{12C-fit-pot}.
Table~\ref{tabl:12C} lists the experimental data presented in the plots.
 }
\label{12C-outside}
\end{figure*}

In Fig.~\ref{12C-fit-pot}, we show the fitted data compared with the obtained theoretical curves corresponding to fit~1.
The resulting agreement is especially good for integrated and differential elastic cross sections for $\theta \le 60^{\circ}$.
Despite the fact that the data for $q> \SI{1.5}{fm^{-1}}$ were not fitted, our model demonstrates a fairly good description of the data even for large angles, except for the dataset at $\SI{100}{MeV}$, which seems to be an outlier.
The obtained differential cross section at \SI{100}{MeV} significantly undershoots the data for $\theta \gtrsim 120^{\circ}$. 
The same discrepancy was also reported in the $\Delta$-hole model analysis of Ref.~\cite{Antonuk:1983gy} and the phenomenological momentum-space potential approach of Ref.~\cite{Gmitro:1987un} with the $\rho^2(r)$-dependent second-order term.

As seen from the top left panel of Fig.~\ref{12C-fit-pot}, the integrated cross sections are well described outside the fitting range denoted by the vertical dashed lines. The predicted differential cross sections based on fit 1 outside the fitting range are plotted in Fig.~\ref{12C-outside}.  
The data measured at 65 and $\SI{200}{MeV}$ are well reproduced. Some deviations are seen at 30 and $\SI{50}{MeV}$, which can be fixed by a more precise treatment of the $s$-wave medium modifications.  
This involves a more intricate energy dependence for $b_0$ with a nonzero real part.
Incorporating nuclear excitations in the propagator $\hat G_{\alpha^*}$ of Eq.~(\ref{U2nd-def}) also becomes particularly crucial at low energies, where the transition strength to collective states becomes significant.

\begin{figure*}[!th]
\includegraphics[width=0.45\textwidth]{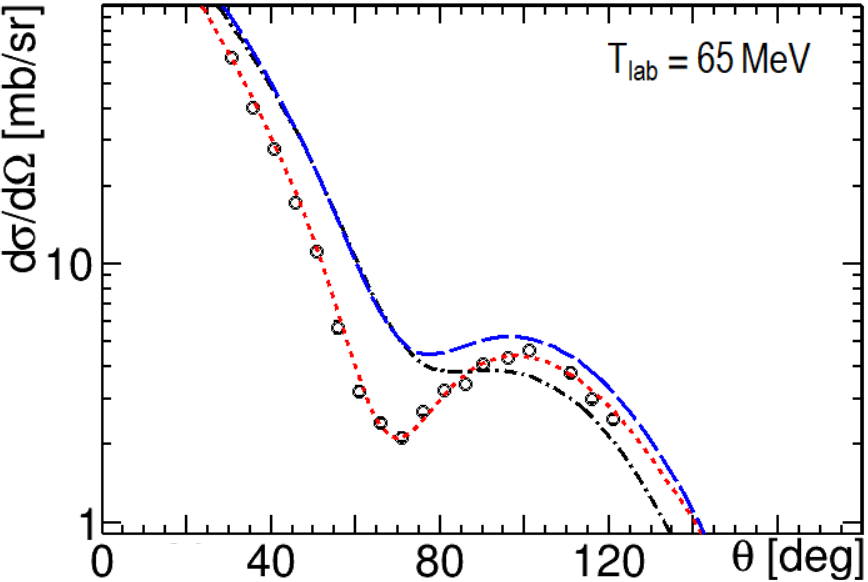}
\includegraphics[width=0.45\textwidth]{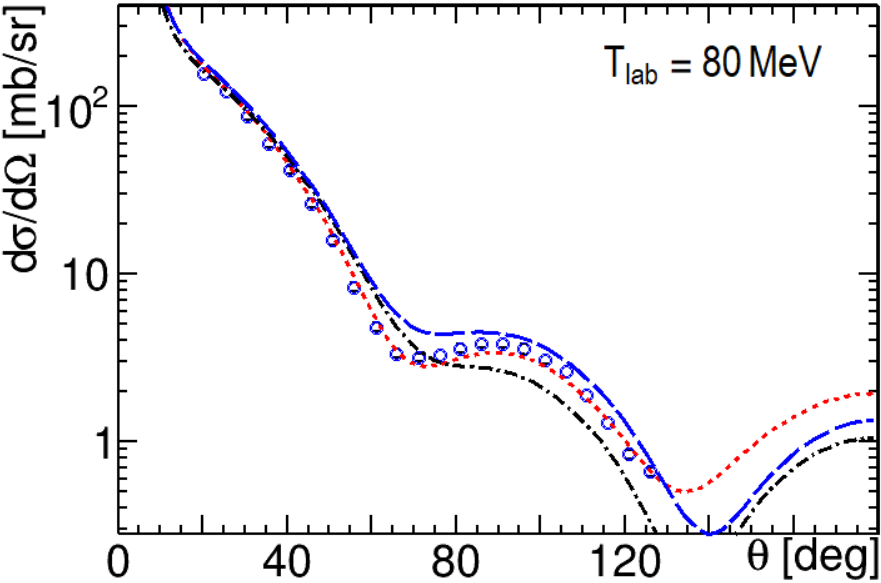}
\includegraphics[width=0.45\textwidth]{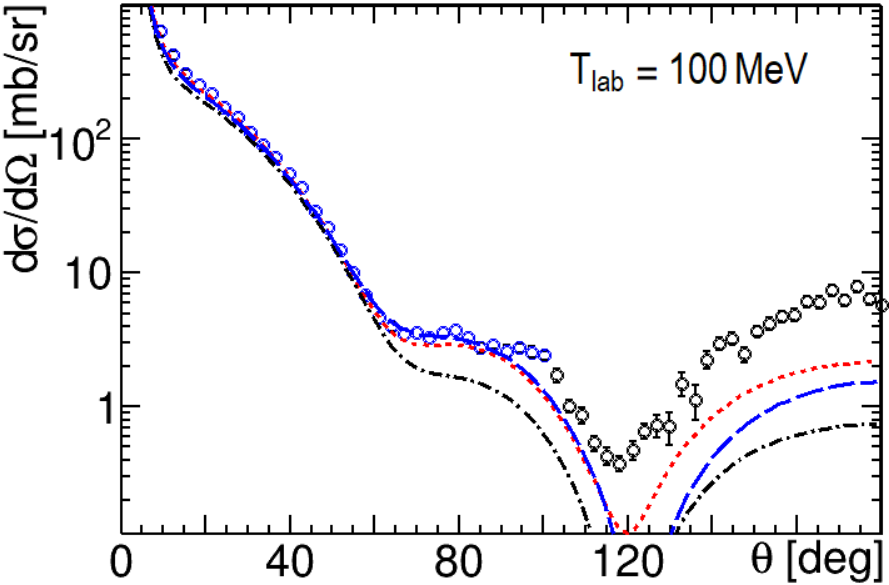}
\includegraphics[width=0.45\textwidth]{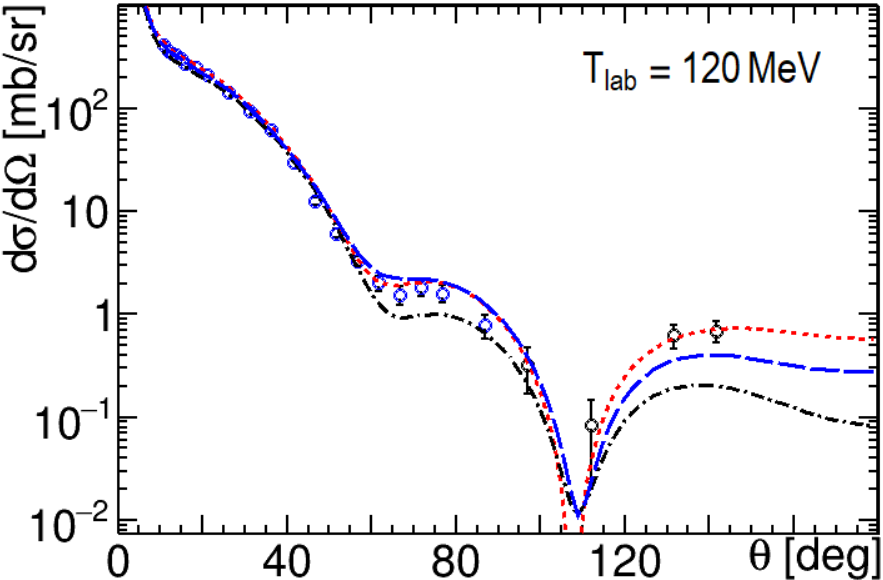}
\includegraphics[width=0.45\textwidth]{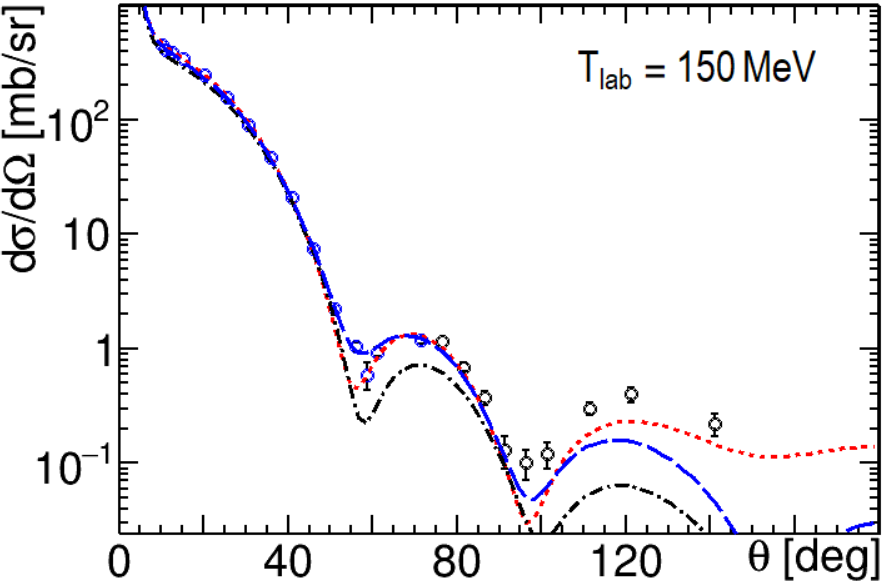}
\includegraphics[width=0.45\textwidth]{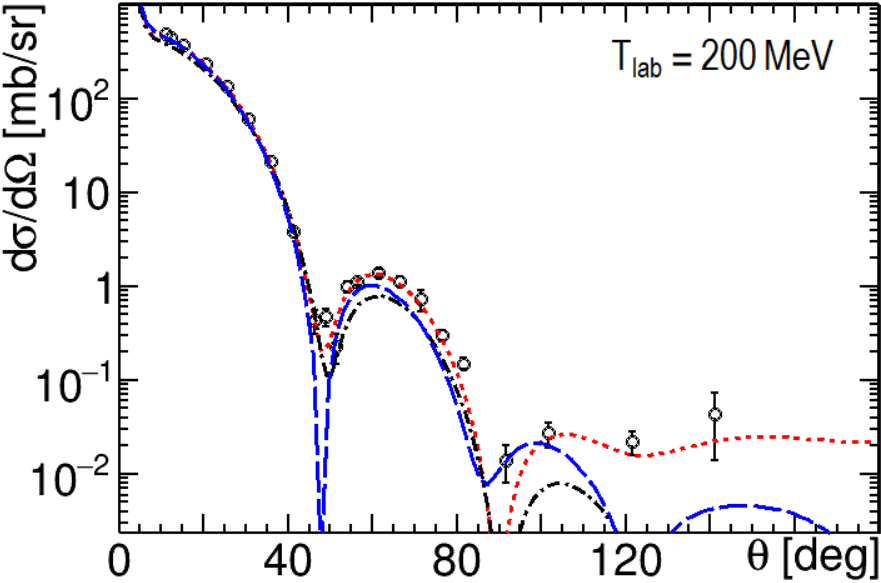}
\caption{
Differential cross sections for $\pi^-$-${}^{12}$C with and without the second-order part of the potential. 
The red short-dashed curves correspond to fit 1 and are identical to those in Figs.~\ref{12C-fit-pot} and~\ref{12C-outside}; the black dash-dotted curves are obtained from fit 1 by switching off the second-order part of the potential, $V^{(2)} = 0$; the blue long-dashed represent the results of fitting $\pi^\pm$-${}^{12}$C data using only the first-order potential $V^{(1)}$.
Table~\ref{tabl:12C} lists the experimental data presented in the plots.
}
 \label{12C-1st-ord}
\end{figure*}

In Fig.~\ref{12C-1st-ord}, we illustrate the impact of the second-order component of the potential on the $\pi^-$-${}^{12}$C differential cross sections. 
We compare theoretical curves obtained using the full potential, i.e., the sum of Eqs.~(\ref{U1st-fin}) and~(\ref{V2-final}), as presented in Figs.~\ref{12C-fit-pot} and \ref{12C-outside}, with results from calculations using only the first-order potential, Eq.~(\ref{U1st-fin}). 
First, we disable the second-order part of the potential, resulting in the black dash-dotted curves in Fig.~\ref{12C-1st-ord}. 
Then, we fit the same dataset using the first-order potential, producing the blue long-dashed curves. 
The minimization process yields $\chi^2/\text{ndf} \approx 10$, indicating a poor fit to the data. 
The plot at $T_\text{lab} = \SI{65}{MeV}$  underscores the substantial significance of the second-order component, $V^{(2)}$, in $\pi$-${}^{12}$C scattering at low energies.
Although the influence of $V^{(2)}$ diminishes as energy increases, it remains substantial even at $T_\text{lab} = \SI{200}{MeV}$ for $\theta > 80^{\circ}$. 
For energies $T_\text{lab} \lesssim \SI{100}{MeV}$, within our parametrization, the first-order potential fails to provide sufficient Coulomb splitting between positive and negative pion differential cross sections. 
Consequently, the differential cross sections for $\pi^+$, for which there is less available experimental data, exhibit notably poorer agreement compared to $\pi^-$.

\subsection{Comparison with \texorpdfstring{\ce{^{16}O}}{Lg} data}

Having the model parameters of the pion-nucleus potential fixed from the $\pi^\pm$-${}^{12}$C data fitting, we can further test the predictive power of our model for another $p$-shell nucleus.
We compare our theoretical predictions based on fit~1 with the data on  $\pi^\pm$-${}^{16}$O scattering.
Table~\ref{tabl:16O} summarizes the experimental data used for the comparison. 

\begin{table}[ht!]
\caption{Summary of the $\pi^\pm$-${}^{16}$O data}
% \begin{tabular*}{\textwidth}[t]{@{\extracolsep{\fill}}lclll@{}}
\begin{tabular}{ccclc}
\hline\hline
Ref.  &   Facility    &    & $T_\text{lab}$ {[}MeV{]} & Observable   \\
\hline
\cite{Seth:1990rp}  &  LAMPF     & $\pi^-$ &  50   &   \\
\cite{Preedom:1981zz}  &  LAMPF    & $\pi^+$ &  50     &  $d\sigma^\text{El}/d\Omega$  \\
\cite{Albanese:1980np}  &  SIN    & $\pi^+$ & 80 - 343                  &    \\
\cite{Mutchler:1975vr} & SREL & $\pi^\pm$ & 155 - 213  \\
\hline
\cite{Ingram:1982bn}  &  SIN     & $\pi^+$   &   114 - 240          &      $\sigma^\text{Tot}, \sigma^\text{El}$    \\
\cite{Clough:1974qt}  &  RAL     &  $\pi^\pm$ & 89 - 342                 &  $\sigma^\text{Tot}$   \\
\hline
\end{tabular}
\label{tabl:16O}
\end{table}

\begin{figure*}[!th]
\includegraphics[width=0.45\textwidth]{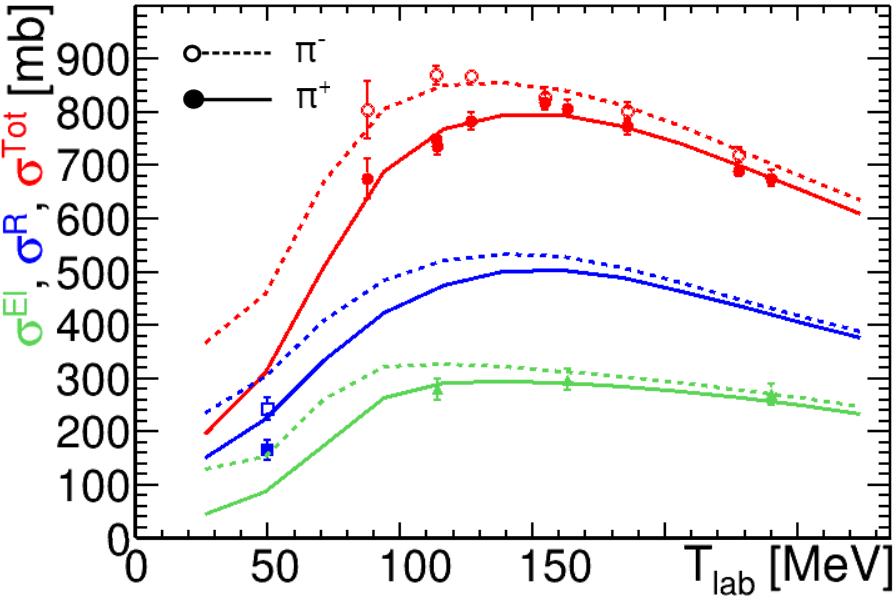} 
\includegraphics[width=0.45\textwidth]{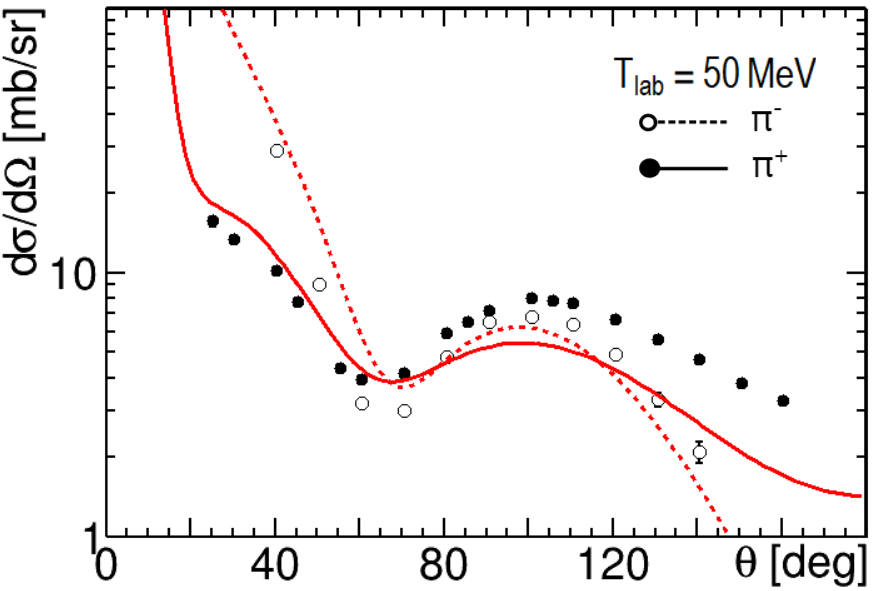}
\includegraphics[width=0.45\textwidth]{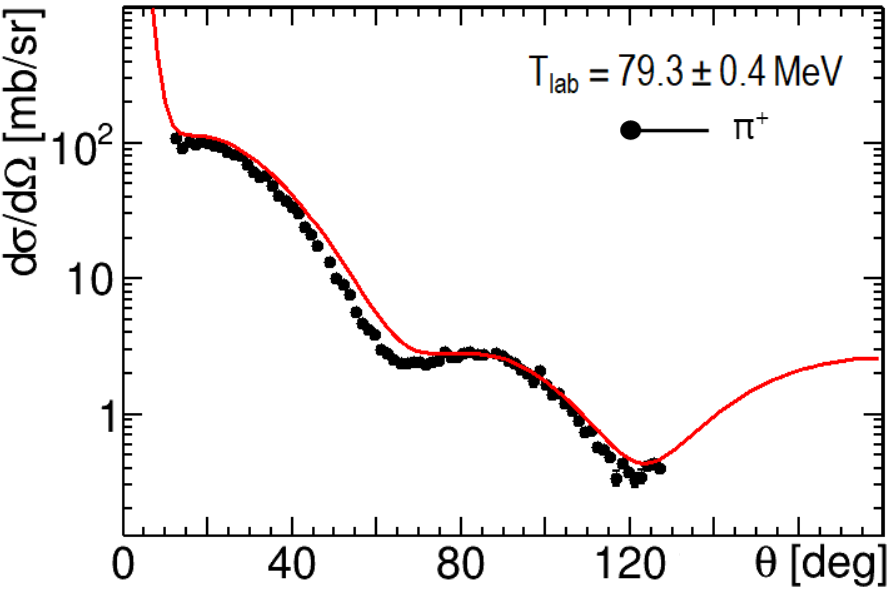}
\includegraphics[width=0.45\textwidth]{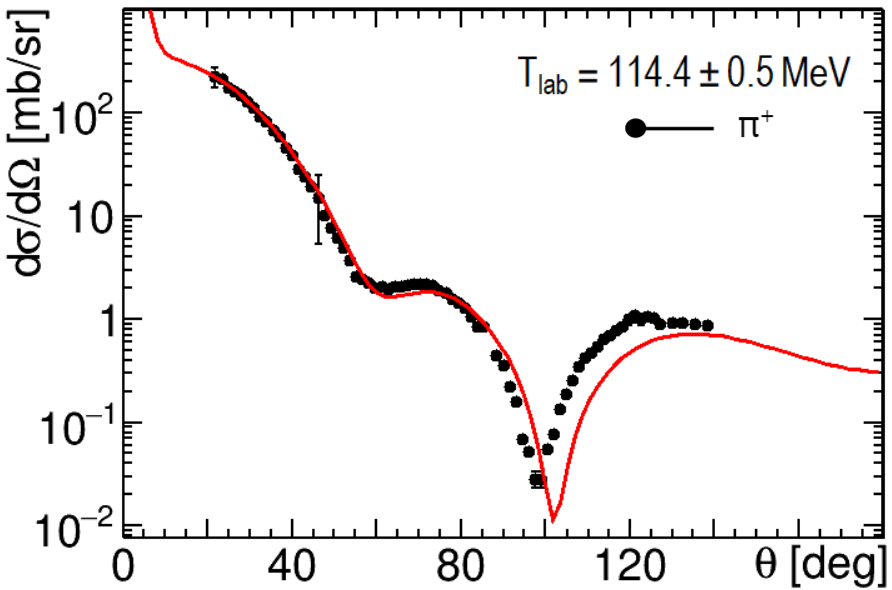}
\includegraphics[width=0.45\textwidth]{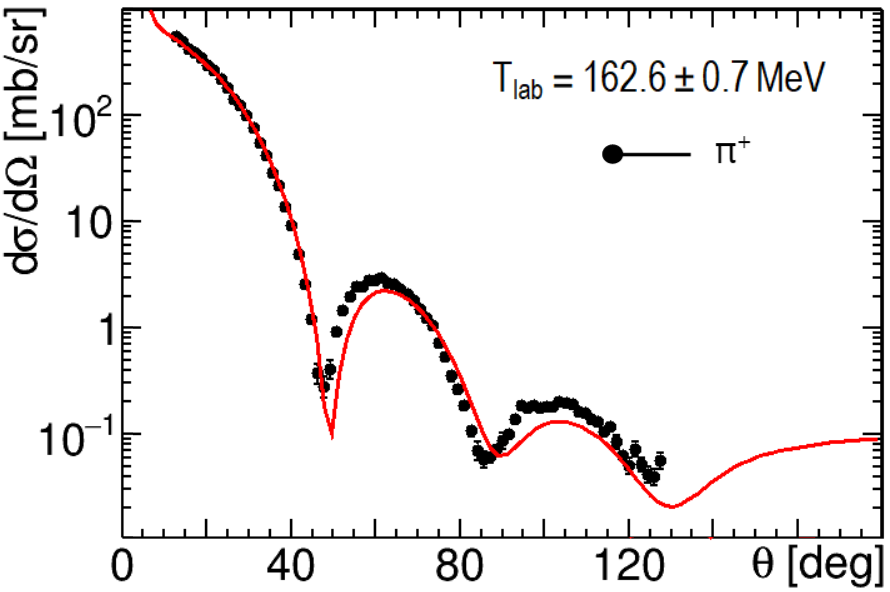}
\includegraphics[width=0.45\textwidth]{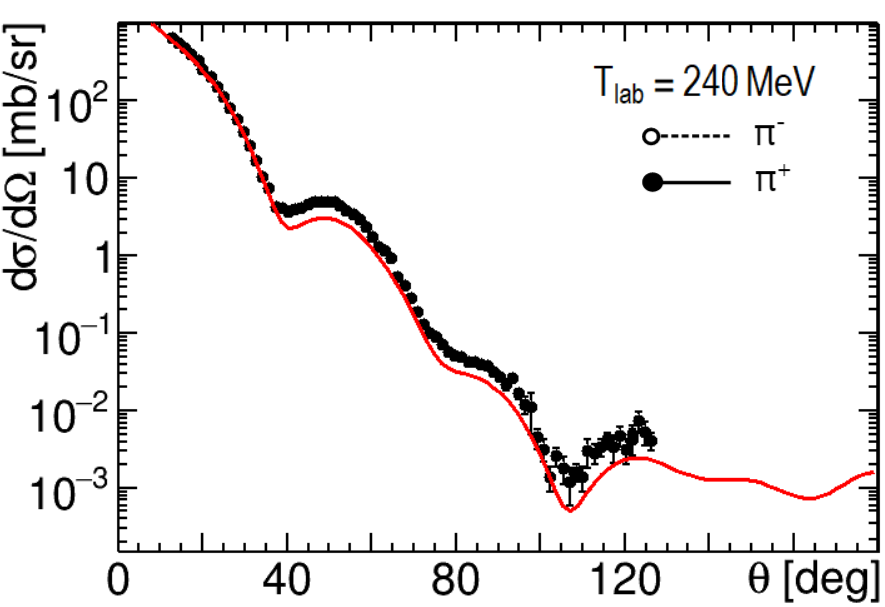} 
\includegraphics[width=0.45\textwidth]{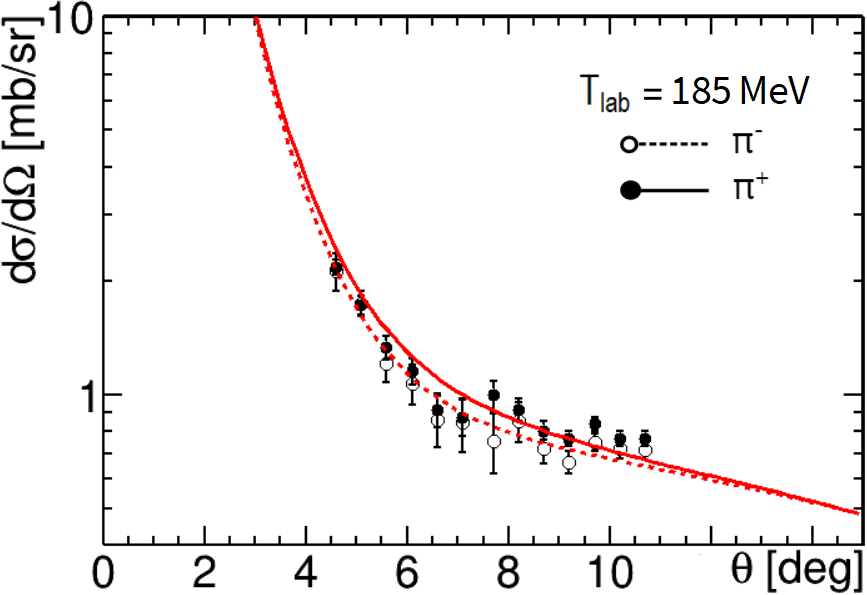}
\includegraphics[width=0.45\textwidth]{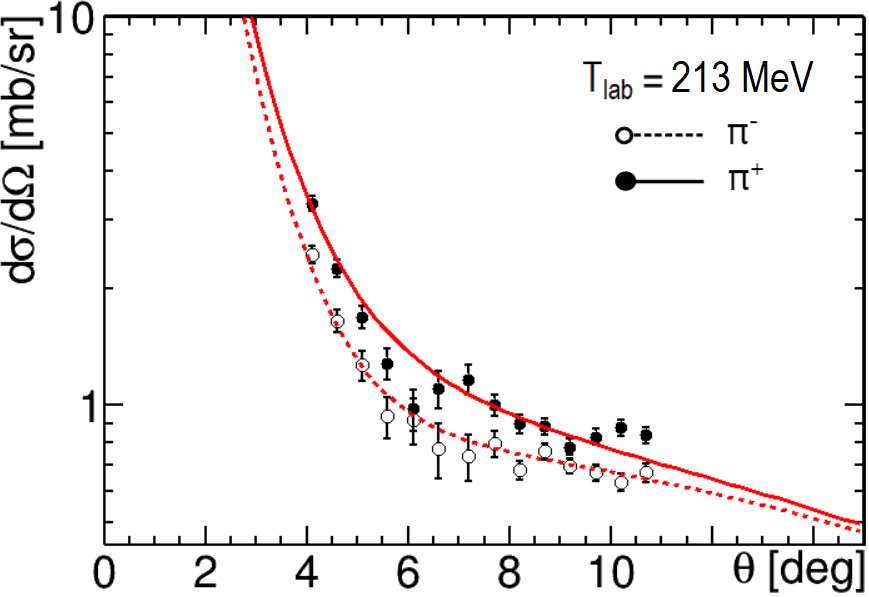} 
\caption{
Comparison of the theoretical calculation based on fit 1 with the data for $\pi^\pm$-${}^{16}$O scattering.
The meaning of the curves is the same as in Fig.~\ref{12C-fit-pot}.
Table~\ref{tabl:16O} lists the experimental data presented in the plots.
\\
\\
 }
 \label{plt:16O}
\end{figure*}

Since ${}^{12}$C and ${}^{16}$O are both spin-isospin-zero closed $p$-subshell nuclei, in our calculation, we replace only the nuclear form factors and apply the correlation functions given by Eqs.~(\ref{C-D-16O}). 
The pion-nucleon scattering amplitudes are kept the same.
The resulting plots with our predictions are presented in Fig.~\ref{plt:16O}, demonstrating a rather good agreement between the model and experimental data.
The small deviations between theoretical curves and the differential cross-section data are similar to those present on the plots for ${}^{12}$C.
The theoretical curves follow the data even for large angles, except for $\SI{114}{MeV}$, where the minimum is shifted by about $5^\circ$.
The small-angle $\pi^\pm$-${}^{16}$O scattering data at 155, 185, and \SI{213}{MeV} from the Space Radiation Effects
Laboratory (SREL)~\cite{Mutchler:1975vr} are well reproduced.
The comparison supports our expectation of the model's universality and demonstrates its predictive power.

\subsection{Comparison with \texorpdfstring{\ce{^{28}Si}}{Lg} and \texorpdfstring{\ce{^{40}Ca}}{Lg} data}

\begin{figure*}[!th]
\includegraphics[width=0.45\textwidth]{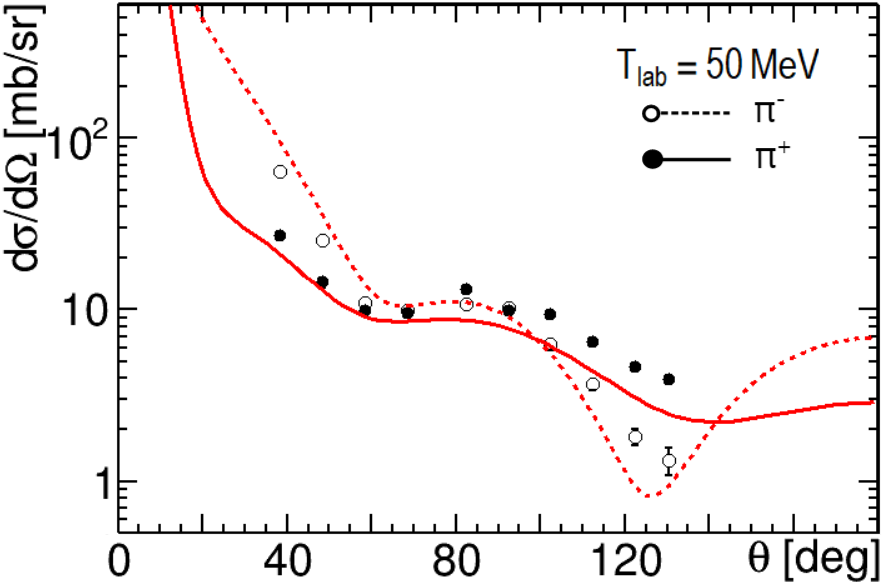} 
\includegraphics[width=0.45\textwidth]{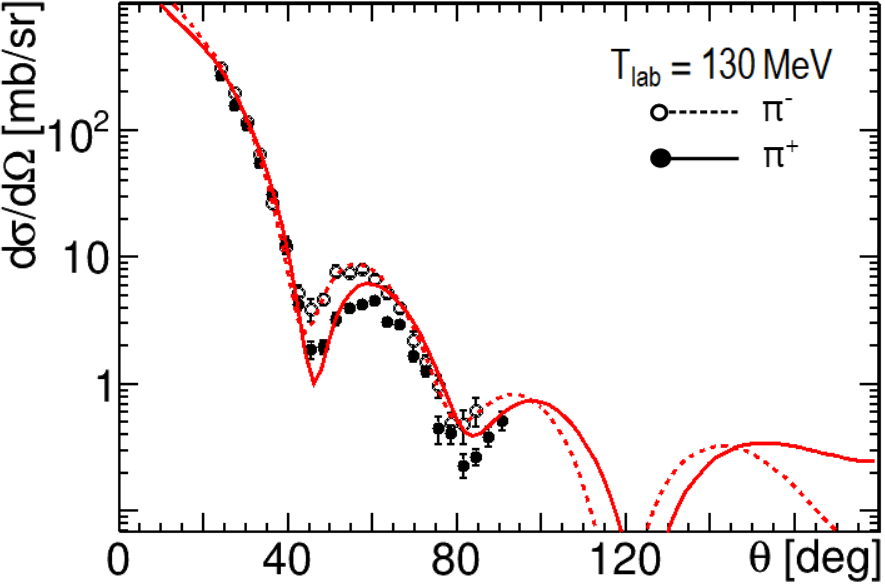}
\includegraphics[width=0.45\textwidth]{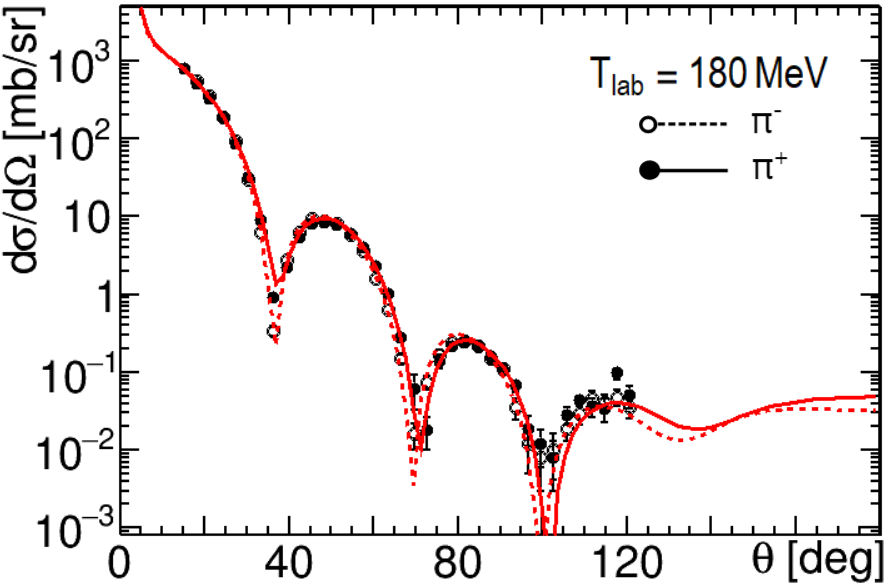}
\includegraphics[width=0.45\textwidth]{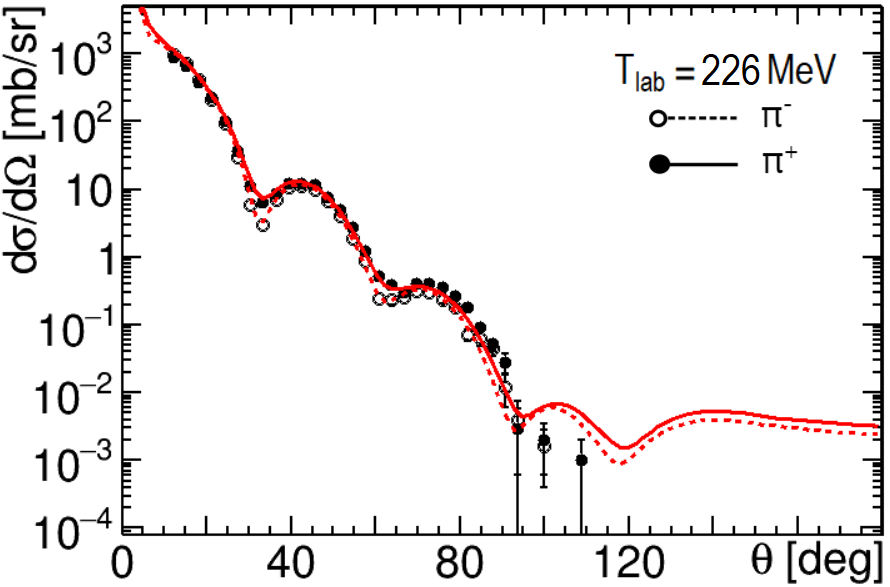}
\caption{
Comparison of the theoretical calculation based on fit 1 with the data for $\pi^\pm$-${}^{28}$Si scattering.
The meaning of the curves is the same as in Fig.~\ref{12C-fit-pot}.
Table~\ref{tabl:40Ca} lists the experimental data presented in the plots.
}
\label{plt:28Si}
\end{figure*}

The described model copes well with describing both  ${}^{12}$C and ${}^{16}$O data using the same set of parameters. 
In general, it can be applied for any spin- and isospin-zero nucleus if the nuclear form factor and correlation functions $C_0$ and $C_\text{ex}$ are known.
However, the calculation of the second-order part of the potential, Eq.~(\ref{V2-final}), becomes involved even for $p$-shell nuclei.
Moreover, the harmonic oscillator shell model used to calculate the correlation functions for ${}^{12}$C and ${}^{16}$O is much less suitable for describing heavier nuclei like ${}^{40}$Ca, requiring more realistic nucleon wave functions.
However, considering that the influence of the second-order correction decreases for heavier nuclei, we can still try applying the harmonic oscillator model to derive $C_0$ and $C_\text{ex}$ for closed $d$-subshell nuclei, as given in Eqs.~(\ref{C-D-28Si}) and~(\ref{C-D-40Ca}).

\begin{table}[ht!]
\caption{Summary of the $\pi^\pm$-${}^{28}$Si and $\pi^\pm$-${}^{40}$Ca differential cross section data}
% \begin{tabular*}{\textwidth}[t]{@{\extracolsep{\fill}}lclll@{}}
\begin{tabular}{ccclc}
\hline\hline
Ref.  &   Facility    &    & $T_\text{lab}$ {[}MeV{]} & Nucleus     \\
\hline
\cite{Wienands:1987ms}  &  TRIUMF     & $\pi^\pm$ & 50   &  ${}^{28}$Si  \\
\cite{Preedom:1979pr}  &  SIN     & $\pi^\pm$ &   130, 180, 226   &   \\
\hline
\cite{Seth:1990rp}  &  LAMPF     & $\pi^-$ &  50   &  \\
\cite{Preedom:1981zz}  &  LAMPF    & $\pi^+$ &  50     &    \\
\cite{Dam:1982xr}  &  LAMPF    & $\pi^\pm$ & 65                  &   ${}^{40}$Ca     \\
\cite{Leitch:1984ym}  &  LAMPF     & $\pi^\pm$   &   80          &   \\
\cite{Gretillat:1981bq}  &  SIN     &  $\pi^\pm$ & 130, 180, 230       &    \\
\hline
\end{tabular}
\label{tabl:40Ca}
\end{table}

\begin{figure*}[!th]
\includegraphics[width=0.45\textwidth]{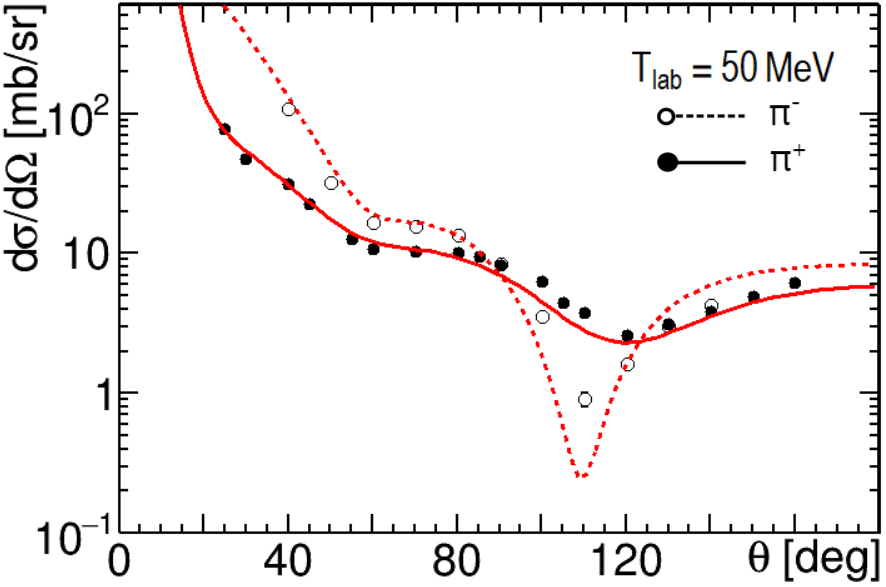} \includegraphics[width=0.45\textwidth]{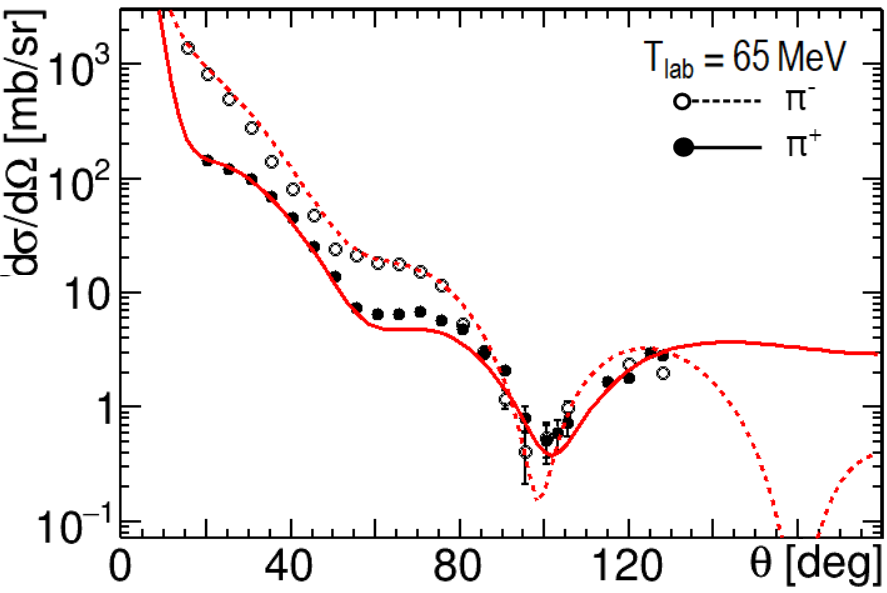}
\includegraphics[width=0.45\textwidth]{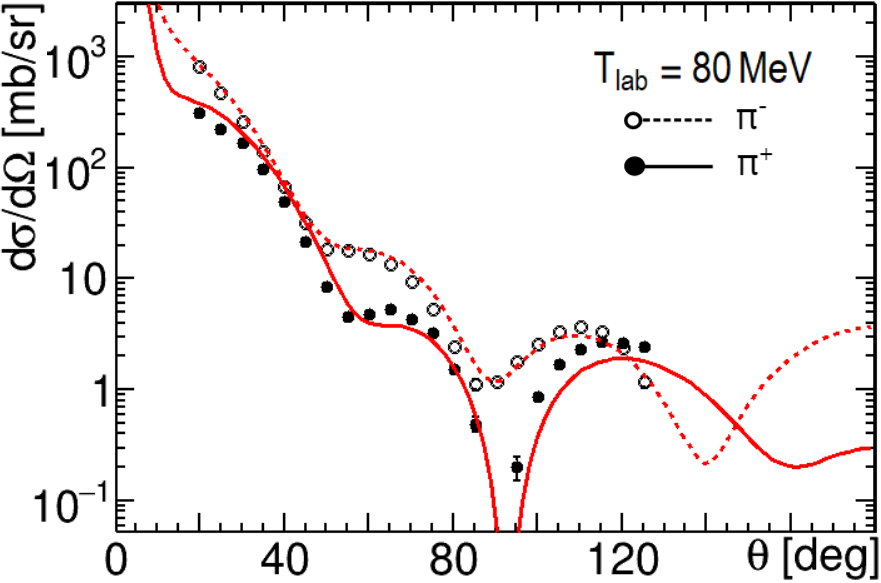}
\includegraphics[width=0.45\textwidth]{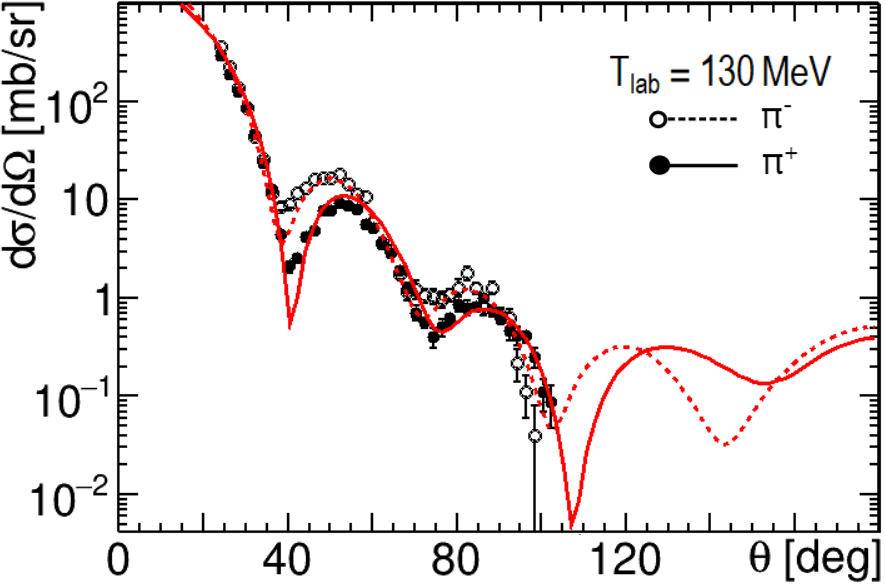}
\includegraphics[width=0.45\textwidth]{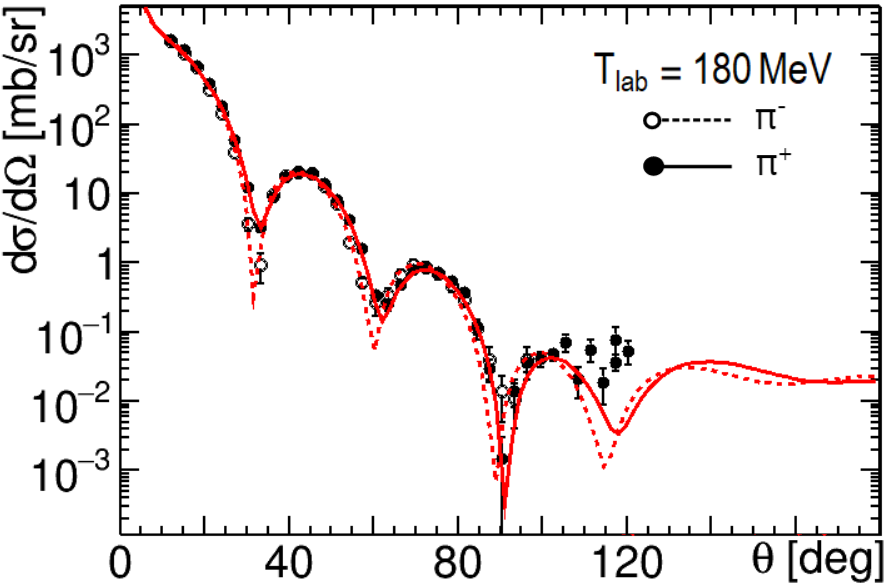}
\includegraphics[width=0.45\textwidth]{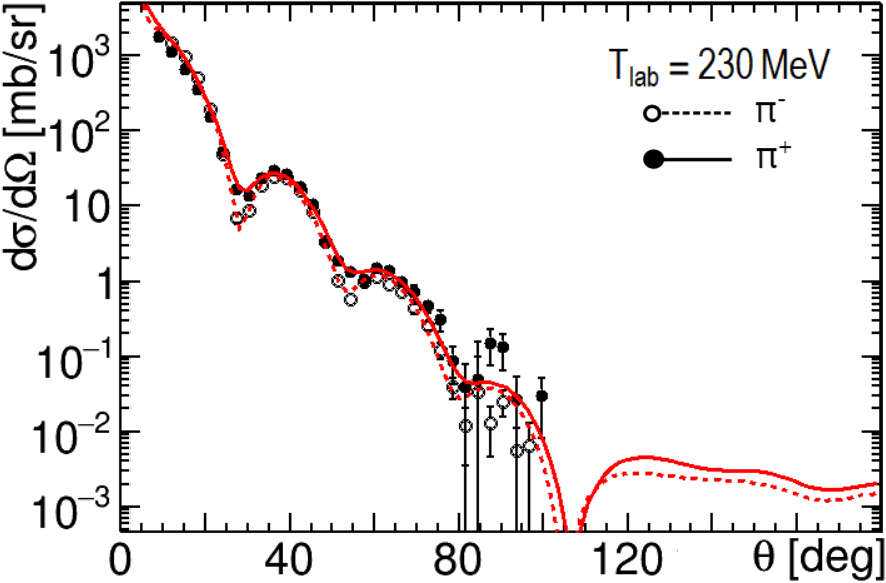}
\caption{
Comparison of the theoretical calculation based on fit 1 with the data for $\pi^\pm$-${}^{40}$Ca scattering.
The meaning of the curves is the same as in Fig.~\ref{12C-fit-pot}.
Table~\ref{tabl:40Ca} lists the experimental data presented in the plots.
}
\label{plt:40Ca}
\end{figure*}

In Figs.~\ref{plt:28Si} and~\ref{plt:40Ca}, we demonstrate our prediction for the $\pi^\pm$-${}^{28}$Si and $\pi^\pm$-${}^{40}$Ca differential cross sections, respectively.
The theoretical model is compared with the experimental differential cross section data listed in Table~\ref{tabl:40Ca}. 
Given that no additional adjustments were made, the agreement between our prediction and the data is surprisingly good, especially at larger energies.
The observed small discrepancy at low energies can be explained by a more decisive influence in heavier nuclei of the $s$-wave part of the potential and stronger Coulomb-nuclear interference.

\begin{figure*}[!th]
\includegraphics[width=0.45\textwidth]{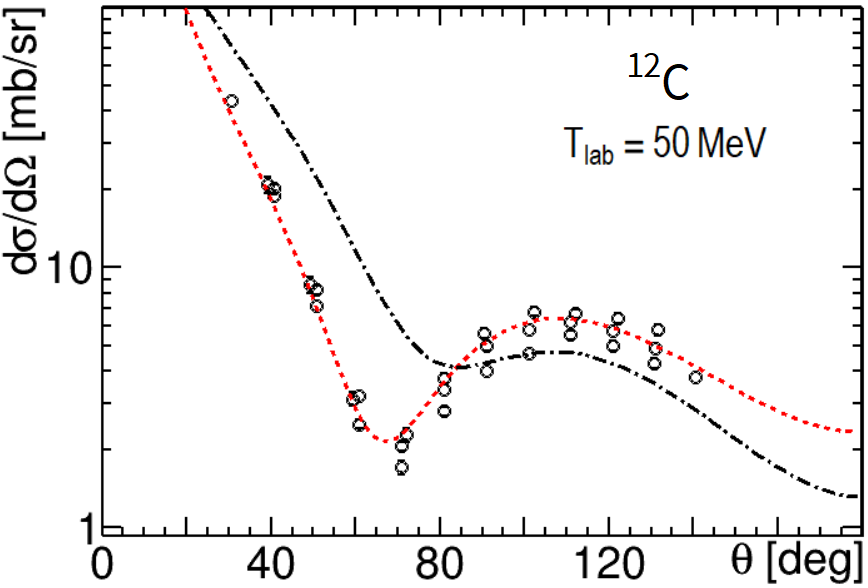} \includegraphics[width=0.45\textwidth]{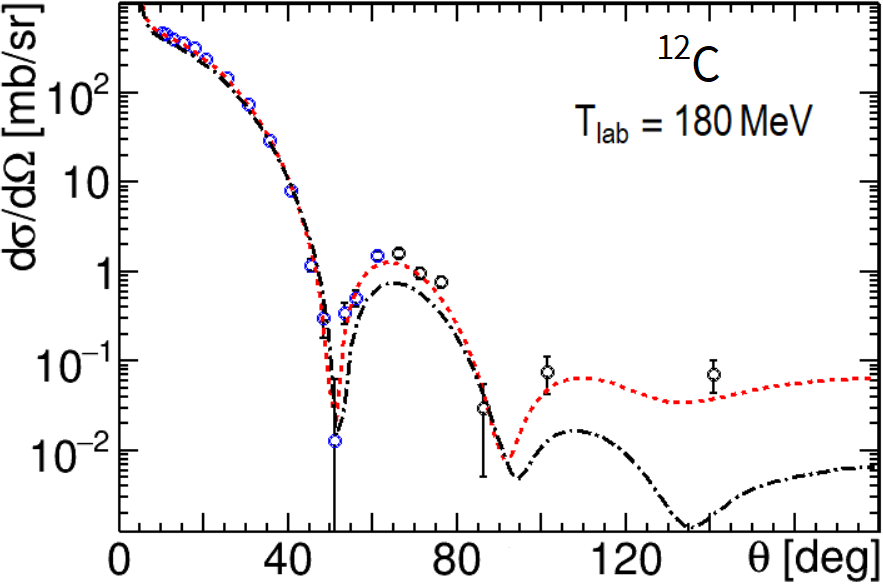}
\includegraphics[width=0.45\textwidth]{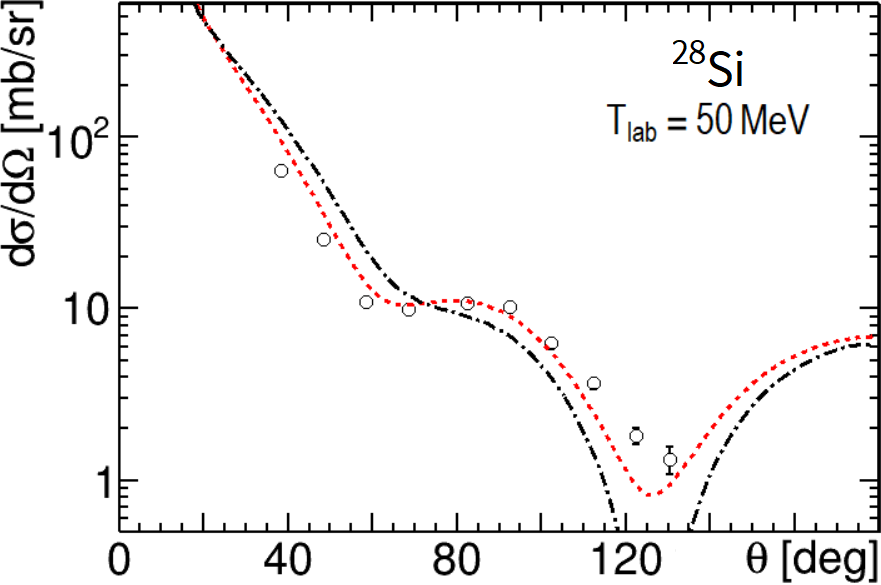}
\includegraphics[width=0.45\textwidth]{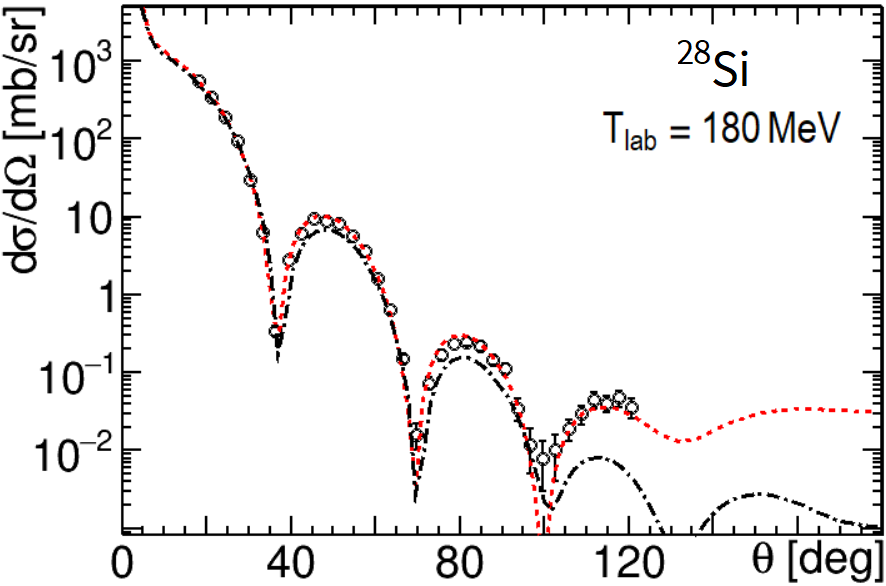}
\includegraphics[width=0.45\textwidth]{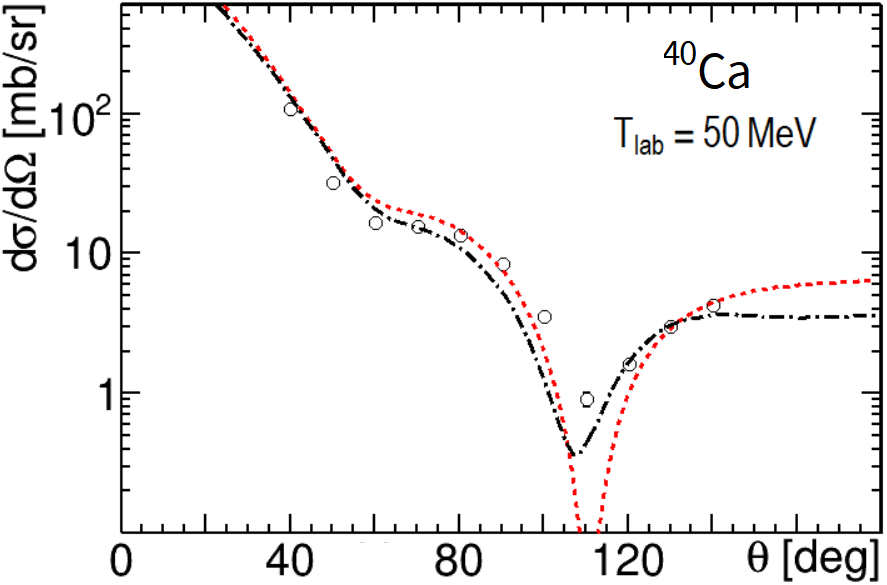}
\includegraphics[width=0.45\textwidth]{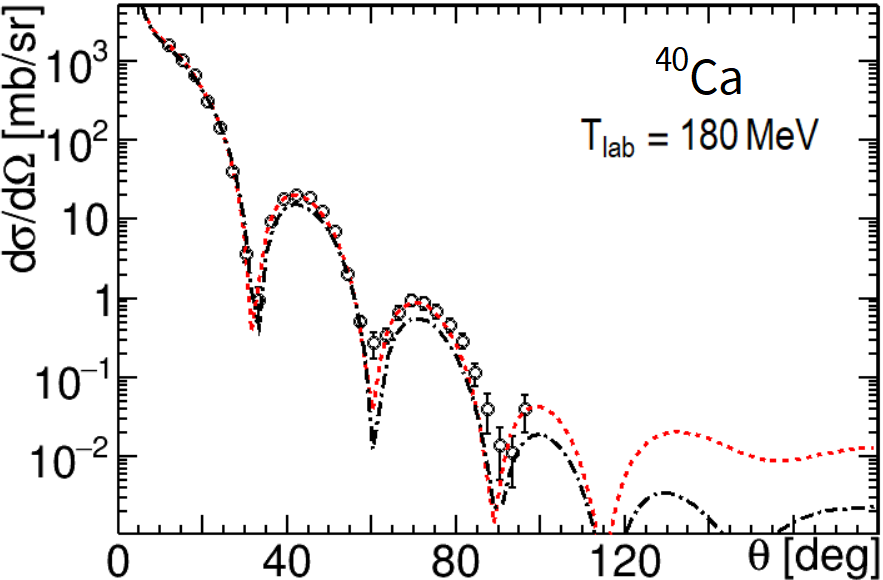}
\caption{
Comparison of the theoretical calculation based on fit 1 with the data for $\pi^-$ scattering on ${}^{12}$C, ${}^{28}$Si and ${}^{40}$Ca at 50 and $\SI{180}{MeV}$ pion laboratory kinetic energy.
The meaning of the curves is the same as in Fig.~\ref{12C-1st-ord}.
Tables~\ref{tabl:12C} and~\ref{tabl:40Ca} list the experimental data presented in the plots.
}
\label{plt:A-dependence}
\end{figure*}

In Fig.~\ref{plt:A-dependence}, we further investigate the influence of the second-order part of the potential, Eq.~(\ref{V2-final}), by comparing the differential cross sections for $\pi^-$ scattering on ${}^{12}$C, ${}^{28}$Si, and ${}^{40}$Ca at 50 and $\SI{180}{MeV}$ pion laboratory kinetic energy.
We present the results of the full calculation, as displayed in Figs.~\ref{12C-fit-pot}, \ref{12C-outside}, \ref{plt:28Si}, and \ref{plt:40Ca}, alongside the theoretical curves obtained after setting $V^{(2)} = 0$.
As can be seen from Figs.~\ref{12C-1st-ord} and~\ref{plt:A-dependence}, the contribution of $V^{(2)}$ diminishes as the nucleus becomes heavier and as the energy increases. 
However, the second-order component remains important for achieving a good agreement with the data, particularly for $\theta > 60^{\circ}$ in the case of $\pi$-${}^{40}$Ca scattering at $\SI{180}{MeV}$.

\section{Conclusion and outlook}
\label{sec:conclusion}

In the present work, we have constructed the second-order pion-nuclear potential in momentum space. 
The potential is based on the individual pion-nucleon scattering amplitudes extracted from SAID. 
The second-order correction to the potential depends on two types of correlation functions and, as a result, is consistent with the Pauli principle.
The many-body medium effects are incorporated in the complex effective $\Delta$ self-energy and the modifications to the $s$-wave scattering parameters.

In our approach, only three fitting parameters are introduced: the real and imaginary parts of the $\Delta$ self-energy and the $s$-wave isoscalar slope parameter. 
The free parameters were determined by fitting the $\pi^\pm$-${}^{12}$C scattering data in the energy range of 80–\SI{180}{MeV} pion laboratory kinetic energy, which shows a strong sensitivity to the $\Delta$-resonance properties.
The developed second-order potential was found to yield a successful description of the total, angle-integrated elastic, reaction, and differential elastic cross-section data, assuming that the model parameters are energy independent.

Furthermore, the model demonstrates that it yields a good description of the $\pi^\pm$-${}^{12}$C data not only in the fitting range but also outside of it. 
To check its predictive power, we have applied the second-order potential to heavier nuclei, using the three parameters which have been fixed by fitting the ${}^{12}$C data. 
The model predictions for ${}^{16}$O, ${}^{28}$Si, and ${}^{40}$Ca nuclei were found to nicely agree with the experimental data, supporting the model's universality and predictive power.

In future work, we plan to provide a more detailed analysis for scattering on heavy nuclei and for the case of nuclei with nonzero isospin. 
As a next step, the presented model can also be applied to analyzing electron- or neutrino-induced pion production processes on nuclei.

\section*{Acknowledgments}

This work was supported by the Deutsche Forschungsgemeinschaft (DFG, German Research Foundation), in part through the Collaborative Research Center [The Low-Energy Frontier of the Standard Model, Projektnummer 204404729 - SFB 1044], and in part through the Cluster of Excellence [Precision Physics, Fundamental Interactions, and Structure of Matter] (PRISMA$^+$ EXC 2118/1) within the German Excellence Strategy (Project ID 39083149).

\appendix

\section{Scattering by nuclear and Coulomb potentials}
\label{sec:Coulomb}

The charged pion that approaches the nucleus,  $\pi^-$ ($\pi^+$), is accelerated (decelerated) due to the influence of the long-range Coulomb field of the nucleus. 
This effect occurs before the pion reaches the range of the strong interaction described by the pion-nucleus potential $\hat U(E)$.
At intermediate energies, the pion-nucleon scattering has a strong energy dependence due to the resonant $P_{33}$ channel and is sensitive to this Coulomb energy shift.
As a result, the potential $\hat U(E)$ in the scattering equations must be replaced with the nuclear-Coulomb potential $\hat U_{NC}(E)$, which can be approximated as
\be
\hat U_{NC}(E) =  \hat U(E - \langle \hat U_C \rangle) + \hat U_C.
\label{nuclear-Coulomb-pot}
\ee
In Eq.~(\ref{nuclear-Coulomb-pot}), besides adding the Coulomb potential, $\hat U_C$, we shift the reaction energy by the value of the Coulomb potential at the root-mean-squared (rms) radius of the nucleon distribution~\cite{Huefner:1973dy,dedonder1980comparison}.
The shift in the energy argument of the nuclear potential describes the intermediate Coulomb rescattering and, in general, is given at the operator level, but assuming the commutativity of $U_C$ with the Green's function and neglecting the nucleus excitation by the Coulomb potential, we arrive at Eq.~(\ref{nuclear-Coulomb-pot}) (see Ref.~\cite{Cannata:1990xt} for details).

We explicitly separate the momentum transfer dependent nuclear structure characteristics, namely the form factor and correlation functions, from the angle- and energy-dependent single-nucleon scattering amplitudes in the pion-nucleus potential.
When dealing with the pion-nuclear potential in coordinate space, applying the Coulomb energy shift in Eq.~(\ref{nuclear-Coulomb-pot}) is straightforward and consists in shifting only the argument of the scattering coefficients in Eqs.~(\ref{f-sp-coefs}).
However, the situation is more complicated for the potential in momentum space due to its dependence on the off-shell momentum.
To address this, we assume that the entire pion-nucleon on-shell transition and scattering amplitudes are calculated as described in Sec.~\ref{sec:pion-nucl-ampl} but with the shifted on-shell momentum $k_0(T_\text{lab} - \langle U_C \rangle)$ in the pion-nucleus c.m. frame, where
\be
k_0^2(T_\text{lab}) = \frac{m_A^2 T_\text{lab} (2m_\pi + T_\text{lab})}{(m_A + m_\pi)^2 + 2 m_A T_\text{lab}}.
\ee
For a smooth off-shell extrapolation, we further assume that the Coulomb-affected off-shell momenta involved in calculating the off-shell vertex factor, Eq.~(\ref{off-vertex-fact}), are replaced by
\be
k^2 \longrightarrow k^2  + k_0^2(T_\text{lab} - \langle U_C \rangle) - k_0^2(T_\text{lab}).
\ee

A direct solution of the scattering equation~(\ref{LSh-pseudo-classical}) involving the long-range Coulomb interaction is difficult due to $1/q^2$ singularity in the momentum space representation of the Coulomb potential.
To address this issue, we apply Vincent and Phatak's method~\cite{Vincent:1974zz} to treat the Coulomb-nuclear interaction in momentum space.
It is assumed that in coordinate space, the nuclear part of the potential vanishes beyond the cutoff radius $R_\text{cut}$.
As a result, at $r \ge R_\text{cut}$, only the point-charge Coulomb potential exists, and the radial part of the coordinate space wave function can be expressed as
\be
u_l(r) \propto \mathscr{F}_l(\eta_c, k_0 r) + k_0 F_l \, \mathscr{H}_l(\eta_c, k_0 r),
\label{Coulomb-wf}
\ee
with $\mathscr{H}_l \equiv \mathscr{H}_l^+ = \mathscr{G}_l + i \mathscr{F}_l$, where $\mathscr{F}_l$ and $\mathscr{G}_l$ are the regular and irregular Coulomb functions~\cite{michel2007precise}.
The amplitude $F_l$ in Eq.~(\ref{Coulomb-wf}) represents the correct Coulomb-modified nuclear partial-wave scattering amplitude that describes the observed cross sections and enters Eqs.~(\ref{F_NC-expansion})–(\ref{sigma-Tot}). 
The asymptotic Coulomb wave function, Eq.~(\ref{Coulomb-wf}), is smoothly matched with the cutoff solution at $r = R_\text{cut}$, which yields:
\be
F_l = \frac1{k_0} \frac{\mathscr{F}'_l(\eta_c, \rho) - \xi_l \mathscr{F}_l(\eta_c, \rho)}{\xi_l \mathscr{H}_l(\eta_c, \rho) - \mathscr{H}'_l(\eta_c, \rho)} ,
\label{delta-Phatac}
\ee
where $\rho = k_0 R_\text{cut}$ and
\be
\xi_l = \frac{\mathscr{F}'_l(0, \rho) + k_0 F_l^\text{cut} \mathscr{H}'_l(0, \rho)}{\mathscr{F}_l(0, \rho) + k_0 F_l^\text{cut} \mathscr{H}_l(0, \rho)}. 
\ee
The partial amplitude $F_l^\text{cut}$ is the solution of the pion-nucleus scattering equation with the short-range potential, which is the sum of the Coulomb potential cut at $R_\text{cut}$ and the strong pion-nuclear potential described in Sec.~\ref{sec:Uopt}.
We derive $F_l^\text{cut}$ from Eq.~(\ref{LSh-pseudo-classical}) using the momentum space representation of the cut Coulomb potential given by
\be
V_C^\text{cut}(q) = -2 \bar \omega \frac{\alpha Z_\pi}{q^2} \left[\rho_\text{ch}(q) \rho_\text{ch}^\pi(q) - Z \cos(q R_\text{cut})\right],
\ee
where $\rho_\text{ch}(q)$ and $\rho_\text{ch}^\pi(q)$ are the charge form factors of the nucleus and pion. 
We use the value $R_\text{cut} = \SI{8}{fm}$.

The original Kerman-McManus-Thaler (KMT) multiple scattering formalism does not explicitly address the Coulomb interaction. 
As a result, the KMT scattering Equations~(\ref{T-series-KMT-ground}) and~(\ref{LSh-pseudo-classical}) in the pure Coulomb scattering limit, $\hat U \rightarrow \hat U_C$,  fail to provide  the correct Coulomb scattering amplitude due to factor $(A-1)/A$.
The treatment of the Coulomb interaction in the KMT formalism was examined in detail in Ref.~\cite{Ray:1980ck}.
To recover the Coulomb scattering amplitude effectively, we follow the "KMT No.~3 prescription" of Ref.~\cite{Ray:1980ck} (Eqs. (48)–(50)) and replace the pure Coulomb KMT $T$ matrix with the analogous quantity in the Watson approach. 
Despite being a minor correction, this approach improves the calculated cross sections by a few percent.

\section{Nuclear form factor and correlation functions}
\label{sec:FF-and-correlation}

The determination of the nuclear charge density, $\rho_\text{ch}(r)$, provides information on the nucleon distribution within nuclei. 
In this work, we use the Fourier-Bessel (FB) series expansion to provide an accurate, model-independent description of the charge distribution \cite{Dreher:1974pqw}. 
The charge density, $\rho_\text{ch}(r)$, is assumed to be zero beyond a certain cutoff radius $R_c$. 
Within the interval $r \le R_c$, we can then expand $\rho_\text{ch}(r)$ into the FB series:
\begin{equation} 
\rho_\text{ch}(r) = \theta(R_c - r) \sum_{n=1}^{n_\text{max}} a_n j_0 \left(q_n r \right), 
\label{rho_ch(r)-FB}
\end{equation}
where $q_n = n \pi / R_c$ are the zeros of the zero-order Bessel function $j_0(x) = \sin x / x$, and the coefficients of the series are determined by fitting experimental data on electron scattering. 
The number of expansion coefficients is determined by the maximal experimentally measured momentum $q_\text{max}$ as $n_\text{max} = q_\text{max} R_c / \pi$.

For spin-zero nuclei, the charge distribution, $\rho_\text{ch}(r)$, and the charge form factor, $\rho_\text{ch}(q)$, are related by the Fourier transform, 
which for spherically symmetric nwuclei is given by
\be
\rho_\text{ch}(q) = 4\pi \int r^2\diff r j_0(q r) \rho_\text{ch}(r).
\label{FT-for-rho_ch}
\ee
Correspondingly, the FB expansion, Eq.~(\ref{rho_ch(r)-FB}), in the momentum space becomes
\be
\rho_\text{ch}(q) = 4\pi \frac{\sin(q R_c)}{q} \sum_{n=1}^{n_\text{max}} a_n \frac{(-1)^n}{q^2 - q_n^2}.
\ee

The nuclear charge density does not correspond to the proton density in the nucleus because of the finite size of the proton.
Moreover, the neutron also possesses a charge distribution with a negative mean square radius.
The nuclear charge distribution, $\rho_\text{ch}(r)$, can be found as the convolution of the distribution $\rho(r)$ of the nucleons in the nucleus with the nucleon charge density.
As a result, the form factor for isospin-zero nuclei is given as
\be
\rho(q) = \frac{2 \rho_\text{ch}(q)}{\rho_\text{ch}^{(p)}(q) + \rho_\text{ch}^{(n)}(q)}, 
\ee
where $\rho_\text{ch}^{(p)}(q)$ and $\rho_\text{ch}^{(n)}(q)$ are the proton and neutron charge form factors, respectively.
We utilize the nucleon charge form factors obtained from the global fits of electron scattering data presented in Ref.~\cite{Ye:2017gyb}.

While the FB expansion is a reliable approach for the first-order potential, Eq.~(\ref{U1st-fin}), the second-order correction, Eq.~(\ref{V2-final}), requires a model for deriving the two-body density and correlation functions, Eqs.~(\ref{rho_ex-and-C(r)}).
Assuming the $A$-body Slater determinant form of the total nuclear wave function,
\be
\Psi_0^\text{SD}(x_1, \ldots, x_A) = \frac1{\sqrt{A!}} \det\{ \phi_{\alpha_i}(x_j)\},
\label{Slater-det-def}
\ee
with $i,j = 1, \ldots,A$ and the multi-index $\alpha \equiv \{n, l, j, m, m_j \}$, we can express the exchange correlation function, Eq.~(\ref{rho_ex(r)-def}), in terms of the shell model single-particle nucleon wave functions $\phi_{\alpha_i}(x_j)$:
\be
C_\text{ex}(x_1, x_2) = \sum_{i,j = 1}^A
\phi_{\alpha_i}^\dag(x_1) \phi_{\alpha_j}^\dag(x_2)  \phi_{\alpha_i}(x_2)  \phi_{\alpha_j}(x_1).
\label{rho_ex-shell}
\ee
The corresponding nuclear density within the shell-model is given by
\be
\rho(r) = \sum_{\sigma \tau} \sum\limits_{i = 1}^A \phi_{\alpha_i}^\dag(x) \phi_{\alpha_i}(x).
\ee

In this work, we use the harmonic oscillator (HO) nuclear shell model~\cite{Heyde1990} to obtain approximate single-particle wave functions of nucleons, $\phi_{nlm}(x)$.
A direct calculation followed by the Fourier transform provides the following HO nuclear form factor for closed $p$-subshell nuclei (${}^{12}$C and ${}^{16}$O):
\be
\rho(q) = \left[ A - \frac{A - 4}{6}  a^2 q^2  \right] e^{-\frac14 \frac{A-1}A a^2 q^2}, 
\label{rho-HO-12C}
\ee
where $a$ is the HO parameter. As in Eq.~(\ref{form-factor-Jacobi}), factor $(A-1)/A$ in the exponential takes into account the center-of-mass motion correction.

Performing a similar calculation with the additional closed $d$ subshell, we arrive at the HO form factors for ${}^{28}$Si,
\be
\rho(q) = \left[ 28 - 6 a^2 q^2 + \frac15 a^4 q^4 \right] e^{-\frac14 \frac{A-1}A a^2 q^2},
\label{rho-HO-28Si}
\ee
and ${}^{40}$Ca, 
\be
\rho(q) = \left[ 40 - 10 a^2 q^2 + \frac12 a^4 q^4 \right] e^{-\frac14 \frac{A-1}A a^2 q^2}.
\label{rho-HO-40Ca}
\ee

\begin{table}[!t]
\caption{
Comparison of the HO parameter $a$ and rms charge radius for ${}^{12}$C, ${}^{16}$O, ${}^{28}$Si, and ${}^{40}$Ca.
The value of $\langle r_\text{ch,FB}^2\rangle^{1/2}$ ($\langle r_\text{ch,HO}^2\rangle^{1/2}$) represents the rms charge radius calculated using the FB (HO) charge density. 
The experimental value from Ref.~\cite{Angeli:2013epw} is provided by $\langle r_\text{ch,exp}^2\rangle^{1/2}$. }
\begin{tabular}{ccccc}
\hline\hline
 & \quad ${}^{12}$C & \quad ${}^{16}$O & \quad ${}^{28}$Si & \quad ${}^{40}$Ca \\
\hline
$a$ [fm] & \quad  1.63  & \quad 1.76 & \quad 1.82  &  \quad 1.98 \\
$\langle r_\text{ch,FB}^2\rangle^{1/2}$ [fm]  & \quad 2.47  &  \quad 2.74 & \quad 3.09 & \quad 3.45 \\
$\langle r_\text{ch,HO}^2\rangle^{1/2}$ [fm]  &  \quad 2.47  & \quad 2.71 & \quad 3.08  & \quad 3.47 \\
% $\langle r_\text{ch,exp}^2 \rangle^{1/2}$ [fm] &  $\SI{2.470\pm0.002}{}$  &  $\SI{2.699\pm0.005}{}$  & $\SI{3.122\pm0.002}{}$ & $\SI{3.478\pm0.002}{}$ \\
$\langle r_\text{ch,exp}^2 \rangle^{1/2}$ [fm] &  \quad 2.47  & \quad 2.70  & \quad 3.12 & \quad 3.48 \\
\hline
\end{tabular}
\label{tabl:density-summary}
\end{table}

The HO form factors, Eqs.~(\ref{rho-HO-12C}) and~(\ref{rho-HO-40Ca}), enable us to determine the corresponding HO model parameters. 
The extracted values of $a$ used in our calculation of the correlation functions are listed in Table~\ref{tabl:density-summary}.
The FB coefficients are taken from Refs.~\cite{Cardman:1980dja} (${}^{12}$C) and~\cite{DeJager:1987qc} (${}^{16}$O, ${}^{28}$Si, and ${}^{40}$Ca).
In each case, $R_c = \SI{8}{fm}$ is used. 
In Table~\ref{tabl:density-summary}, we also compare the rms charge radius for HO and FB analyses with experimental values from Ref.~\cite{Angeli:2013epw}.

To obtain the two-body correlation functions $C_0$ and $C_\text{ex}$ in momentum space within the HO shell model, we generalize the derivation presented in  Refs.~\cite{Jackson:1970zz,murugesu1971optical} to the $\bm q \ne \bm q'$ case.
Starting from Eq.~(\ref{rho_ex-shell}), followed by spin-isospin summation and the Fourier transform, Eq.~(\ref{D-and-C(q)}), we arrive at the two-body correlation functions
\begin{subequations}
\begin{align}
&C_\text{ex}(\bm q_1, \bm q_2) = 
\sum_{\sigma_{1,2}} \sum_{\tau_{1,2}}
C_\text{ex}(\bm q_1, \sigma_1, \tau_1, \bm q_2, \sigma_2, \tau_2)
\label{C-sppin-iso-sum},
\\
&C_0(\bm q_1, \bm q_2) = C_\text{ex}(\bm q_1, \bm q_2) - \frac1A \rho(q_1) \rho(q_2),
\end{align}
\end{subequations}
which yields the following forms.

\begin{widetext}
For ${}^{12}\text{C}$:
\begin{subequations}
\begin{align}
&C_\text{ex}(\bm q_1, \bm q_2) = \left( 12 - \frac43 a^2 (q_1^2 + q_2^2) -
  4 \sqrt{\frac23} a^2 \bm q_1 \cdot \bm q_2 + \frac23 a^4 (\bm q_1 \cdot \bm q_2)^2 \right)
  \exp\left[ - \frac14 \frac{A-1}A a^2 \left(q_1^2 + q_2^2 \right) \right],\\
&C_0(\bm q_1, \bm q_2) = \left(
  -  4 \sqrt{\frac23} a^2 \bm q_1 \cdot \bm q_2 + \frac23 a^4 (\bm q_1 \cdot \bm q_2)^2  - \frac4{27} a^4 q_1^2 q_2^2 \right)
  \exp\left[ - \frac14 \frac{A-1}A  a^2 \left(q_1^2 + q_2^2 \right) \right],
\end{align}
\label{C-D-12C}%
\end{subequations}
for ${}^{16}\text{O}$,
\begin{subequations}
\begin{align}
&C_\text{ex}(\bm q_1, \bm q_2) = \left( 16 - 2 a^2 (\bm q_1 + \bm q_2)^2 + a^4 (\bm q_1 \cdot \bm q_2)^2 \right)
  \exp\left[ - \frac14 \frac{A-1}A  a^2 \left(q_1^2 + q_2^2 \right) \right],\\
&C_0(\bm q_1, \bm q_2) = \left(
  - 4 a^2 \bm q_1 \cdot \bm q_2 + a^4 (\bm q_1 \cdot \bm q_2)^2  - \frac14 a^4 q_1^2 q_2^2 \right)
  \exp\left[ - \frac14 \frac{A-1}A  a^2 \left(q_1^2 + q_2^2 \right) \right],
\end{align}
\label{C-D-16O}%
\end{subequations}
for ${}^{28}\text{Si}$,
\begin{subequations}
\begin{align}
&C_\text{ex}(\bm q_1, \bm q_2) = 
\left( 
28 - 2a^2 \left( 3(\bm q_1+\bm q_2)^2 + 4 \left( \sqrt{5/3} - 1 \right) \bm q_1 \cdot \bm q_2 \right)
+ \frac1{240} a^8 q_1^4 q_2^4 \left(1-3x^2 \right)^2
\right. \notag
\\
& \qquad\qquad\qquad + \left. 
\frac1{15}a^4 \left( 3 \left(q_1^4+q_2^4 \right) + 4\sqrt{15} \bm q_1 \cdot \bm q_2 \left(q_1^2+q_2^2 \right) + q_1^2q_2^2 \left(13 - \sqrt{15} + 3(12 + \sqrt{15})x^2 \right)
\right)\right. \notag \\
& \qquad\qquad\qquad - \left. 
\frac1{30} a^6 q_1^2 q_2^2 \left(
\sqrt{15} \bm q_1 \cdot \bm q_2 \left(3x^2 - 1 \right) + \left(q_1^2 + q_2^2 \right) (3x^2 + 1)
\right)
\right) \exp\left[ - \frac14 \frac{A-1}A a^2 \left(q_1^2 + q_2^2 \right) \right],
\\
&C_0(\bm q_1, \bm q_2) = a^2 q_1 q_2 
\left( 
-\frac43 \left(3 + 2\sqrt{15} \right)x
+ \frac1{105} a^2 \left( 28\sqrt{15} \left(q_1^2 + q_2^2 \right) x - \left( 44 + 7 \sqrt{15} - 21 \left(12 + \sqrt{15} \right) x^2 \right) q_1 q_2
\right) 
\right. \notag \\
&\left. \qquad\qquad\qquad\qquad\qquad
- \frac1{210} a^4 q_1q_2 \left(
7\sqrt{15} \bm q_1 \cdot \bm q_2 \left( 3x^2 - 1 \right) + \left( q_1^2 + q_2^2\right) \left(21x^2 -2 \right)
\right) 
\right. \notag \\ 
&\left. \qquad\qquad\qquad\qquad\qquad
+\frac1{8400} a^6 q_1^3 q_2^3 \left(
23 - 210x^2 + 315x^4 \right)
\right) 
\exp\left[ - \frac14 \frac{A-1}A a^2 \left(q_1^2 + q_2^2 \right) \right],
\end{align}
\label{C-D-28Si}%
\end{subequations}
with $x = \bm q_1 \cdot \bm q_2 / (q_1 q_2)$, and for ${}^{40}\text{Ca}$,
\begin{subequations}
\begin{align}
&C_\text{ex}(\bm q_1, \bm q_2) = \left( 40 - 10 a^2 (\bm q_1 + \bm q_2)^2 + \frac12 a^4 \left((\bm q_1 + \bm q_2)^4 + 10 (\bm q_1 \cdot \bm q_2)^2 \right) 
- \frac12 a^6 (\bm q_1 \cdot \bm q_2)^2 \left(q_1^2 + q_2^2 + \bm q_1 \cdot \bm q_2 \right) 
\right.  \notag\\  
&\qquad\qquad\qquad\left.
+ \frac1{16} a^8 (\bm q_1 \cdot \bm q_2)^4
\right)
  \exp\left[ - \frac14 \frac{A-1}A a^2 \left(q_1^2 + q_2^2 \right) \right],
  \\
&C_0(\bm q_1, \bm q_2) = \left(
    -20 a^2 \bm q_1 \cdot \bm q_2 + \frac12 a^4 
    \left( 4 (\bm q_1 + \bm q_2)^2 \bm q_1 \cdot \bm q_2 + 6 (\bm q_1 \cdot \bm q_2)^2 - 3 q_1^2 q_2^2  \right)
+ \frac1{160} a^8 \left( 10 (\bm q_1 \cdot \bm q_2)^4 - q_1^4 q_2^4
\right)
\right.  \notag\\  
&\qquad\qquad\qquad\left.
- \frac1{8} a^6 \left( 4 ( q_1^2 + q_2^2 + \bm q_1 \cdot \bm q_2) (\bm q_1 \cdot \bm q_2)^2 - (q_1^2 + q_2^2) q_1^2 q_2^2
\right)
\right)
\exp\left[ - \frac14 \frac{A-1}A a^2 \left(q_1^2 + q_2^2 \right) \right].
\end{align}
\label{C-D-40Ca}%
\end{subequations}

By accounting for the difference in normalization conventions, we find that the obtained correlation functions at $\bm q_2 = - \bm q_1$ are consistent with the results reported in Ref.~\cite{murugesu1971optical}.\footnote{We compare $\bm q_2 = - \bm q_1$ instead of $\bm q_2 = \bm q_1$ due to using different Fourier transform definitions with Ref.~\cite{murugesu1971optical}.}

\section{The second-order part of the potential}
\label{sec:2nd-explicit}

The pion angular distribution results from the interplay between partial scattering amplitudes and the Coulomb phase shifts, as is described by Eq.~(\ref{F-l-def}).
In Fig.~\ref{fig-U-l-expansion}, we demonstrate the two lowest partial waves for the first- and second-order on-shell potential for pion scattering on ${}^{12}$C.
As depicted in the plot, the second-order contribution, Eq.~(\ref{V2-final}), substantially affects both $s$ and $p$ waves in the low-energy region.
This impact allows us to generate enough splitting through the nuclear-Coulomb interference to describe both positive and negative pions accurately. 
This was impossible to achieve using only the first-order potential and the same minimal number of free parameters, as discussed in Sec.~\ref{sec:fit}.
Note that the second-order contribution to the potential has the most pronounced effect on the $s$ and $p$ waves, and its influence decreases as the angular momentum increases.

\begin{figure}[!th]
\begin{minipage}[h]{0.49\linewidth}
\center{\includegraphics[width=1\linewidth]{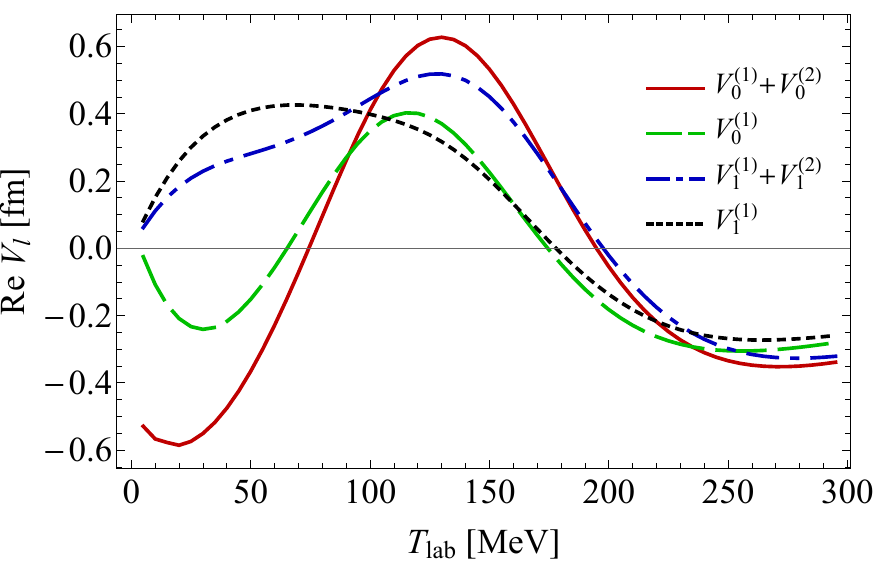}} \\
\end{minipage}
\hfill
\begin{minipage}[h]{0.49\linewidth}
\center{\includegraphics[width=1\linewidth]{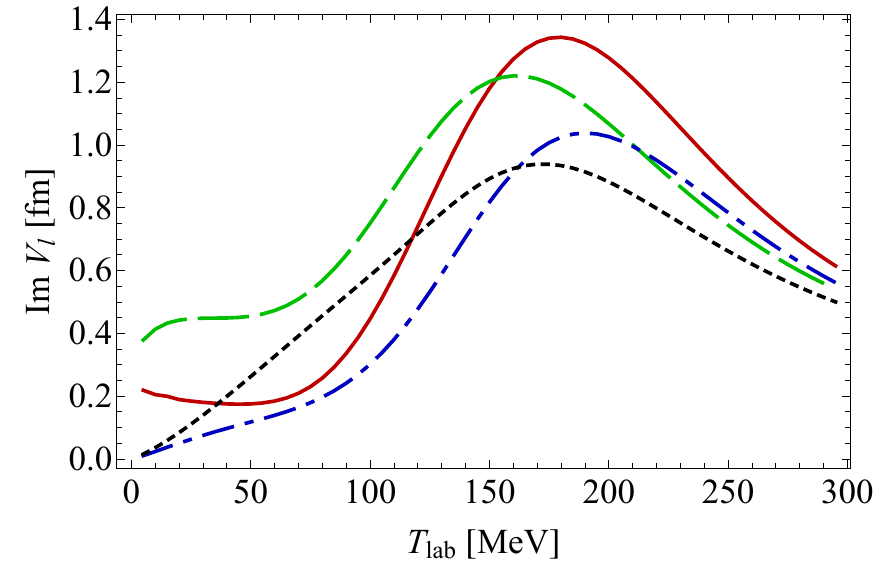}} \\
\end{minipage}
\caption{
The $s$- and $p$-wave components of the first- and second-order potential for on-shell pion scattering on ${}^{12}$C as a function of pion laboratory kinetic energy.
The potential parameters are given by fit~1 in Table~\ref{tabl:fit-Sigma}.
The left and right panels are for real and imaginary parts, respectively.  
The solid red (dot-dashed blue) curve represents the $s$-wave ($p$-wave) component of the full potential, i.e., the sum of Eqs.~(\ref{U1st-fin}) and~(\ref{V2-final}), while the long-dashed green (short-dashed blue) curve corresponds to the first-order potential, Eq.~(\ref{U1st-fin}).
}
\label{fig-U-l-expansion}
\end{figure}

To derive the explicit form of the second-order part of the pion-nucleus potential, we consider the $s$- and $p$-wave pion-nucleon scattering amplitude, Eq.~(\ref{f-low-energy}), with the off-shell momentum dependence given by Eq.~(\ref{f-off-shell}).
The unit vectors normal to the pion-nucleon scattering planes entering Eqs.~(\ref{U2-isospin-struct})–(\ref{V2-final}) are
$\bm n_1 = (\bm k_\text{2cm} \times \bm k_\text{2cm}'')/|\bm k_\text{2cm} \times \bm k_\text{2cm}''|$ and $\bm n_2 = (\bm k_{\text{2cm}'}'' \times \bm k_{\text{2cm}'}')/|\bm k_{\text{2cm}'}'' \times \bm k_{\text{2cm}'}'|$. 
The subscript "2cm" ("$\text{2cm}'$") corresponds to the c.m. system of the pion and the first (second) nucleon.  
Collecting all the components, the second-order part of the potential, Eq.~(\ref{V2-final}), can be written as a sum of four terms:
\be
V^{(2)}(\bm k' ,\bm k) 
= 
V_{ss}+ V_{sp} +  V_{pp} + V_{pp}^{(s)}, 
\label{U2nd-sum}
\ee
where
\begin{subequations}
\begin{align}
&V_{ss} =  \int \frac{\diff \bm k^{\prime\prime}}{2\pi^2}
\tilde{\mathscr{W}}(\bm k', \bm k'') \tilde{\mathscr{W}}(\bm k'', \bm k) 
\frac{1}{k_0^2 - {k^{\prime\prime}}^2+ i\varepsilon} 
\left[ 
b_0^2 C_0(\bm k^\prime - \bm k^{\prime\prime},  \bm k^{\prime\prime} - \bm k) +2 b_1^2 C_\text{ex}(\bm k^\prime - \bm k^{\prime\prime},  \bm k^{\prime\prime} - \bm k)
\right],
\label{Uopt-2nd-ss}
\\
&V_{sp} = \int \frac{\diff \bm k^{\prime\prime}}{2\pi^2}
\tilde{\mathscr{W}}(\bm k', \bm k'') \tilde{\mathscr{W}}(\bm k'', \bm k)
\frac{\bm k'_{\text{2cm}'} \cdot \bm k''_{\text{2cm}'} + \bm k_\text{2cm} \cdot \bm k''_\text{2cm}}{k_0^2 - {k^{\prime\prime}}^2+ i\varepsilon} \left[ 
b_0 c_0 C_0(\bm k^\prime - \bm k^{\prime\prime},  \bm k^{\prime\prime} - \bm k) +2 b_1 c_1 C_\text{ex}(\bm k^\prime - \bm k^{\prime\prime},  \bm k^{\prime\prime} - \bm k)
\right],
\label{Uopt-2nd-sp}
\\
&V_{pp} =  \int \frac{\diff \bm k^{\prime\prime}}{2\pi^2}
\tilde{\mathscr{W}}(\bm k', \bm k'') \tilde{\mathscr{W}}(\bm k'', \bm k)
\frac{(\bm k'_{\text{2cm}'} \cdot \bm k''_{\text{2cm}'}) (\bm k''_\text{2cm} \cdot \bm k_\text{2cm})}{k_0^2 - {k^{\prime\prime}}^2+ i\varepsilon} 
\left[ 
c_0^2 C_0(\bm k^\prime - \bm k^{\prime\prime},  \bm k^{\prime\prime} - \bm k) +2 c_1^2 C_\text{ex}(\bm k^\prime - \bm k^{\prime\prime},  \bm k^{\prime\prime} - \bm k)
\right],
\label{Uopt-2nd-pp}
\\
&V_{pp}^{(s)} = - \int \frac{\diff \bm k^{\prime\prime}}{2\pi^2} 
\tilde{\mathscr{W}}(\bm k', \bm k'') \tilde{\mathscr{W}}(\bm k'', \bm k)
\frac{[\bm k'_{\text{2cm}'} \times \bm k''_{\text{2cm}'}] \cdot [\bm k''_\text{2cm} \times \bm k_\text{2cm}]}{k_0^2 - {k^{\prime\prime}}^2+ i\varepsilon} 
\left[ 
s_0^2 +2 s_1^2 
\right]C_\text{ex}(\bm k^\prime - \bm k^{\prime\prime},  \bm k^{\prime\prime} - \bm k).
\label{Uopt-2nd-pp-spin}
\end{align}
\label{Uopt-2nd-comp}
\end{subequations}

\begin{figure}[!th]
\begin{minipage}[h]{0.49\linewidth}
\center{\includegraphics[width=1\linewidth]{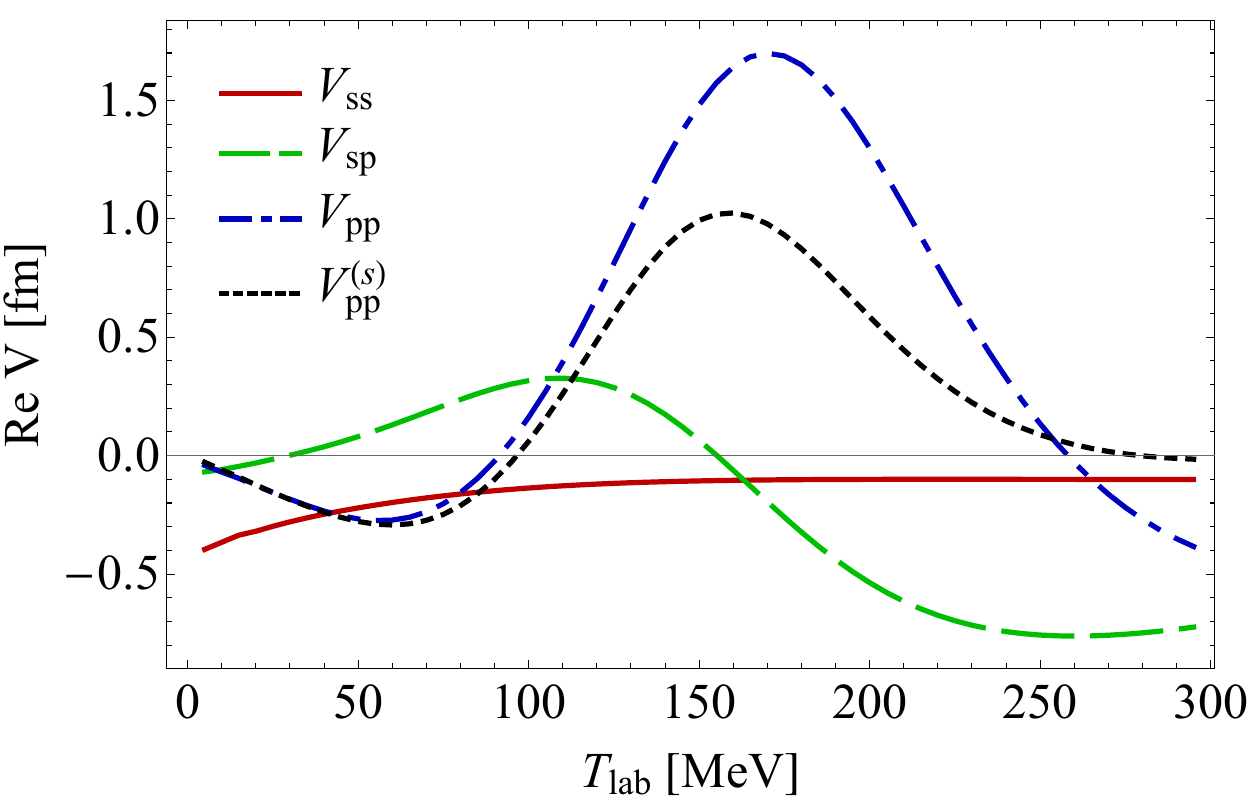}} \\
\end{minipage}
\hfill
\begin{minipage}[h]{0.49\linewidth}
\center{\includegraphics[width=1\linewidth]{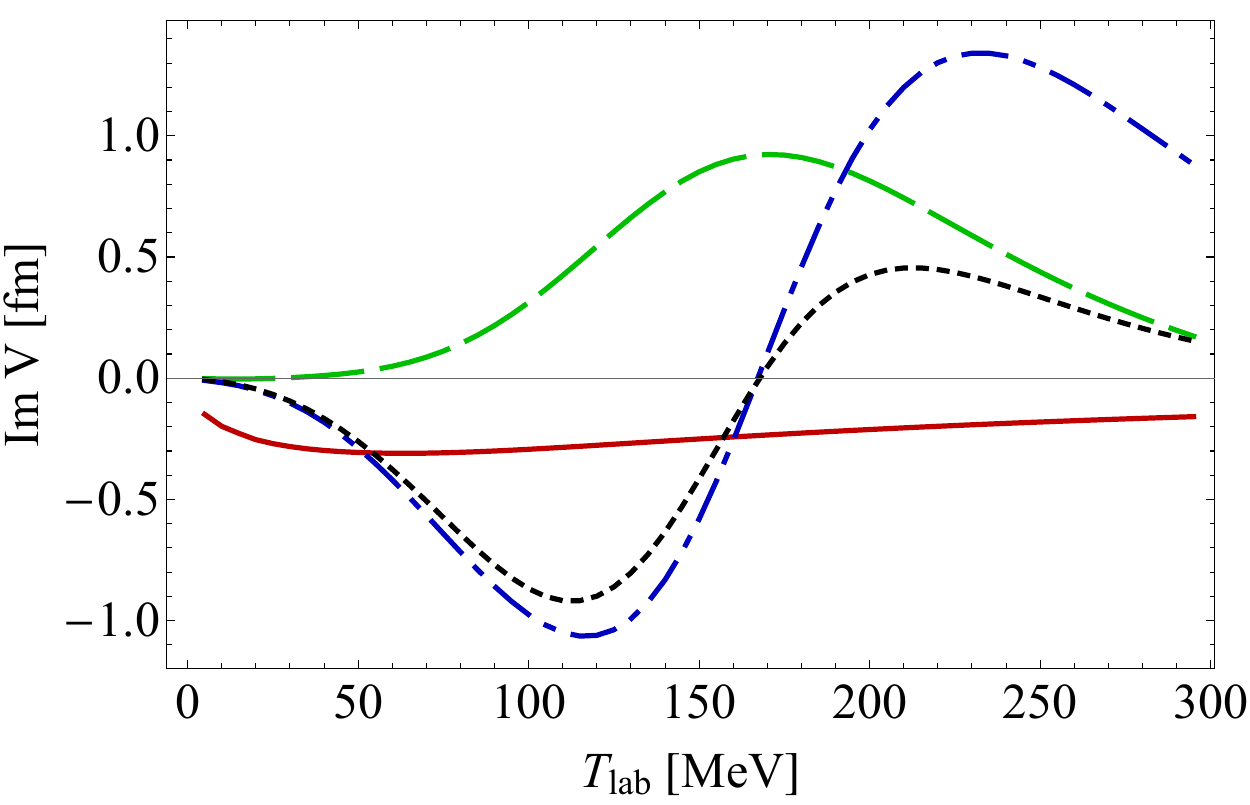}} \\
\end{minipage}
\caption{
The components of the on-shell forward pion-nucleus potential, Eqs.~(\ref{Uopt-2nd-comp}), for ${}^{12}$C as a function of pion laboratory kinetic energy for parameters given by fit~1 in Table~\ref{tabl:fit-Sigma}. 
The left and right panels are for real and imaginary parts, respectively.  
The solid red, dashed green, and dash-dotted blue curves correspond to the second-order $s$-$s$-, $s$-$p$-, and $p$-$p$-wave  interference of the spin-independent pion scattering, Eqs.~(\ref{Uopt-2nd-ss})–(\ref{Uopt-2nd-pp}).
The short-dashed curves represent the contribution from the spin-dependent part of the pion-nucleon amplitude,  Eq.~(\ref{Uopt-2nd-pp-spin}).
 }
\label{fig-Uss-sp-pp-comp}
\end{figure}

\end{widetext}

Each second-order contribution described in Eqs.~(\ref{Uopt-2nd-comp}) represents the interference between the $s$- and $p$-wave parts of the pion-nucleon amplitude.
Figure~\ref{fig-Uss-sp-pp-comp} demonstrates the second-order components for on-shell forward scattering on ${}^{12}$C. 
Generally, the scattering parameters $b_{0,1}$, $c_{0,1}$, and $s_{0,1}$ in Eqs.~(\ref{Uopt-2nd-comp}) depend modestly on the angle between the corresponding momenta. 
For the purpose of evaluating the second-order correction, we assume these parameters to be angle independent and fixed at the forward scattering angle.

The peculiarity of our approach is the presence of two correlation functions in the second order.
However, in the $s$-$s$-wave interference term $V_{ss}$, Eq.~(\ref{Uopt-2nd-ss}), the first term with the $C_0$ correlation function is negligible due to the smallness of $b_0$ compared to the real part of $b_1$.
This enables us to compare our approach with the $s$-wave potential originally derived in Ref.~\cite{Ericson:1966fm}.
With the second-order correction, the $s$-wave coordinate space potential given by Eq.~(\ref{U-absorption}) acquires the form
\be
U^{(s)}(r) \propto \left( b_0 - \left(b_0^2 + 2 b_1^2 \right) \left\langle \frac1r \right\rangle  \right) \rho(r) + B_0 \rho^2(r),
\label{U(r)-2nd}
\ee
where $\langle 1/r\rangle$ is the so-called inverse nucleon correlation length, which within the Fermi gas model for zero pion kinetic energy becomes
$
\left\langle 1/r \right\rangle = {3p_F}/({2\pi}) \approx \SI{0.65}{fm^{-1}}.
$
Performing the integration in Eq.~(\ref{Uopt-2nd-ss}) in the limit $k_0 \rightarrow 0$, we obtain $V_{ss}(0,0) = 2b_1^2 \langle C_\text{ex} \rangle$, with $\langle C_\text{ex} \rangle / A$ acquiring the values $\SI{0.61}{fm^{-1}}$ and $\SI{0.56}{fm^{-1}}$ for ${}^{12}$C and ${}^{40}$Ca, respectively.
The approximate agreement between $V_{ss}$ and $U^{(s)}(r)$ at the threshold allows us to directly apply the results of the pionic atom analyses in Secs.~\ref{subsec-b0} and~\ref{subsec-b1}.

The $p$-$p$-wave interference term $V_{pp}$, Eq.~(\ref{Uopt-2nd-pp}), corresponds to the second-order term,
\be
U_{pp}(\bm r) \propto - \frac13 \frac{A-1}A \left( 4\pi c_0 \right)^2  \bm \nabla \rho^2(r) \bm \nabla,
\label{U(r)-LLEE-2nd}
\ee
in the coordinate space $p$-wave  potential describing the Lorentz-Lorenz-Ericson-Ericson effect~\cite{Ericson:1966fm}:
\be
U^{(p)}(\bm r) \propto  \bm \nabla \frac{c_0 \rho(r)}{1 + \frac{4\pi}3  \frac{A-1}A  c_0 \rho(r)} \bm \nabla .
\label{U(r)-LLEE}
\ee
The kinematic factors are omitted in Eqs.~(\ref{U(r)-LLEE-2nd}) and~(\ref{U(r)-LLEE}) for simplicity.
While our model does not account for effects beyond second order, unlike Eq.~(\ref{U(r)-LLEE}), we expect $V_{pp}$ to be much more realistic. 
The reason for this is that Eq.~(\ref{U(r)-LLEE-2nd}) is obtained from Eq.~(\ref{Uopt-2nd-pp}) in the limit of zero pion kinetic energy by setting $C_\text{ex}(\bm q_1, \bm q_2) = 0$ and $C_0(\bm q_1, \bm q_2) = \rho(q_1) \rho(q_2)$, which may be a crude approximation in the resonance energy region.

The term $V_{sp}$, Eq.~(\ref{Uopt-2nd-sp}), characterizes the $s$-$p$-wave interference.
It is nonzero in our approach since we perform the computation within the nuclear shell model without resorting to the Fermi gas model.
As seen from Fig.~\ref{fig-Uss-sp-pp-comp}, this term is not negligible and is important both at high and low energies.
Similarly to the case of $V_{ss}$, the term proportional to $C_0$ gives a much smaller contribution due to the ratio of $b_0$ and $b_1$.

The term $V_{pp}$ describes $p$-$p$-wave interference accounting for processes with (term proportional to $C_\text{ex}$) and without (term proportional to  $\propto C_0$) the isospin exchange.
Similarly, the term $V_{pp}^{(s)}$, Eq.~(\ref{Uopt-2nd-pp-spin}), characterizes the spin exchange.
This term has similar energy dependence as $V_{pp}$ (see Fig.~\ref{fig-Uss-sp-pp-comp}), because both $c_{0,1}$ and $s_{0,1}$ are proportional to the $P^1_{33}$ partial amplitude.
However, $V_{pp}$ and $V_{pp}^{(s)}$ have different angle-dependent structure.

\bibliographystyle{apsrevM}
\bibliography{Bibliography}

\end{document}